\documentclass[aps,prx,reprint,twocolumn,superscriptaddress,longbibliography]{revtex4-2}

\usepackage[T1]{fontenc}
\usepackage[utf8]{inputenc}
\usepackage{amsmath,amssymb,amsthm,bm,mathtools}
\usepackage{graphicx}
\usepackage{booktabs}
\usepackage{cancel}
\usepackage[hidelinks,hypertexnames=false]{hyperref}
\usepackage{xcolor}
\usepackage{float}
\usepackage{tikz}
\usetikzlibrary{shapes,arrows.meta,positioning,fit,backgrounds,calc}
\graphicspath{{../../figures/}{./figures/}{./tikz/}}

\begin{document}

\title{The Phases of Hierarchy}

\author{Gerard McCaul}
\email{eigen@gmccaul.co.uk}
\affiliation{EigenEngine}
\date{\today}

\begin{abstract}We develop a dynamical phase theory of hierarchy, through which the stability and responsiveness of a hierarchical adaptive system is classified.  To do so, we develop a class of models we term \emph{adaptive normal forms}, where a global macroscale dynamics is coupled to a population of stochastic agents. Each system is parametrised by a function of the other, forming a dynamic feedback loop.~Electorates, markets and institutions may all be described by such a loop, where a population compares the state of its environment with its expectations, and its collective verdict revises the rules that produce the next state. We show that the same statistics responsible for `the wisdom of the crowd' also supplies a loop gain in an adaptive system that determines its dynamic stability. From this we derive the critical gain determining dynamical stability analytically, and demonstrate its dependence on the population size $N$. When this critical gain is exceeded, the system loses stability through a flip bifurcation. We then show that this instability can be avoided by hierarchialising the system, with a required depth that scales with $\log N$, providing a dynamical justification to the logarithmic depth bounds employed in organisational economics. We further demonstrate that hierarchialisation is bought at the cost of the normal form's responsiveness, as signals travelling up the hierarchy are suppressed exponentially with layer depth.  Treating each layer as a real-space renormalisation step then leads to a general classification scheme for the dynamical phases of a hierarchical adaptive system. Through this, it is possible to draw connections between adaptive normal forms and machine learning architectures such as Transformers and generative diffusion. Finally, we show the predicted phase transitions can be observed at finite sampling, and propose measurements that could place a real adaptive system in its phase.\end{abstract}

\maketitle

\section{Introduction: Hierarchy as a Phase Variable}

Hierarchies pervade nature and society. From the fractal branching of cardiovascular networks~\cite{West1997} to the nested trophic levels of ecological food webs~\cite{Pimm1982}, multiscale organisation is the default architecture of natural systems. This structural mandate holds equally true for the divisional structures of the modern corporate firm~\cite{Chandler1962} and the layers of state bureaucracies~\cite{Weber1922}. More recently, layered architectures have become central to artificial intelligence, where the organisation of deep networks affects how signals and gradients propagate~\cite{He2016, Liu2020}. The representations built by these layers can themselves organise hierarchically~\cite{Doimo2020Nucleation}. 

Despite their ubiquity, hierarchies in adaptive systems are typically treated as a fixed architectural fact \cite{Simon1962} or an exogenous institutional constraint \cite{Tsebelis2002, North1990}. In organisational economics, bounded parallel information processing imposes a minimum depth $L_{\min}\sim\log N$ \cite{Radner1993, Bolton1994, Garicano2000} on hierarchy, while sociophysics has investigated its role in bias amplification and polarisation \cite{Galam1986, Galam1990}. Such formulations obscure an underlying \emph{dynamic} instability however - when a lower-level population reacts to a shifting environment, its aggregate response scales non-linearly and sharpens. If this collective signal is allowed to directly reparametrise higher-level constraints without mediation, large populations induce expectation cycling across the system's phase space \cite{Holling1973, Scheffer2009, Goodwin1967, Ezekiel1938}. These considerations identify hierarchy not merely as an architectural feature for managing complexity, but as a critical phase variable regulating feedback.

Here we develop this idea formally, and demonstrate that a hierarchical adaptive system possesses three distinct dynamical phases (Fig.~\ref{fig:phases}) governing both its stability and response to change. To do so, we first develop in Sec.~\ref{sec:normal_form} a formal model of adaptive systems we term an \emph{adaptive normal form}. In Section \ref{sec:gain} we demonstrate that majority-like aggregation creates a $\sqrt{N}$ gain [Eq. \eqref{eq:gain-population}] in system response, which raises the susceptibility of the system to perturbation. If this gain crosses the boundary in Eq.~\eqref{eq:critical-gain}, the system undergoes what we term an \textit{ultraepistemic catastrophe (UEC)}, a closed-loop phase in which epistemic aggregation improves local discrimination while destroying global dynamical stability. This is structurally analogous to the ultraviolet catastrophe in physics, where a locally reasonable rule for the partition of energy became pathological when summed over too many modes. Here, hierarchical structures collapse because locally useful aggregation exhibits the same pathology when inserted into an adaptive feedback loop with too many effectively independent agents. 

We then demonstrate in Sec.~\ref{sec:hierarchy} that hierarchialisation can attenuate this gain, and that stability can be restored to the system with a hierarchical depth that scales with $\log N$, rescuing the system from the UEC. This is not without cost however. Section \ref{sec:inertia} models these costs, and demonstrates that excess hierarchy leads to a third \emph{inertial} phase, characterised by delay and signal attenuation, in which the system loses its adaptive capacity. Section \ref{sec:rg_universality} establishes the full breadth of this result by linking the adaptive normal form to the real space renormalisation group. This is then exploited both to produce a general classification of the stability of hierarchies under different modes of aggregation, and establish a structural correspondence to architectures in machine learning. Finally, Section \ref{sec:empirical} tests these predictions under finite sampling and outlines the empirical signatures required to track this transition in real systems, before Section \ref{sec:conclusion} concludes.

\section{The Normal Form}
\label{sec:normal_form}

\begin{figure*}[t]
    \centering
    \includegraphics[width=\textwidth]{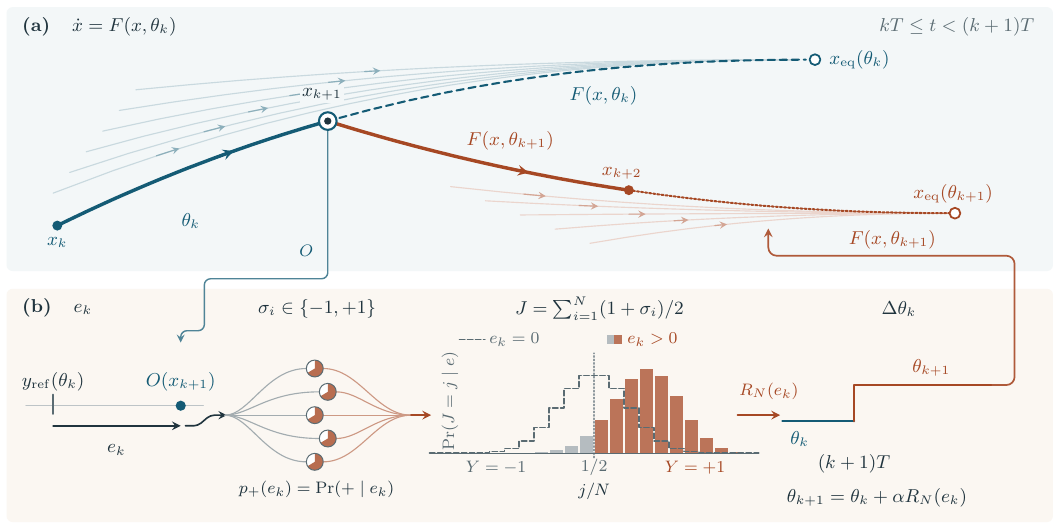}
    \caption{\textbf{A mismatch changes the distribution of the collective response.} (a) Retaining the parameter at $x_{k+1}$ would continue the dashed trajectory. A revision changes the field without displacing the state. Open circles mark the equilibria approached if the respective parameter values are held. (b) In the majority example, each disc shows the probability of a positive response (filled sector), $p_+(e_k)=[1+\tanh(\beta e_k)]/2$, where the signed distance from the reference mark to the observed point is $e_k=O(x_{k+1})-y_{\rm ref}(\theta_k)$. The positive count $J$ among $N$ independent agents is then binomial. At the same odd $N$, compare the distribution of the fraction of positive responses for zero mismatch (dashed outline) and positive mismatch (bars). Probability moves from the negative-majority side to the positive-majority side. Their probability difference, positive minus negative, is $R_N(e_k)$. The fraction has mean $p_+$ and standard deviation $\sqrt{p_+(1-p_+)/N}$: the same bias is more decisive in a larger population. The resulting parameter increment is applied at the census point $(k+1)T$.}
    \label{fig:adaptive-normal-form}\label{fig:equation_schematic}\label{fig:schematic}
\end{figure*}

We begin by outlining the model from which we draw our results, which we term an \emph{adaptive normal form}. In it, a population of $N$ agents is coupled to the macroscopic system whose behaviour they experience. The agent population then exerts its own influence by periodically updating the parameterisation of the dynamical generator of the macroscopic system. In return, the macroscopic system acts on the agent population through an observable of its state. In terms of physics, we therefore have a discrete system whose generator depends on the evolution of a functional, which itself depends on an order parameter of the discrete system. This kind of abstract dynamical pairing can be made concrete by example. For instance, a public mandate and a policy state can be described in this manner - the mandate sets the conditions under which policy evolves, while the outcome of that policy informs the next mandate. Figure~\ref{fig:adaptive-normal-form} follows one circuit of this feedback loop.

We call the instants $t_k=kT$ \emph{census points}: at each one the lower-scale population aggregates its responses and delivers a mandate for the next interval. The census interval $T$ is the time for which the mandate remains in force. Throughout $t_k\leq t<t_{k+1}$ the control parameter $\theta_k$ is held fixed and the macroscopic state vector $\mathbf x$ evolves continuously according to $\dot{\mathbf x}=\mathbf F(\mathbf x,\theta_k)$. During that interval the state is allowed to equilibrate. By this we mean that if the parameter were left in place, the state would approach an attracting operating point $\mathbf x_{\rm eq}(\theta_k)$. This is a weak assumption that a steady state for the unperturbed dynamics at fixed mandate exists, but we do not require the system to have relaxed to this state before the next update. 

At the next census point, $t_{k+1}$, the population observes $O(\mathbf x_{k+1})$, where $\mathbf x_k=\mathbf x(t_k)$. Such a reading in isolation does little however. The behaviour of the lower-scale population will change based on its \emph{expectation} of the dynamics.  We therefore write $y_{\rm ref}(\theta_k)$ for the reference reading against which the population judges the outcome of the previous period of macroscale dynamics. If the mandate is itself a forecast in the units of the observable, then $y_{\rm ref}(\theta_k)=\theta_k$. More generally the reference is specified by the response rule, not by the equilibrium of the macroscopic dynamics. This is where different models of aggregation enter the system. If $Y$ is the signal returned by aggregating the $N$ individual responses, then we write $R_N(e)=\mathbb{E}[Y\mid e]$ for its mean at a given mismatch $e$. Taking the example of majority aggregation, $Y$ is simply a sign, so $R_N$ is the probability of a positive majority minus the probability of a negative one. 

Let $\Phi_T$ be the flow over one census interval at fixed mandate. At $t_{k+1}$ the population responds to the state reached under $\theta_k$ and delivers the next mandate $\theta_{k+1}$. The complete system is
\begin{align}
    \mathbf x_{k+1} &= \Phi_T(\mathbf x_k, \theta_k), \label{eq:normal_x} \\
    e_k &= O(\mathbf x_{k+1}) - y_{\rm ref}(\theta_k), \label{eq:normal_mismatch} \\
    \theta_{k+1} &= \theta_k + \alpha R_N(e_k). \label{eq:normal_theta}
\end{align}
The index on $e_k$ identifies the mandate being evaluated: the reading is taken at $t_{k+1}$, before $\theta_k$ is replaced. The slope of $R_N$ fixes how strongly the aggregate responds near zero mismatch, while $\alpha$ is a unit conversion of the response into a parameter increment. If only one collective signal is sampled at each census, the increment then fluctuates about $\alpha R_N(e_k)$, with probabilities given for majority aggregation in Appendix~\ref{app:majority}. We call Eqs. \eqref{eq:normal_x}--\eqref{eq:normal_theta} an \emph{adaptive normal form} because they capture the essential characteristics of a multiscale adaptive system where feedback forms a closed loop.

Table~\ref{tab:domains} gives several candidate identifications for the formal elements of the normal form. The last column distinguishes types of aggregation, between a threshold decision (such as majority rule) and a proportional response. Both can close the loop, but need not result in the same gain. This matters because averaging and thresholding respond differently to population size - averaging more independent inputs improves the precision of the mean without changing its expectation, while taking the sign of their sum instead makes a small bias more decisive. It is this sharpening of the collective response that the model must capture.

\begin{table*}[t!]
\centering
\caption{Domains of periodic reparametrisation}
\label{tab:domains}
\small
\setlength{\tabcolsep}{4pt}
\begin{tabular}{@{}lllll@{}}
\toprule
\textbf{Domain} & \textbf{$\theta$ (Lower scale)} & \textbf{$x$ (Higher scale)} & \textbf{Mismatch} & \textbf{Threshold?} \\
\midrule
Democracy & Public mandate & Policy state & Experienced policy mismatch & Yes \\
Platforms & Model/user expectation & Recommendation environment & Behavioural shift & Yes \\
Organisations & Target regime & Operational state & Target-performance error & Yes \\
Ecology & Trait/niche expectation & Niche/selection state & Realised fitness mismatch & No \\
\bottomrule
\end{tabular}
\end{table*}

These mechanisms capture well-documented phenomena across fields, ranging from macroeconomics \cite{Muth1961, EvansHonkapohja2001, Mankiw2002, Carroll2003, Coibion2018, Sims2003, Woodford2002, Malmendier2016}, democratic societies \cite{Wlezien1995, Stimson1995, Ferejohn1986}, AI model prediction and collapse \cite{Perdomo2020, Shumailov2024}, cybernetics \cite{Beer1972, ConantAshby1970}, and ecology \cite{OdlingSmee1996, Lenton2008, Boettiger2012, Clements2016}. In this last context, the loop explicitly formalises rapid eco-evolutionary feedback \cite{Fussmann2007}, where trait adaptation at the lower level destabilises the slowly shifting environmental niche at the higher level. Despite the differing underlying mechanisms - whether voters punishing politicians, agents learning inflation, or populations adapting traits - the mathematical structure of the closed loop is identical. In all cases, we shall show stability of the system depends profoundly on how the aggregate response $R_N$ scales with the population $N$.

In the main text we will illustrate our results with the following scalar model. The mandate $\theta_k$ is held fixed over each census interval, and we allow the scalar state $x$ (the one-component case of $\mathbf x$ in Eq.~\eqref{eq:normal_x}) to relax towards $s\theta_k$ on a time scale $\tau$. The population then judges the sampled state against $b\theta_k$, with $b>s>0$. Both $x$ and $\theta$ are deviations from a baseline operating point, so the matched state sits at the origin, and $b>s$ means the state delivers less than the population expects of it (in the inflation example of Sec. VII, this is incomplete pass-through). To specify how this mismatch changes a verdict, take $N$ odd and let the signs $\sigma_i\in\{-1,+1\}$ be independent conditional on $e$ at each census. Writing $p_+(e)$ for the probability of a positive sign, we employ a logistic model,
\begin{equation}
 p_+(e)=\Pr(\sigma_i=+1\mid e)=\frac{1}{1+\exp(-2\beta e)},\qquad \beta>0.
 \label{eq:scalar-sign-probability}
\end{equation}
With this choice either sign is equally likely at zero mismatch. The sensitivity $\beta$ controls how rapidly that balance shifts with $e$. Since each sign has unit magnitude, its conditional mean and variance are
\begin{align}
 m(e)&=\mathbb E[\sigma_i\mid e]=2p_+(e)-1=\tanh(\beta e),\label{eq:agent-mean}\\
 \operatorname{Var}(\sigma_i\mid e)&=1-m(e)^2=4p_+(e)[1-p_+(e)].\label{eq:agent-variance}
\end{align}
Independence gives a binomial count of positive signs, with mean $Np_+(e)$ and variance $Np_+(e)[1-p_+(e)]$. The majority responds to whether the count is above or below $N/2$. Using its expected sign for the mandate revision gives the full census map:
\begin{align}
 x_{k+1}&=r x_k+(1-r)s\theta_k, \qquad r=e^{-T/\tau},\label{eq:held-scalar-state}\\
 e_k&=x_{k+1}-b\theta_k,\label{eq:scalar-mismatch}\\
 \theta_{k+1}&=\theta_k+\alpha R_N(e_k), \qquad \alpha>0,\label{eq:scalar-revision}\\
 R_N(e)&=\mathbb{E}\!\left[\operatorname{sgn}\!\left(\sum_{i=1}^N\sigma_i\right)\middle|\,e\right].\label{eq:mean-majority}
\end{align}
Finally, it is useful to describe the behaviour of adaptive systems in terms of their \emph{loop gain} $K_N$. A small mismatch $e$ produces a revision of size $K_N e$ to first order, where
\begin{equation}
 K_N=\alpha R_N'(0).
 \label{eq:loop-gain}
\end{equation}
Many elaborations to this model are possible, but our claim is that the distinct phases exhibited by adaptive systems are \emph{universal}. That is, all the truly essential features of adaptive systems are captured in this construction. To support this, the technical appendices demonstrate how various generalisations affect the results to follow. In Appendix~\ref{app:majority} we allow all agents to share a common input on a fixed fraction of censuses. We allow the lower scale population to feel a continuous influence in Appendix~\ref{app:scalar_boundary}, and consider a system with multiple dynamical modes in Appendix~\ref{app:robustness}. With this, we are ready to establish the quantity all three calculations require, the \emph{critical loop gain} $K_c$.

\section{Result I: Aggregation Drives Cycling}
\label{sec:gain}\label{sec:scalar}

The collective response of a large population is not simply a smoothed version of an individual's response. Under majority-like aggregation rules, increasing the population size fundamentally changes the shape of the update map \cite{Granovetter1978, Schelling1971, HolmeNewman2006}. Consider a population of $N$ individuals, each receiving the same broadcast signal or mismatch $e_k = O(x_{k+1}) - y_{\rm ref}(\theta_k)$ (e.g., the difference between realised inflation and their prior expectation) [Eq.~\eqref{eq:normal_mismatch}]. If individuals signal or act probabilistically based on this mismatch with the sensitivity $\beta$ of Eq.~\eqref{eq:scalar-sign-probability}, their aggregate response can be modelled as the expected signed majority. For independent binary responses, the slope in Eq.~\eqref{eq:majority-influence} and the gain definition in Eq.~\eqref{eq:loop-gain} give
\begin{equation}
    K_N \sim \alpha \beta \sqrt{\frac{2N}{\pi}}.\label{eq:gain-population}
\end{equation}

Larger populations fundamentally sharpen the collective reparametrisation map under majority-like aggregation, steepening the response curve near the origin. The exact same statistical scaling that grants large populations collective accuracy (such as in the Condorcet Jury Theorem \cite{Arrow1951, Estlund2008, Landemore2012, ListGoodin2001, Ladha1992, Grofman1983, AustenSmith1996, Kao2014, Conradt2003, Sumpter2010}) also systematically increases the loop gain. As $N$ grows, the system is driven towards over-responsiveness, predisposing it to instability unless the higher-level dynamics can relax fast enough to absorb the sharp correction. Consequently, consensus mechanisms designed to filter individual noise simultaneously amplify the macroscopic feedback gain.

\begin{figure*}[t]
    \centering
    \includegraphics[width=\linewidth]{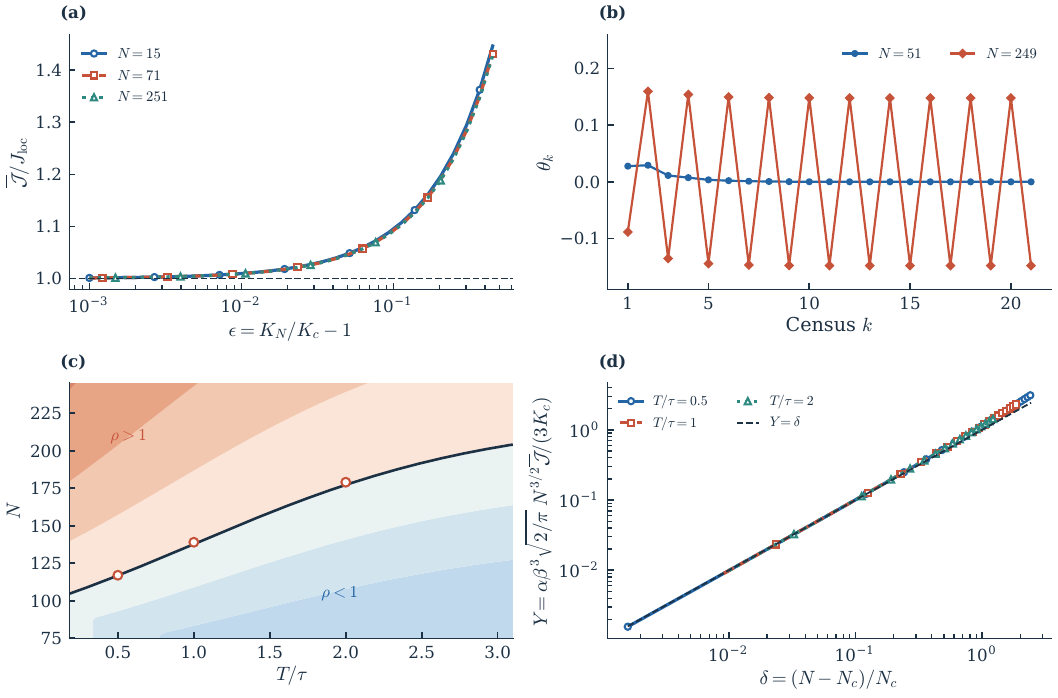}
    \caption{\textbf{Loss at the onset of alternating mandates.} (a) Varying $\alpha$ at fixed $N$ and $T=\tau$ brings the mean squared mismatch towards $J_{\rm loc}=(K_N-K_c)/(\alpha c_N)$ as $\epsilon=K_N/K_c-1$ approaches zero from above. Appendix~\ref{app:majority}, Eq.~\eqref{eq:majority-cubic}, gives the cubic response coefficient $c_N$. Appendix~\ref{app:scalar_boundary}, Eq.~\eqref{eq:scalar-local-loss}, gives the loss law. (b) Below threshold the mandate decays towards zero. Above it the mandate alternates. (c) The matrix $J$ in Eq.~\eqref{eq:scalar-census-jacobian} advances a small disturbance by one census. Its spectral radius $\rho(J)$ [Eq.~\eqref{eq:census-spectral-radius}] is the largest absolute value of its eigenvalues. The contour $\rho(J)=1$ marks onset. Open circles locate it independently from trajectories (Appendix~\ref{app:simulations}). (d) At fixed $\alpha=0.35$ we vary the odd population size. The loss averages the last 2,000 of 24,000 censuses. The reduced population $\delta$ uses $N_c$ interpolated between the two exact odd-$N$ gains bracketing $K_c$. The dashed line is the leading prediction $Y=\delta$ for the rescaled loss.}
    \label{fig:scalar_phase}
\end{figure*}

By formalising the structural loop into a discrete-time scalar model of state relaxation and threshold error observation, it is possible to isolate the threshold of macroscopic divergence. In particular, it is possible to show there exists a \emph{critical gain} $K_c$ at which the mandate reverses at every census, and the system becomes unstable. For the scalar census map in Eqs.~\eqref{eq:held-scalar-state}--\eqref{eq:mean-majority}, Appendix \ref{app:scalar_boundary} derives this critical gain as
\begin{equation}
    0 < K_N < K_c(T) = \frac{2(1+e^{-T/\tau})}{(1+e^{-T/\tau})b - (1-e^{-T/\tau})s}.\label{eq:critical-gain}
\end{equation}
Figure~\ref{fig:scalar_phase}(b) shows the effect of this critical gain. At $T=\tau$ the disturbance dies away for $N=51$, whereas for $N=249$ the mandate settles into a two-cycle. Panel (c) follows the same boundary as the census interval changes. The onset sizes measured from trajectories agree with those predicted by the gain threshold, and under a large-population approximation we obtain a \emph{critical population}:
\begin{equation}
    N_c(T) \simeq \frac{\pi}{2\alpha^2\beta^2} K_c(T)^2.\label{eq:critical-size}
\end{equation}

Longer update periods $T$ strictly increase the stable parameter volume, i.e., $d K_c / d T > 0$ (see Appendix \ref{app:scalar_boundary} for proof). In practical terms, slowing the update cycle restabilises the system. For instance, decreasing the frequency of elections or extending the term limits of representatives allows the institutional state to relax before the next burst of aggregate feedback, pulling the system back inside the stability boundary without needing to artificially filter the signal itself.

If the population $N$ exceeds $N_c(T)$, the fixed point undergoes a flip bifurcation, and the system enters an unstable phase characterised by expectation cycling \cite{BrockHommes1997, BrockHommes1998, Nerlove1958}. In this unstable regime, the persistent cyclic mismatch generates an asymptotic mean loss $\overline{\mathcal{J}}$. This loss is defined as the long-time average of the squared mismatch from Eq.~\eqref{eq:scalar-mismatch}, measured before the mandate changes:
\begin{equation}
 \overline{\mathcal J}=\lim_{n\to\infty}\frac{1}{n}\sum_{k=0}^{n-1}e_k^2 .
 \label{eq:census-loss}
\end{equation}
By Eq.~\eqref{eq:scalar-local-loss}, just above the critical threshold the loss turns on continuously, scaling with the positive part of the excess gain, $(q)_+=\max(q,0)$:
\begin{equation}
    \overline{\mathcal{J}} \propto (K_N - K_c)_+.\label{eq:excess-loss}
\end{equation}
With $\alpha$ and $T$ fixed, let $\delta=(N-N_c)/N_c$. The majority gain grows as $\sqrt N$, so $K_N/K_c-1\simeq\sqrt{1+\delta}-1\simeq\delta/2$ near onset in the large-population approximation. The loss law in Eq.~\eqref{eq:scalar-local-loss} then gives a cyclic mismatch amplitude $\sqrt{\overline{\mathcal J}}\propto (N-N_c)^{1/2}$ just above threshold. Its prefactor depends on $N$ because the cubic coefficient grows as $c_N\simeq\beta^3\sqrt{2/\pi}\,N^{3/2}/6$ [Eq.~\eqref{eq:majority-cubic}]. Combining these gives the near-onset scaling
\begin{equation}
    Y\equiv\frac{\alpha\sqrt{2/\pi}}{3K_c(T)}\,\beta^3 N^{3/2}\overline{\mathcal J}\simeq \delta,\qquad 0<\delta\ll1.\label{eq:reduced-population-loss}
\end{equation}
Figure~\ref{fig:scalar_phase}(d) tests this population scaling at three census intervals, where the ten sampled points with $0<\delta<0.05$ agree with the predicted line to within $1.1\%$. Panel (a) approaches the same onset continuously by varying $\alpha$ at fixed $N$. Panel (c) gives the \emph{phase diagram} for stability as $T$ and $N$ are varied. Below the contour the system relaxes to a stable state, whereas above it the gain exceeds criticality and a period-two limit cycle occurs. Two examples of the mandate $\theta_k$ above and below $N_c$ are plotted in panel (b), illustrating the behaviour of each regime.

\section{Result II: Hierarchy Postpones the UEC}
\label{sec:hierarchy}

To prevent macroscopic divergence as the population $N$ grows, the system must attenuate the macroscopic gain before it crosses $K_c(T)$ in Eq.~\eqref{eq:critical-gain}. How can this be achieved? One reason to insert layers is the cost of representation. As we show in forthcoming companion work, a distinction at one level selects a register at the next, so finer distinctions can reuse the same organising structure. Write $K_N^{(L)}$ for the change in mandate per unit small mismatch when $N$ inputs pass through $L$ layers. A single flat majority gives $K_N^{(1)}=K_N=\alpha R_N'(0)$ by Eq.~\eqref{eq:loop-gain}. How does that gain change when the same inputs are grouped into smaller blocks? Such hierarchies naturally exhibit \emph{self-similarity}, and may be modelled as a self-similar chain of redescription layers \cite{Simon1962, Ashby1956, Beer1972, Ostrom1990, Ostrom2010, Mesarovic1970, Salthe1985}:
\begin{equation}
    \theta^{(0)} \to \theta^{(1)} \to \cdots \to \theta^{(L)} \to x.
\end{equation}
\begin{figure*}[t]
    \centering
    \includegraphics[width=\linewidth]{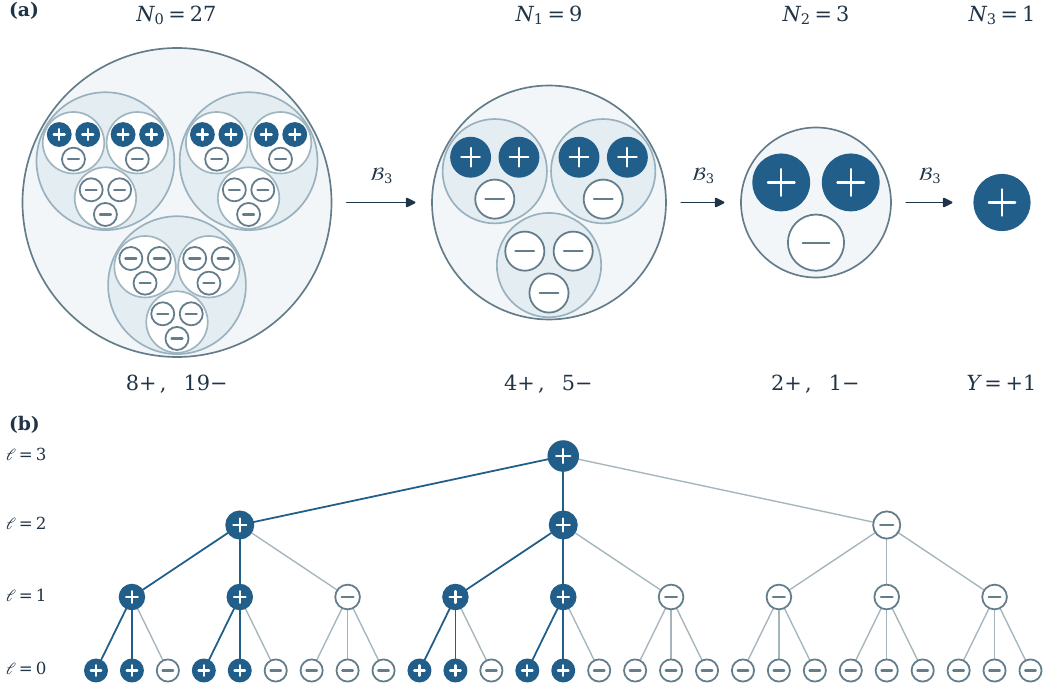}
    \caption{\textbf{One hierarchy in two views.} (a) At each step $\mathcal B_3$ replaces every innermost triple by the sign of its sum. The number of units $N_\ell$ falls by a factor of three, but the arrangement of the remaining blocks is unchanged. (b) The same hierarchy read from its leaves to its root. The leftmost nine leaves form the upper-left block in (a). Four positive leaves give two positive parents and hence one positive output at $\ell=2$. The middle branch does the same. The root is positive although only eight of the twenty-seven inputs are positive.}
    \label{fig:recursive-ballot}
\end{figure*}

Figure~\ref{fig:recursive-ballot} illustrates such a hierarchy, where a block sends its aggregated signal upwards to the next layer. For three inputs $\sigma_i\in\{-1,+1\}$ the block operator $\mathcal B_3$ returns their majority sign (its mean response is derived in Eq.~\eqref{eq:ternary-mean}):
\begin{equation}
    \mathcal B_3(\sigma_1,\sigma_2,\sigma_3)=\operatorname{sgn}(\sigma_1+\sigma_2+\sigma_3).\label{eq:ternary-block}
\end{equation}
Each parent receives three such outputs and applies the same rule. The output of a particular triple is a sign, not its expectation over repeated samples. The nested circles show which distinctions disappear at each step, while the tree shows which signs reach the root and parametrise the macrodynamics. Structurally, this aggregation constitutes a real-space renormalisation group (RG) transformation on a hierarchical lattice (a Migdal-Kadanoff block-spin decimation with Gaussian compression) \cite{Migdal1975, Kadanoff1976}. In arbitrary higher-order networks, such coarse-graining is rigorously formulated via Laplacian diffusion dynamics \cite{Villegas2023, Nurisso2025}, allowing the definition of geometric block-nodes across multiple scales without demanding strong community structure assumptions. Section~\ref{sec:rg_universality} returns to this point in further detail.

In Fig.~\ref{fig:recursive-ballot} there are $N=27$ agents, but each block counts only $n=3$ signs. This number of inputs to one majority decision is the \emph{span} of the block: at the bottom they are agents' signs, and higher up they are the signs returned by the preceding layer. Suppose that at a given mismatch the inputs are independent and each sign has a small mean $m$, so that a positive sign has probability $(1+m)/2$. For a triple, the majority sign then has mean $(3m-m^3)/2$ [Eq.~\eqref{eq:ternary-mean}]. A small bias is therefore amplified by a factor of $3/2$. Because the next layer receives signs of precisely this new bias, the factors multiply from one layer to the next, as in Galam's iterated majority~\cite{Galam1986, Galam1990}.

The comparison is with counting the \emph{same} $N$ signs in one flat majority. For a block of odd span $n$, the amplification factor is the slope $a_n$ of its mean output at $m=0$ [Eq.~\eqref{eq:majority-influence}]. When each span is large, $a_n\simeq\sqrt{2n/\pi}$, and multiplying the slopes over $L$ layers gives a gain roughly $(2/\pi)^{(L-1)/2}$ times the flat gain [Eq.~\eqref{eq:disjoint-block-gain}]. Relative to a flat majority over the same $N$ signs, each layer therefore \emph{attenuates} the gain by a fixed fraction, here $a\simeq\sqrt{2/\pi}$. Writing $K_{\rm flat}=K_N$ for the gain without hierarchialisation, Eq.~\eqref{eq:layer_attenuation} gives $K_N^{(L)}=K_{\rm flat}a^{L-1}$, and as $N$ increases, keeping the gain below $K_c$ requires a number of layers that grows as $\log N$ [Eq.~\eqref{eq:minimum-depth}]. This has the same logarithmic scaling as organisational economics' information-processing bound, but here arises directly from a feedback stability condition \cite{Radner1993, Bolton1994, Garicano2000}. 

\begin{figure*}[t]
    \centering
    \includegraphics[width=\textwidth]{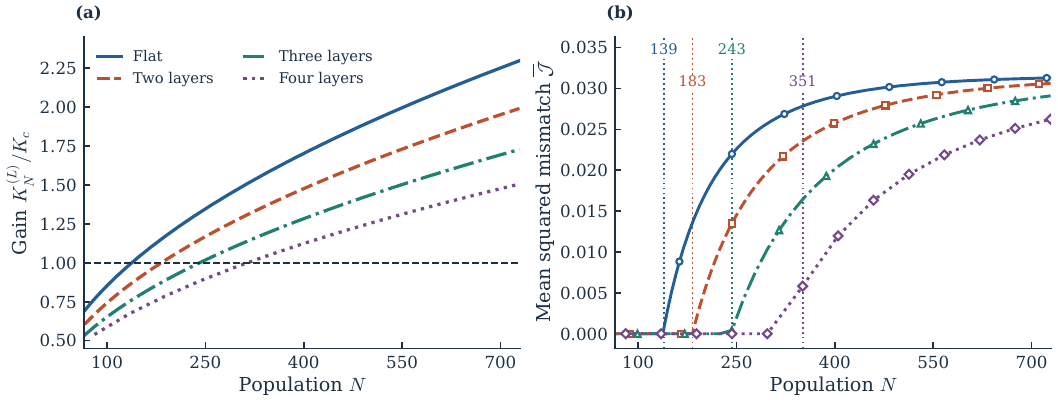}
    \caption{\textbf{Each extra majority round postpones the onset of loss.} (a) At depth $L$, disjoint blocks of $n$ inputs each return one sign. Another $L-1$ rounds of ternary majority reduce those signs to one mandate, so $N=n3^{L-1}$. The gain is $K_N^{(L)}=\alpha\beta a_n(3/2)^{L-1}$ (Appendix~\ref{app:hierarchy}). The horizontal line marks the critical gain $K_c$. (b) Mean squared census mismatch from the iterated nonlinear map. Each vertical line marks the first population allowed by that architecture for which the predicted gain exceeds $K_c$, not a fit to the loss. The curves join discrete odd block sizes to guide the eye. Parameters and sampling windows are given in Appendix~\ref{app:simulations}.}
    \label{fig:hierarchy_postpones}
\end{figure*}

Such attenuation depends on the form of aggregation however. In the majority case we have considered up to this point, the layers of the tree are built from its own agent population. Every block needs at least three inputs, so its depth cannot exceed $\log_3 N$, and the required depth eventually outgrows it (Appendix~\ref{app:hierarchy}). Majority regrouping therefore postpones the instability (Fig.~\ref{fig:hierarchy_postpones}), while stability at large $N$ needs layers whose attenuation does not consume population, such as the transmission stages of Sec.~\ref{sec:inertia}.

Figure~\ref{fig:hierarchy_postpones} illustrates this phenomenon under a hierarchy of ternary majority blocking. Panel (a) shows the multiplicative attenuation of $K_N^{(L)}$, shifting the critical population $N_c$ at which the asymptotic loss switches on. Thus, hierarchialisation can maintain stability as populations grow (or census periods shorten). 

The model considered here treats all aggregations as instantaneous, but this assumes that information transmission is both instantaneous and free. In Appendix~\ref{app:robustness} (Fig.~\ref{fig:hier-delayed-spectrum}) we consider a scenario where aggregation up a hierarchical layer itself takes a census period, and show why a gain comparison alone then no longer decides stability. Even when the feedback is stable, a correction may arrive too late to be useful. These considerations are critical, because - as we shall show in the next section - the stability conferred by hierarchialisation is not free.

\section{Result III: Layer Costs and the Inertial Phase}
\label{sec:inertia}

\begin{figure}[tbp]
    \centering
    \includegraphics[width=\linewidth]{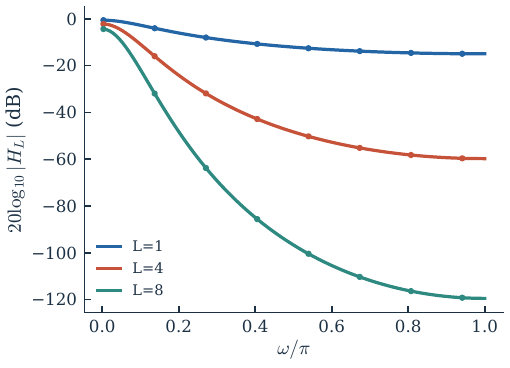}
    \caption{\textbf{Transmission through a leaky cascade.} Solid curves give $20\log_{10}|H_L(\omega)|$ for uptake fraction $\gamma=0.32$ and link transmission $a=0.94$. Markers are calculated from the first 512 samples of the impulse response. Rapid changes are attenuated more strongly than a sustained input.}
    \label{fig:transmission_inertia}
\end{figure}

\begin{figure*}[t]
    \centering
    \includegraphics[width=\linewidth]{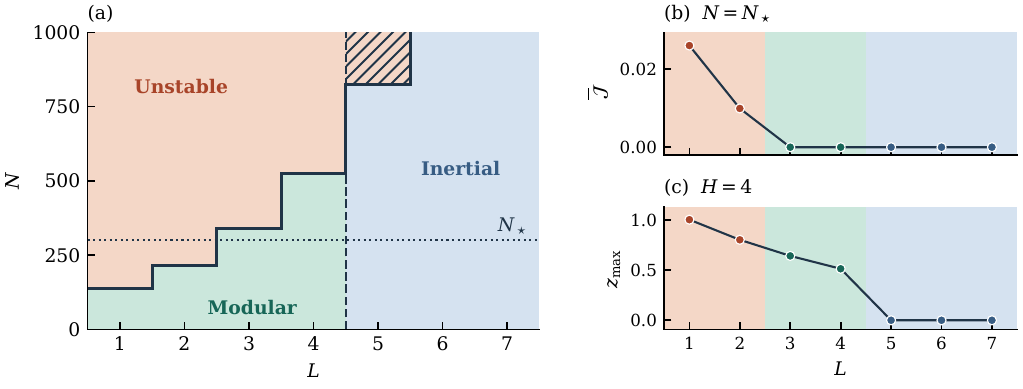}
    \caption{\textbf{Two boundaries delimit the modular phase.} (a) At each integer depth $L$, the staircase separates odd population sizes whose fixed point is stable (below) from those whose feedback gain exceeds $K_c$ (above). Each added static stage transmits a fraction $a=0.8$ of its input. The dashed line marks the last depth reachable in $H=4$ census intervals by the shift chain of Eq.~\eqref{eq:shift-chain} with $\Delta=T$. Hatching marks failure of both tests. (b) Along the dotted cut at $N_\star=301$, mean squared mismatch over the last 2,000 of 12,000 census points falls to numerical zero at $L=3$. (c) The largest top-layer signal $z_{\max}$ that an input of unit energy can produce by census point $H$, starting from rest [Eq.~\eqref{eq:top-displacement}], falls to zero at $L=5$. Only depths three and four meet both requirements.}
    \label{fig:phases}
\end{figure*}

The preceding result invites a naive conclusion: if hierarchy postpones the UEC, one might assume infinite hierarchy is optimal. However, the same layers that suppress  instability also attenuate signal, delay correction, and increase the energy required to reconfigure the system \cite{Pierson1993, Tsebelis2002, Goodhart1984, Campbell1976}. These costs do not arise from the vote itself - a majority block returns just one sign [Eq.~\eqref{eq:odd-majority-block}], but that fixes neither the physical signal nor its cost. The previous sections considered only the informational aspect of the operation that must be performed at every layer - the \emph{aggregation} of verdicts. The \emph{transmission} of those verdicts requires a physical signal to do so, and as such carries a cost. The distinction is vital as while the content of a verdict may be binary, its carrier signal may be continuous. For example, the polarity of a pulse can encode the verdict while its amplitude varies with transmission. Even for the same verdict, the time and energy spent collecting the inputs and delivering the pulse can differ. A hierarchy in which every block computes the correct verdict can therefore still fail if the signal carrying that verdict arrives too weak or too late. To formalise the resulting trade-off, we introduce a \emph{total organisational loss}:
\begin{equation}
    \mathcal{J}_{\rm tot}(L) = \overline{\mathcal{J}}_{\rm dyn}(L) + C(L) + w_{\rm tr}\mathcal{I}_{\rm tr} + w_{\rm ctrl}\mathcal{I}_{\rm ctrl}.\label{eq:total-layer-loss}
\end{equation}
Here, $\overline{\mathcal{J}}_{\rm dyn}(L)$ is the mean squared mismatch of Eq.~\eqref{eq:census-loss} at depth $L$, $C(L)$ is the direct administrative cost of maintaining layers, and the nonnegative weights $w_{\rm tr}$ and $w_{\rm ctrl}$ convert the transmission and control penalties to the same loss units. An additive linear combination for this total loss is assumed to explicitly reflect that these penalties operate through independent physical channels - administrative overhead, frequency attenuation and control resistance - none of which can cleanly substitute for another. As derived in Appendix \ref{app:cost_functions}, the two forms of inertia are specifically selected to capture distinct failure modes of deep hierarchies. These are:

\textbf{Transmission inertia:} Chosen to isolate the loss of high-fidelity environmental tracking, it is defined as the frequency-domain attenuation of signals passing through the layers, given by
\begin{equation}
 \mathcal I_{\rm tr}(L,\omega)=-\log|H_L(\omega)|,
 \label{eq:transmission-inertia}
\end{equation} 
where $H_L(\omega)$ is the cascade transfer function for a signal with angular frequency $\omega$ (radians per census)~\cite{SmithDigitalFilters}, given by Eq.~\eqref{eq:cascade-transfer}. Fig.~\ref{fig:transmission_inertia} shows examples of this transmission through layers of the hierarchy. Because the stages multiply, $\mathcal I_{\rm tr}=-L\log|H_1(\omega)|$ is linear in depth, so with a per-layer administrative cost $C(L)=\mu L$ it simply renormalises $\mu$ to $\mu_{\rm eff}(\omega)=\mu-w_{\rm tr}\log|H_1(\omega)|$. The control inertia defined next behaves differently, acting as a hard barrier at a finite depth, and it is this barrier that draws the inertial boundary in Fig.~\ref{fig:phases}.

\begin{figure*}[t]
    \centering
    \includegraphics[width=\textwidth]{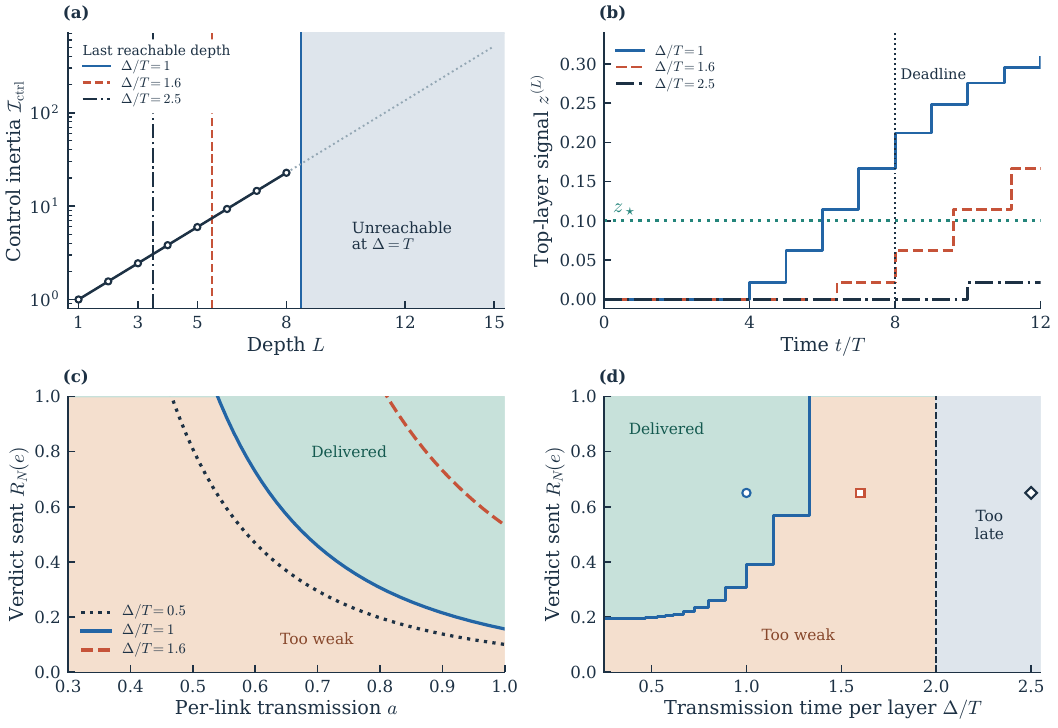}
    \caption{\textbf{A correction can be too weak or too late.} The deadline is $8T$. (a) Control inertia [Eq.~\eqref{eq:control-inertia}] of the shift chain in Eq.~\eqref{eq:shift-chain} with $a=0.8$. Reachable depths cost $a^{-2(L-1)}$ whatever the transmission time $\Delta$, but no input reaches the top beyond the last reachable depth $\lfloor HT/\Delta\rfloor$, marked by vertical lines for the three values of $\Delta$ used in (b). Shading marks the unreachable depths at $\Delta=T$. (b--d) Delivery of a bounded verdict: four layers start from rest, take up a fraction $\gamma=0.5$ per step, and receive a sustained verdict $R_N(e)$. The top layer must reach $z_\star=0.1$. (b) Top-layer signal for $R_N(e)=0.65$ and $a=0.8$ at three transmission times $\Delta$. Dotted lines mark $z_\star$ and the deadline. (c) Delivery requires a verdict on or above the curve for the chosen $\Delta/T$. Shading is for $\Delta=T$. (d) At $a=0.8$ the boundary is a staircase because only completed transmission steps contribute. In the grey region the first response arrives after the deadline. The three markers locate the traces in (b).}
    \label{fig:control_inertia}
\end{figure*}

\textbf{Control inertia:} Chosen using the reachability Gramian~\cite{Hespanha2009} to quantify the energetic resistance to perturbation, it defines the minimum energy required to reach the top layer from the bottom layer before a \emph{deadline} of $H$ census intervals, that is, by time $HT$. Collecting the layer signals into the state $\mathbf z_j=(z^{(1)}_j,\ldots,z^{(L)}_j)^\top$, each step then moves the attenuated signal up one layer:
\begin{equation}
 \mathbf z_{j+1}=A\mathbf z_j+\mathbf q_1 z^{(0)}_j,\qquad \mathbf z_0=\mathbf 0 .
 \label{eq:shift-chain}
\end{equation}
Here $A$ has $a$ on its first subdiagonal and zeros elsewhere, and $\mathbf q_\ell$ is the unit vector with a one in position $\ell$ and zeros elsewhere. The verdict enters the bottom layer directly, as in Eq.~\eqref{eq:layer_attenuation}, and crosses $L-1$ links on its way to the top. We measure the effort spent on an input history $z^{(0)}_0,z^{(0)}_1,\ldots$ by its energy $\sum_j(z^{(0)}_j)^2$. Now let each transmission step take a time $\Delta$, which need not equal the census period. The deadline then allows $m_H=\lfloor HT/\Delta\rfloor$ steps for a perturbation to travel up the chain, and $\mathbf q_L^\top\mathbf z_{m_H}$ is the top-layer signal when it expires. With $W_m$ the Gramian of Eq.~\eqref{eq:reachability-gramian} after $m$ steps, the minimum energy for a unit top-layer signal is exactly
\begin{equation}
 \mathcal I_{\rm ctrl}(L,H)=
 \begin{cases}
  1/(\mathbf q_L^\top W_{m_H} \mathbf q_L), & \mathbf q_L^\top W_{m_H} \mathbf q_L>0,\\
  +\infty, & \mathbf q_L^\top W_{m_H} \mathbf q_L=0 ,
 \end{cases}
  \label{eq:control-inertia}
 \end{equation}
This closed form reveals a parallel phase for an \emph{excess} of hierarchy. For $L\Delta>HT$, $\mathcal{I}_{\rm ctrl}$ diverges to infinity, meaning the top is literally unreachable from the bottom before the deadline. For $L\Delta\le HT$, the signal is still attenuated by $a^{-2(L-1)}$ for link transmission $0<a<1$. This exponential attenuation is why deep hierarchies physically wall off rapid reconfiguration \cite{Hespanha2009, Kalman1960, Kalman1963, LeeMarkus1967, Moore1981} (Fig.~\ref{fig:control_inertia}a). With one step per census, $\Delta=T$, the top is reachable only when $L\le H$. A slower carrier moves this wall to shallower depths without changing the cost of the depths that remain reachable. The same construction gives the quantity plotted in Fig.~\ref{fig:phases}(c). Starting from rest, the largest top-layer signal that an input of unit energy can produce by the deadline is
\begin{equation}
 z_{\max}(L,H)=
 \max_{\sum_{j=0}^{m_H-1}(z^{(0)}_j)^2\le1}|\mathbf q_L^\top\mathbf z_{m_H}|
 =\sqrt{\mathbf q_L^\top W_{m_H} \mathbf q_L}
 =\mathcal I_{\rm ctrl}^{-1/2}.
 \label{eq:top-displacement}
\end{equation}

Equation~\eqref{eq:control-inertia} separates two ways in which depth defeats a signal. Beyond the causal limit $L\Delta>HT$ no input arrives in time, however large. Within it, $\mathcal I_{\rm ctrl}$ prices attenuation by allowing the input to be as large as necessary. We emphasise again that this is for the \emph{signal  $z$} carrying the verdict response (which itself satisfies $|R_N(e)|\le1$). Suppose then that the mandate signal must change by at least $\alpha z_\star$ before the deadline, so that the top layer must hold $z^{(L)}\ge z_\star>0$. Arrival alone is then insufficient to ensure delivery, and we must also account for the magnitude of signal attenuation.

Figure~\ref{fig:control_inertia}(b) illustrates the distinction with one sustained verdict $R_N(e)=0.65$ sent into four layers that each take up half their input per step ($\gamma=0.5$). The top-layer signal exceeds $z_\star=0.1$ by the deadline when $\Delta=T$. At $\Delta=1.6T$ the top has begun to respond by the deadline but falls short, while at $\Delta=2.5T$ it has not moved at all. Panels (c) and (d) give the delivery boundary as the verdict sent, the fraction $a$ transmitted per link and the transmission time are varied. A larger verdict can compensate for attenuation only if there is time for the signal to arrive.  Since $|R_N(e)|$ increases with $|e|$, a smallest deliverable verdict is also a smallest correctable mismatch: a deep or slow hierarchy corrects in time only those errors large enough to produce a decisive verdict at the bottom. Appendix~\ref{app:cost_functions} derives both boundaries in Eq.~\eqref{eq:deadline-amplitude-test}.

When systems accumulate excess layers $L \gg L_{\min}$, the inertial penalties dominate the total loss in Eq.~\eqref{eq:total-layer-loss}, and the system enters an \textit{inertial phase}. In this phase, the system is technically stable, but it is so heavily filtered and delayed that it cannot adequately respond to perturbations. This maps  onto the conservation phase (the $K$-phase or ``rigidity trap'', where $K$ is a phase name, not the loop gain) of the Holling Adaptive Cycle \cite{GundersonHolling2002}, where systems accumulate structure but lose their fundamental adaptive capacity, rendering them vulnerable to external shocks. The hierarchical structures required to prevent explosive feedback eventually compromise the system through structural inertia. Both boundaries are plotted in Fig.~\ref{fig:phases} at fixed population. 

\section{Result IV: Real-Space Renormalisation and the Classification Theorem}
\label{sec:rg_universality}
In the preceding sections we have established the role hierarchialisation can play in both the dynamic stability and responsiveness of an adaptive system. It is worth pausing a moment to consider the implications of this, as it suggests that hierarchy will emerge  in \emph{any} system bound by efficiency and stability constraints. A good test of this prediction can be found in the field of machine learning, where emergent hierarchies have been observed. One direct example is to be found in convolutional neural networks, where in deeper layers the probability density peaks structure themselves in a hierarchy that mirrors the semantic hierarchy of the concepts they are classifying \cite{Doimo2020Nucleation}.   

To make contact with this result,  we need a description of hierarchy that applies equally to networks as to institutions. The hierarchy of Sec.~\ref{sec:hierarchy} already provides one once it is read as a coarse-graining. Each layer of a majority tree replaces a block of signs by a single sign, and the question at every layer is how much of an incoming bias survives the replacement. The mathematical generalisation of this process is captured by the \emph{real-space renormalisation group} (RG). In Appendix \ref{app:rg_universality_proofs} we apply this technique to generalise our previous results, and obtain a more general classification for the phases of hierarchy, based on the form of upward aggregation the system adopts. If $g(m)$ is the expected output bias of one block with input bias $m$, its amplification over a layer of the hierarchy is captured by the RG eigenvalue $\lambda_0=g'(0)$.  Composing the layers together  [Eq.~\eqref{eq:block-multiplier}] then classifies the local transmission of bias under repeated blocking, and the stability of the closed loop depends on whether the gain transmitted through the hierarchy exceeds the stability threshold of the coupled census map [Eq. \eqref{eq:critical-gain}].  As such, we may \emph{classify} the behaviour of an adaptive normal form based on the RG eigenvalue $\lambda_0$ that its mode of aggregation produces. Consequently we may generalise the majority aggregation employed in previous sections into one case of a general classification of hierarchical stability.
\begin{enumerate}
    \item \textbf{Amplifying ($\lambda_0>1$)}: Each layer increases a small bias.
    \item \textbf{Marginal ($\lambda_0=1$)}: The bias is unchanged to first order.
    \item \textbf{Attenuating ($0\le\lambda_0<1$)}: Each layer reduces a small bias.
\end{enumerate}

The $\sqrt{N}$ law in Eq.~\eqref{eq:gain-population} applies to the flat independent majority considered to this point, while continuous averaging or a single dominant input can remove that growth without ensuring stability (Appendices~\ref{app:majority} and~\ref{app:rg_universality_proofs}). This leads to four corollaries on our previous results based on the form of aggregation applied, generating the observed taxonomy of collective behaviours:

\paragraph{Corollary 1: Simple Majority (The UEC).}
For an odd majority block the exact multiplier is $a_n$ in Eq.~\eqref{eq:majority-influence}, with $a_n\sim\sqrt{2n/\pi}$. It exceeds one for odd $n\ge3$.

\paragraph{Corollary 2: Continuous Averaging (The Marginal Class).}
A proportional average with unit calibration preserves the input bias, so $\lambda_0=1$. Its gain does not grow with population, but it must still be compared with the critical gain of the census map (Appendix~\ref{app:rg_universality_proofs}).

\paragraph{Corollary 3: Super-Majority (The Frozen Branch).}
A super-majority rule requires a fraction $\frac{1}{2} + \Delta p$ to trigger a state change, where $\Delta p>0$ is the excess vote fraction required above a simple majority. Because the variance of the sample mean shrinks as $1/N$, the probability of crossing this threshold decays exponentially. As derived in Appendix~\ref{app:rg_universality_proofs}, for a unanimous block of size $n=3$, the eigenvalue evaluates to $\lambda_0 = \frac{3}{4} < 1$. With a fair binary fallback on mixed outcomes, repeated blocking attenuates small biases.

\paragraph{Corollary 4: Weighted Majority (The Self-Stabilising Oligarchy).}
Highly unequal weights can invalidate the Gaussian approximation used for an equally weighted majority. In the limit where one input outweighs all the others combined, the exact bias slope is one [Eq.~\eqref{eq:lambda_oligarchy}]. The gain then stops growing with $N$, but its magnitude must still be compared with the critical gain of the census map.

\begin{table}[tbp]
\centering
\caption{Universality Classes of Aggregation}
\label{tab:universality}
\renewcommand{\arraystretch}{1.5}
\setlength{\tabcolsep}{4pt}
\begin{tabular}{p{0.28\linewidth}p{0.17\linewidth}p{0.19\linewidth}p{0.2\linewidth}}
\toprule
\textbf{Aggregation Rule} & \textbf{Scaling $K_N$} & \textbf{RG Eigenvalue} & \textbf{Fixed Point} \\
\midrule
Simple Majority & $\sim\sqrt{N}$ (flat) & $a_n>1$ (odd $n\ge3$) & Consensus \\
Continuous Average & $\mathcal{O}(1)$ & $1$ & Marginal ($m=m$) \\
Super-Majority ($\Delta p>0$) & $\to 0$ & $< 1$ & Frozen ($0$) \\
Dominant input & $\mathcal{O}(1)$ & $1$ & Bias transmitted \\
\bottomrule
\end{tabular}
\end{table}

\subsection*{Laplacian coarse-graining}

This RG picture also supplies the link to the emergent hierarchies observed in the context of machine learning. To find the same structure in networks of weights, the aggregated blocks must arise from the couplings between weights. This can be analysed using the Laplacian formalism \cite{Villegas2023}. Let $\omega_{ij}\ge0$ be the weight of the link between vertices $i$ and $j$. A quantity $\mathbf u$ held on the vertices then diffuses according to
\begin{equation}
 \frac{d\mathbf u}{dt}=-\mathcal L\,\mathbf u,\qquad
 \mathcal L=I-A,\qquad
 A_{ij}=\frac{\omega_{ij}}{\sum_l\omega_{il}},
 \label{eq:graph-diffusion}
\end{equation}
with solution $\mathbf u(t)=e^{-t\mathcal L}\mathbf u(0)$. An eigenmode of $\mathcal L$ with eigenvalue $\lambda$ retains a fraction $e^{-t\lambda}$ of its amplitude after a time $t$. This factor plays the part of the RG multiplier $\lambda_0^L$ in Eq.~\eqref{eq:block-multiplier}, with diffusion time in place of the number of layers. The constant mode has $\lambda=0$ and is carried unchanged, like the continuous average of Corollary~2, while every other mode is attenuated, the fastest first.

\begin{figure*}[t]
\centering
\begin{tikzpicture}[x=1cm,y=1cm,font=\small,>=Stealth]
\definecolor{heatink}{HTML}{235568}
\definecolor{pathink}{HTML}{AA4926}
\node[anchor=west] at (0,4.62) {\textbf{a}\quad $e^{-t\mathcal L}$};
\draw[->,heatink,line width=.8pt] (4.2,4.62)--(13.1,4.62) node[midway,above] {$t$};
\filldraw[fill=heatink!100,draw=black!18,line width=.25pt] (1.16,4.12) rectangle ++(.42,-.42);
\filldraw[fill=heatink!0,draw=black!18,line width=.25pt] (1.58,4.12) rectangle ++(.42,-.42);
\filldraw[fill=heatink!0,draw=black!18,line width=.25pt] (2.00,4.12) rectangle ++(.42,-.42);
\filldraw[fill=heatink!0,draw=black!18,line width=.25pt] (2.42,4.12) rectangle ++(.42,-.42);
\filldraw[fill=heatink!0,draw=black!18,line width=.25pt] (1.16,3.70) rectangle ++(.42,-.42);
\filldraw[fill=heatink!100,draw=black!18,line width=.25pt] (1.58,3.70) rectangle ++(.42,-.42);
\filldraw[fill=heatink!0,draw=black!18,line width=.25pt] (2.00,3.70) rectangle ++(.42,-.42);
\filldraw[fill=heatink!0,draw=black!18,line width=.25pt] (2.42,3.70) rectangle ++(.42,-.42);
\filldraw[fill=heatink!0,draw=black!18,line width=.25pt] (1.16,3.28) rectangle ++(.42,-.42);
\filldraw[fill=heatink!0,draw=black!18,line width=.25pt] (1.58,3.28) rectangle ++(.42,-.42);
\filldraw[fill=heatink!100,draw=black!18,line width=.25pt] (2.00,3.28) rectangle ++(.42,-.42);
\filldraw[fill=heatink!0,draw=black!18,line width=.25pt] (2.42,3.28) rectangle ++(.42,-.42);
\filldraw[fill=heatink!0,draw=black!18,line width=.25pt] (1.16,2.86) rectangle ++(.42,-.42);
\filldraw[fill=heatink!0,draw=black!18,line width=.25pt] (1.58,2.86) rectangle ++(.42,-.42);
\filldraw[fill=heatink!0,draw=black!18,line width=.25pt] (2.00,2.86) rectangle ++(.42,-.42);
\filldraw[fill=heatink!100,draw=black!18,line width=.25pt] (2.42,2.86) rectangle ++(.42,-.42);
\node at (2,2.16) {$t=0$};
\filldraw[fill=heatink!50,draw=black!18,line width=.25pt] (6.16,4.12) rectangle ++(.42,-.42);
\filldraw[fill=heatink!50,draw=black!18,line width=.25pt] (6.58,4.12) rectangle ++(.42,-.42);
\filldraw[fill=heatink!0,draw=black!18,line width=.25pt] (7.00,4.12) rectangle ++(.42,-.42);
\filldraw[fill=heatink!0,draw=black!18,line width=.25pt] (7.42,4.12) rectangle ++(.42,-.42);
\filldraw[fill=heatink!50,draw=black!18,line width=.25pt] (6.16,3.70) rectangle ++(.42,-.42);
\filldraw[fill=heatink!50,draw=black!18,line width=.25pt] (6.58,3.70) rectangle ++(.42,-.42);
\filldraw[fill=heatink!0,draw=black!18,line width=.25pt] (7.00,3.70) rectangle ++(.42,-.42);
\filldraw[fill=heatink!0,draw=black!18,line width=.25pt] (7.42,3.70) rectangle ++(.42,-.42);
\filldraw[fill=heatink!0,draw=black!18,line width=.25pt] (6.16,3.28) rectangle ++(.42,-.42);
\filldraw[fill=heatink!0,draw=black!18,line width=.25pt] (6.58,3.28) rectangle ++(.42,-.42);
\filldraw[fill=heatink!50,draw=black!18,line width=.25pt] (7.00,3.28) rectangle ++(.42,-.42);
\filldraw[fill=heatink!50,draw=black!18,line width=.25pt] (7.42,3.28) rectangle ++(.42,-.42);
\filldraw[fill=heatink!0,draw=black!18,line width=.25pt] (6.16,2.86) rectangle ++(.42,-.42);
\filldraw[fill=heatink!0,draw=black!18,line width=.25pt] (6.58,2.86) rectangle ++(.42,-.42);
\filldraw[fill=heatink!50,draw=black!18,line width=.25pt] (7.00,2.86) rectangle ++(.42,-.42);
\filldraw[fill=heatink!50,draw=black!18,line width=.25pt] (7.42,2.86) rectangle ++(.42,-.42);
\node at (7,2.16) {$\lambda_{\rm f}^{-1}\ll t\ll\lambda_{\rm s}^{-1}$};
\filldraw[fill=heatink!25,draw=black!18,line width=.25pt] (11.16,4.12) rectangle ++(.42,-.42);
\filldraw[fill=heatink!25,draw=black!18,line width=.25pt] (11.58,4.12) rectangle ++(.42,-.42);
\filldraw[fill=heatink!25,draw=black!18,line width=.25pt] (12.00,4.12) rectangle ++(.42,-.42);
\filldraw[fill=heatink!25,draw=black!18,line width=.25pt] (12.42,4.12) rectangle ++(.42,-.42);
\filldraw[fill=heatink!25,draw=black!18,line width=.25pt] (11.16,3.70) rectangle ++(.42,-.42);
\filldraw[fill=heatink!25,draw=black!18,line width=.25pt] (11.58,3.70) rectangle ++(.42,-.42);
\filldraw[fill=heatink!25,draw=black!18,line width=.25pt] (12.00,3.70) rectangle ++(.42,-.42);
\filldraw[fill=heatink!25,draw=black!18,line width=.25pt] (12.42,3.70) rectangle ++(.42,-.42);
\filldraw[fill=heatink!25,draw=black!18,line width=.25pt] (11.16,3.28) rectangle ++(.42,-.42);
\filldraw[fill=heatink!25,draw=black!18,line width=.25pt] (11.58,3.28) rectangle ++(.42,-.42);
\filldraw[fill=heatink!25,draw=black!18,line width=.25pt] (12.00,3.28) rectangle ++(.42,-.42);
\filldraw[fill=heatink!25,draw=black!18,line width=.25pt] (12.42,3.28) rectangle ++(.42,-.42);
\filldraw[fill=heatink!25,draw=black!18,line width=.25pt] (11.16,2.86) rectangle ++(.42,-.42);
\filldraw[fill=heatink!25,draw=black!18,line width=.25pt] (11.58,2.86) rectangle ++(.42,-.42);
\filldraw[fill=heatink!25,draw=black!18,line width=.25pt] (12.00,2.86) rectangle ++(.42,-.42);
\filldraw[fill=heatink!25,draw=black!18,line width=.25pt] (12.42,2.86) rectangle ++(.42,-.42);
\node at (12,2.16) {$t\gg\lambda_{\rm s}^{-1}$};
\foreach \i/\lab in {0/A,1/B,2/C,3/D}{
\node[font=\scriptsize] at (0.88,{3.91-.42*\i}) {$\lab$};
\node[font=\scriptsize] at ({1.37+.42*\i},4.29) {$\lab$};}
\draw[black!20] (.05,1.82)--(13.5,1.82);
\node[anchor=west] at (0,1.48) {\textbf{b}\quad $\kappa$};
\draw[<-,pathink,line width=.8pt] (4.55,1.48)--(13.1,1.48) node[midway,above] {$\kappa$};
\begin{scope}[line width=1.15pt,draw=black!45]
\draw (12,0)--(9.7,0)--(8.6,.55)--(4.9,.55)--(3.8,.86)--(2,.86);
\draw (4.9,.55)--(3.8,.28)--(2,.28);
\draw (9.7,0)--(8.6,-.55)--(4.9,-.55)--(3.8,-.28)--(2,-.28);
\draw (4.9,-.55)--(3.8,-.86)--(2,-.86);
\end{scope}
\draw[pathink,line width=2pt] (12,0)--(9.7,0)--(8.6,.55)--(4.9,.55)--(3.8,.86)--(2,.86);
\foreach \y/\lab in {.86/A,.28/B,-.28/C,-.86/D}{
\node[left] at (2,\y) {$\lab$};}
\draw[black!60,dashed] (7,-.93)--(7,.99);
\foreach \x/\y in {2/.86,7/.55,12/0}{\fill[pathink] (\x,\y) circle (2.8pt);}
\draw[fill=white,black!55] (7,-.55) circle (2.6pt);
\node[anchor=west] at (7.18,.8) {$\{A,B\}$};
\node[anchor=west] at (7.18,-.81) {$\{C,D\}$};
\end{tikzpicture}
\caption{\textbf{Which distinctions survive at a given scale?} (a) Entries of $e^{-t\mathcal L}$ for two strongly coupled pairs $AB$ and $CD$ with weak symmetric links between pairs. Each column is the heat from one initial vertex. From left to right the panels show the identity, the pair-averaging limit and global equilibration. Ink density represents the entry on a common linear scale. The fast rate $\lambda_{\rm f}$ mixes partners before the slow rate $\lambda_{\rm s}$ mixes pairs. (b) Read from right to left. In this four-leaf sketch of a Parisi tree, increasing noise precision $\kappa$ resolves first the pair and then the pure state. The coloured path follows one choice. At the dashed cut, $A$ can be distinguished from $C$ but not yet from $B$.}
\label{fig:diffusion-resolution}
\end{figure*}
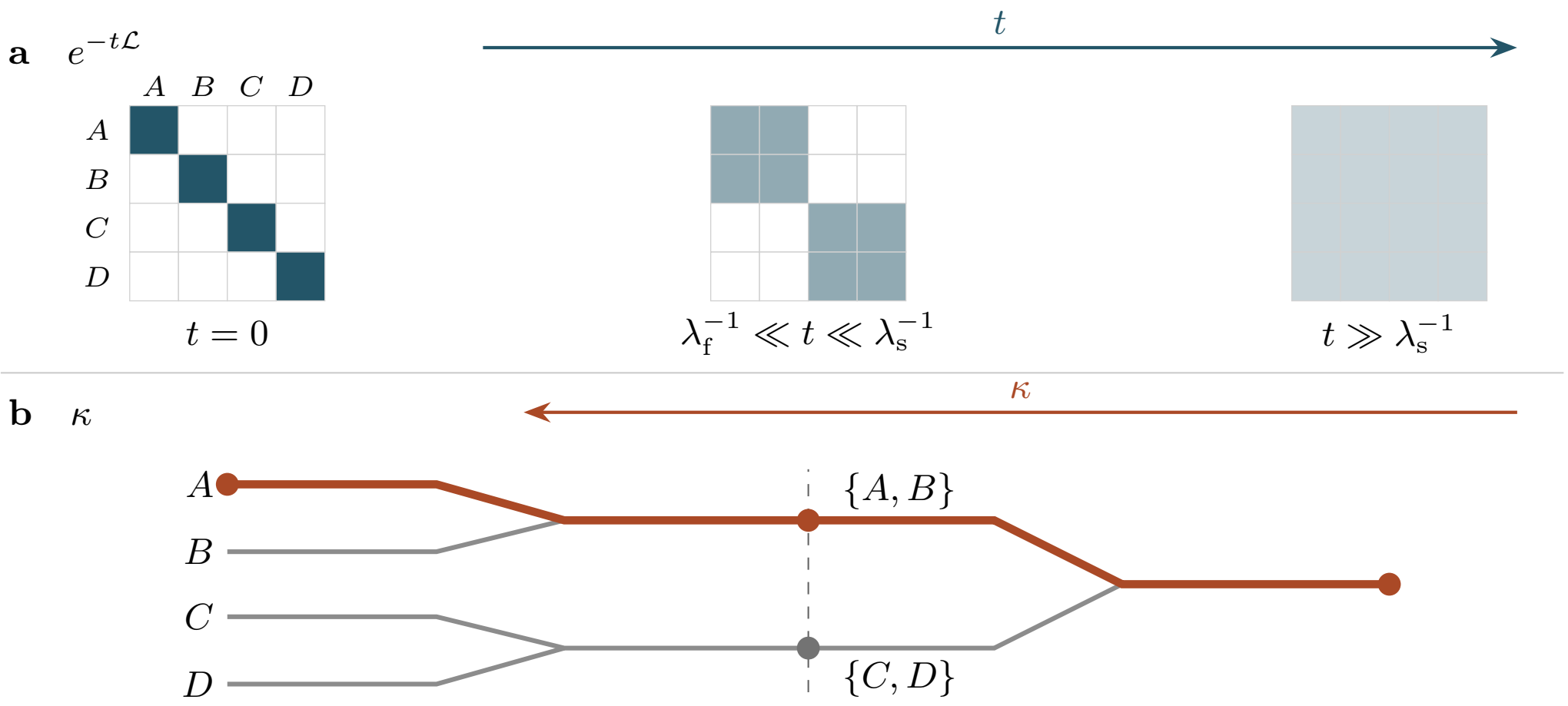

That ordering is what produces a hierarchy. Consider two pairs of vertices with strong links within each pair and weak links between them (Fig.~\ref{fig:diffusion-resolution}a). For $\lambda_{\rm f}^{-1}\ll t\ll\lambda_{\rm s}^{-1}$ one can still tell which pair received the heat, but not how it is distributed \emph{within} the pair. On that time scale each pair behaves as a single block variable, whose effective relaxation follows from projecting the Laplacian onto the blocks [Eq.~\eqref{eq:block-projected-relaxation}]. This separation of relaxation rates is the basis of Laplacian renormalisation \cite{Villegas2023}, and has since been extended to higher-order networks \cite{Nurisso2025}. Diffusion time plays the role of scale, and a block is defined by the distinctions that survive to it.

Generative diffusion runs the same process in reverse. Its forward process is heat flow in data space, and generation follows a noisy sample back towards the data distribution,~with recent work exposing the hierarchical structure that develops from this process \cite{GarnierBrun2026Speciation}. In the regime of full replica symmetry breaking its pure states form a Parisi tree \cite{Mezard1984Nature, MezardParisi1991}, and the distance between leaves is set by the level at which their branches meet: if $A$ and $B$ are closer to one another than to $C$, then $C$ is equally far from both. This ultrametric property \cite{Panchenko2013Ultrametricity} means that a sample can identify the branch containing $A$ and $B$ before it distinguishes the two leaves. As the noise precision $\kappa$ increases, the reverse process commits to ever finer branches (Fig.~\ref{fig:diffusion-resolution}b). The two panels of Fig.~\ref{fig:diffusion-resolution} are therefore the same statement read in opposite directions: forward diffusion erases the finest distinctions first, and denoising resolves them last.

This picture can be extended to the Transformer architecture, which performs the same coarse-graining with depth in place of diffusion time. Arrange the token representations entering layer $\ell$ as the rows of $X_\ell$, and let $A_\ell$ be the attention matrix, whose rows sum to one. With the value map set to the identity, a residual attention update of step $\delta t$ is
\begin{equation}
 X_{\ell+1}=X_\ell+\delta t\,(A_\ell X_\ell-X_\ell)=(I-\delta t\,\mathcal L_\ell)X_\ell ,
 \label{eq:attention-laplacian}
\end{equation}
which is an Euler step of Eq.~\eqref{eq:graph-diffusion} on a graph whose vertices are tokens and whose edge weights are the attention weights. Tokens joined by strong attention therefore merge into a block before the slower differences between groups decay, and a stack of attention layers performs a real-space coarse-graining of the token graph. In continuous-depth models the same picture appears as interacting particles, with the attention weights setting how strongly one token moves towards another \cite{Geshkovski2025, Castin2025, Bruno2025, Karagodin2025}. With $A_\ell$ fixed and symmetric, the block construction of Appendix~\ref{app:rg_universality_proofs} applies directly. Token-dependent attention and value projections add terms to Eq.~\eqref{eq:token-attention-example}, and Eq.~\eqref{eq:token-angular-speed} follows the direction of each token separately from its length. Transformers and diffusion models therefore both hierarchialise without any explicit design, making them natural targets for the phase theory developed here. A forward pass through these layers supplies the hierarchy alone. The loop of the adaptive normal form closes once the output is scored and that score revises the parameters, as in training, reinforcement learning or a meta-level objective.

\section{Finite Sampling and Empirical Signatures}
\label{sec:empirical}

Every result so far has fed the expected verdict $R_N(e_k)$ back into the mandate, and consequently described the \emph{mean} dynamics. In a finite population the actual response will fluctuate about this expectation, and we now investigate the effect of finite sampling on the phase boundary predicted by Eq.~\eqref{eq:critical-gain}. To separate the noise from the feedback gain, take $M$ independent majority verdicts at the same mismatch and average them before updating. Each verdict is a separate majority of $N$ inputs, drawn afresh from the response law of Eq.~\eqref{eq:scalar-sign-probability}. To make this concrete, we can supply a natural institutional reading of this sampling as an an electorate divided into $M$ districts of $N$ voters, with the mandate revised in proportion to the resulting seat share. The census then samples $MN$ responses in all. Regrouping the same $N$ signs into different majorities would not serve, because verdicts that share inputs are correlated and their average does not lose variance as $1/M$. Districting is a majority layer followed by an averaging layer, and averaging is marginal (Corollary~2), so the gain is set by the district size $N$ rather than by the total electorate $MN$. Increasing $M$ reduces the conditional variance by a factor of $M$, without changing its mean [Eq.~\eqref{eq:sampled-increment-variance}]. The number $M$ therefore controls the sampling noise without changing the gain, and $M\to\infty$ recovers the mean map.

\begin{figure*}[!t]
    \centering
    \includegraphics[width=\textwidth]{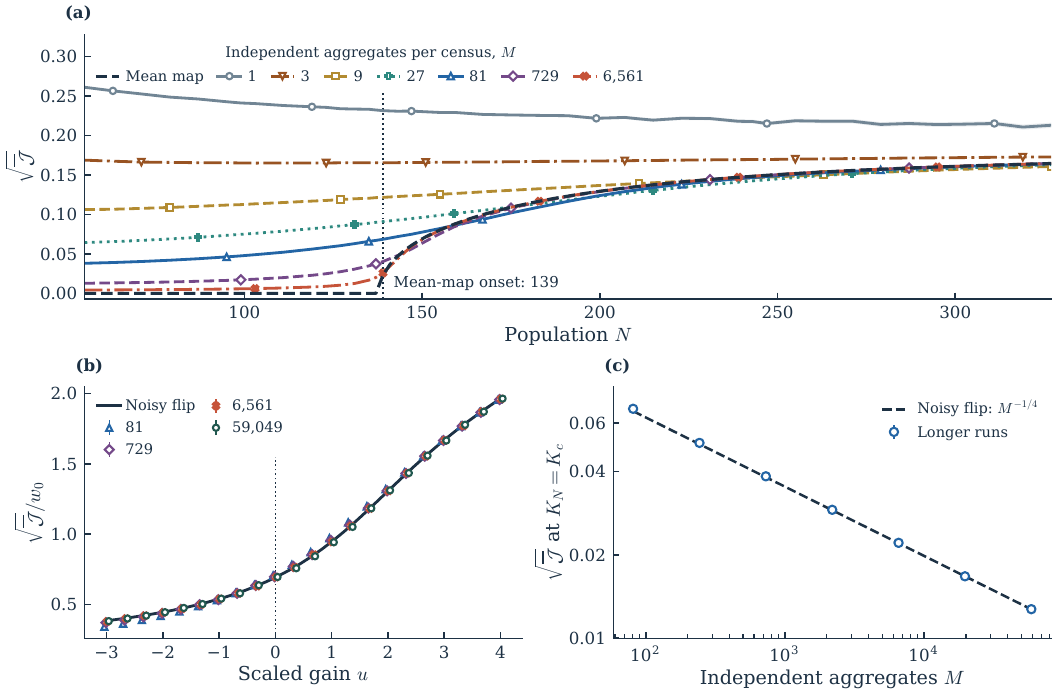}
    \caption{\textbf{Sampling rounds the onset of alternating mandates.} (a) Root mean squared mismatch as the odd population $N$ grows, for seven values of $M$, the number of independent majority verdicts per census. The dark dashed curve is the deterministic two-cycle amplitude and is zero below onset. The vertical dotted line marks the first unstable odd size of the mean map. (b) At fixed $N$ at the mean-map onset, varying the update step $\alpha$ resolves the smaller gain window near the flip. The axes use $w_0$ and $u$ from Eq.~\eqref{eq:noisy-flip-scales}. The solid line is the quartic diffusion prediction without fitting. (c) At the deterministic threshold, longer independent runs test the predicted $M^{-1/4}$ amplitude. }
    \label{fig:empirical_signatures}
\end{figure*}

Figure~\ref{fig:empirical_signatures}(a) follows the root mean squared mismatch $\sqrt{\overline{\mathcal J}}$ [Eq.~\eqref{eq:census-loss}] as the population grows. The mean map goes from zero loss to sustained alternation at a single population, while a sampled loop has nonzero mismatch even below that size. As sampling is increased, the measured mismatch converges towards the mean map.  At large $M$ sampling therefore rounds the onset without moving it.

To resolve the change close to onset, we hold $N$ fixed and vary the step size, so that the excess gain $\epsilon=K_N/K_c-1$ can be changed continuously. Near the flip the linear growth rate of the alternating component is proportional to $\epsilon$, and a cubic term limits its amplitude. Balancing this cubic restoring term against sampling noise predicts a rounding width proportional to $M^{-1/2}$ in $\epsilon$ and a residual mismatch proportional to $M^{-1/4}$ at the deterministic threshold [Eqs.~\eqref{eq:noisy-flip-step} and~\eqref{eq:noisy-flip-scales}]. Panels (b) and (c) test these predictions. At large $M$ the scaled mismatch collapses onto the predicted curve, where $u$ measures excess gain in noise-width units [Eq.~\eqref{eq:noisy-flip-scales}], and both the measured gain width and critical mismatch follow the predicted $M^{-1/2}$ and $M^{-1/4}$ scalings.

The predictions of the preceding sections therefore survive finite sampling in a definite form. The onset sits where the mean map places it, its width in gain narrows as $M^{-1/2}$, and the residual alternation at threshold decays as $M^{-1/4}$ (Appendix~\ref{app:stochastic_validation}). Real data rarely offer these controls - an electorate or a market supplies a single history at whatever gain and noise level it happens to have, so any test must compare a measured gain with its critical value directly. To apply these predictions to a time series, let $K_{\rm eff}(t)$ be the measured small-mismatch slope of the mandate revision, including any intervening layers, and $\tau_t$ the measured relaxation time. Substituting $\tau_t$ into the scalar threshold of Eq.~\eqref{eq:critical-gain} then gives the dimensionless stability ratio
\begin{equation}
    \Xi_t=\frac{K_{\rm eff}(t)}{K_c(T,\tau_t)}.\label{eq:operating-ratio}
\end{equation}
Variance and serial correlation are often used as warnings of a loss of resilience \cite{Scheffer2009, Dakos2008, Carpenter2006, Kefi2014, Boettiger2012, Clements2016, Ives1995, Carpenter2011}. Under sampling noise the variance is nonzero even far below the deterministic boundary, so on its own it does not locate the boundary. The ratio $\Xi_t$ instead requires two separate measurements: how strongly the revision responds to mismatch, and how quickly the macroscopic state relaxes between censuses. A possible public-data testbed for this theory is the dynamics of inflation expectations \cite{EvansHonkapohja2001, CoibionGorodnichenko2015, Mankiw2002, Carroll2003, Coibion2018}. In this domain, $\theta_t$ is the surveyed inflation forecast, so $y_{\rm ref}(\theta_t)=\theta_t$, $O(x)$ corresponds to realised inflation, and $e_t$ is the forecast error. The reference slope is then $b=1$, and the assumption $b>s$ of Sec.~\ref{sec:normal_form} reads $s<1$: realised inflation moves less than one-for-one with expectations, so pass-through is incomplete. Complete pass-through, $b=s$, would give a whole line of matched states rather than a single one, but the flip boundary survives transverse to that line at $K_c=1+e^{T/\tau}$ [Eq.~\eqref{eq:critical-gain}]. Sudden increases in media salience or structural changes that align public updates can rapidly increase the effective gain $K_{\rm eff}$, driving $\Xi_t$ towards $1$. Empirical validation against survey and macroeconomic time series is deferred to a separate work, but the variables map directly onto the theory developed here (Table~\ref{tab:domains}). Forecast errors and subsequent revisions can be measured separately~\cite{CoibionGorodnichenko2015}. Together with an estimate of how quickly realised inflation relaxes, they would place a real adaptive system on one side or the other of $\Xi_t=1$.

\section{Conclusion}
\label{sec:conclusion}

We have isolated the minimal structure required for a population to reparametrise its environment without tearing the system apart.  The standard view that hierarchy is merely an exogenous architectural constraint for managing complexity hides the necessary cost of collective adaptation. Using the adaptive normal form, we have demonstrated the existence of a critical gain determining stability, which can be controlled via hierarchialisation. Such structure carries its own penalties  however, and we have shown that each layer of a hierarchy delays and attenuates population signals, forcing a trade-off between stability and responsiveness. From this it was possible to categorise the behaviour of a broad class of adaptive systems, and propose the \emph{phases of hierarchy}. Aspects of our results may be derived from well-known constituent materials \cite{Grofman1983, Ladha1992,Galam1986, Galam1990, Kalman1960}, but our contribution here is to develop a unified account under the umbrella of the adaptive normal form. This synthesis demonstrates how the techniques of statistical physics - specifically real-space renormalisation, finite-size scaling, and bifurcation theory - can represent and answer intractable questions of institutional design. By mapping social and organisational structures to block-spin decimations and phase boundaries, the trade-offs in institutional design that are usually argued qualitatively become phase boundaries that can be calculated. We have also shown that the same renormalisation language carries over to a stack of attention layers and to the reverse process of generative diffusion.

This means the results derived here also have direct application to fields such as machine learning. This is particularly relevant to the phenomenon known as the \emph{edge of stability}~\cite{Cohen2021EdgeOfStability}. Recent work has attacked this problem and identified period-two orbits that appear on either side of the critical learning rate~\cite{Litman2026EdgeOfStability}. This is the flip boundary of Eq.~\eqref{eq:critical-gain} in another guise, with the learning rate in the role of the loop gain. Such parallels are at minimum suggestive, and the Laplacian picture of Sec.~\ref{sec:rg_universality} offers a route to deepening this correspondence and enhancing the interpretability of deep neural networks.

Another potential extension concerns heavy-tailed noise. The $\sqrt{N}$ concentration rests on the central limit theorem, but real networks often have power-law degree distributions and oligarchic topologies. Correlation alone does not cap the gain (Appendix~\ref{app:majority}), and a single dominant input removes its growth (Corollary~4). Between these limits the generalised central limit theorem for stable laws could change the scaling exponent itself, either deepening the ultraepistemic catastrophe or opening new stable phases in scale-free architectures.

The framework also offers a sobering perspective on democratic societies. Digital communication can bypass traditional hierarchical filters, raising the effective population $N$ that takes part directly in institutional feedback while shortening the update period $T$. In the model both changes push a majority-driven loop towards the unstable phase. Polarised electorates and whiplash in policy mandates are precisely the signals one might expect when a system has crossed into its unstable phase. The measurements proposed in Sec.~\ref{sec:empirical} set a signpost for further investigation.

A hierarchy need not occupy several layers at once. If we read the index $\ell$ in Eq.~\eqref{eq:disjoint-block-gain} as a generation in a sequence, this work becomes applicable to the nascent field of autoresearch. In this case the output passed on by one generation becomes the input inherited by the next. Along a lineage, a small change in an inherited quantity is multiplied by the local response slope at each handover. This gives a dynamical counterpart to a question we take up in forthcoming companion work: what must a system retain, and at what cost, when the procedure making the next improvement can itself be revised? Self-improving agents from the Gödel machine~\cite{schmidhuber2006godel} to the Darwin Gödel Machine~\cite{zhang2025darwin} and HyperAgents~\cite{zhang2026hyperagents} all implement revisionary lineages of this kind, but at the present moment lack a full dynamical accounting of their stability and efficiency. 

 We close with a general observation. In the present account hierarchy acts as a stability control with a price. Viable systems are neither completely flat nor so deep they cannot be steered. The optimal structure is therefore one with the minimal hierarchy necessary to keep adaptive feedback inside its stability basin. Unfortunately the historic trend has been for bureaucracy to expand to meet the needs of the expanding bureaucracy, far beyond the point at which it is necessary. A little hierarchy is necessary. The difficulty - as anyone who has worked in a large institution will recognise - is stopping at a little.

\begin{acknowledgments}GM wishes to thank Aldo Glielmo for their insightful commentary on an early draft.
Formal derivations, figures and manuscript drafting were assisted by \emph{EigenEngine}, an integrated
research environment and model harness, using ChatGPT and Claude. The author
remains responsible for the ideas, interpretation and correctness of the results.
More information is available at \url{https://gmccaul.co.uk/eigenengine}.
\end{acknowledgments}

\bibliography{../../literature/references}

\begin{thebibliography}{106}%
\makeatletter
\providecommand \@ifxundefined [1]{%
 \@ifx{#1\undefined}
}%
\providecommand \@ifnum [1]{%
 \ifnum #1\expandafter \@firstoftwo
 \else \expandafter \@secondoftwo
 \fi
}%
\providecommand \@ifx [1]{%
 \ifx #1\expandafter \@firstoftwo
 \else \expandafter \@secondoftwo
 \fi
}%
\providecommand \natexlab [1]{#1}%
\providecommand \enquote  [1]{``#1''}%
\providecommand \bibnamefont  [1]{#1}%
\providecommand \bibfnamefont [1]{#1}%
\providecommand \citenamefont [1]{#1}%
\providecommand \href@noop [0]{\@secondoftwo}%
\providecommand \href [0]{\begingroup \@sanitize@url \@href}%
\providecommand \@href[1]{\@@startlink{#1}\@@href}%
\providecommand \@@href[1]{\endgroup#1\@@endlink}%
\providecommand \@sanitize@url [0]{\catcode `\\12\catcode `\$12\catcode
  `\&12\catcode `\#12\catcode `\^12\catcode `\_12\catcode `\%12\relax}%
\providecommand \@@startlink[1]{}%
\providecommand \@@endlink[0]{}%
\providecommand \url  [0]{\begingroup\@sanitize@url \@url }%
\providecommand \@url [1]{\endgroup\@href {#1}{\urlprefix }}%
\providecommand \urlprefix  [0]{URL }%
\providecommand \Eprint [0]{\href }%
\providecommand \doibase [0]{https://doi.org/}%
\providecommand \selectlanguage [0]{\@gobble}%
\providecommand \bibinfo  [0]{\@secondoftwo}%
\providecommand \bibfield  [0]{\@secondoftwo}%
\providecommand \translation [1]{[#1]}%
\providecommand \BibitemOpen [0]{}%
\providecommand \bibitemStop [0]{}%
\providecommand \bibitemNoStop [0]{.\EOS\space}%
\providecommand \EOS [0]{\spacefactor3000\relax}%
\providecommand \BibitemShut  [1]{\csname bibitem#1\endcsname}%
\let\auto@bib@innerbib\@empty
\bibitem [{\citenamefont {West}\ \emph {et~al.}(1997)\citenamefont {West},
  \citenamefont {Brown},\ and\ \citenamefont {Enquist}}]{West1997}%
  \BibitemOpen
  \bibfield  {author} {\bibinfo {author} {\bibfnamefont {G.~B.}\ \bibnamefont
  {West}}, \bibinfo {author} {\bibfnamefont {J.~H.}\ \bibnamefont {Brown}},\
  and\ \bibinfo {author} {\bibfnamefont {B.~J.}\ \bibnamefont {Enquist}},\
  }\bibfield  {title} {\bibinfo {title} {A general model for the origin of
  allometric scaling laws in biology},\ }\href
  {https://doi.org/10.1126/science.276.5309.122} {\bibfield  {journal}
  {\bibinfo  {journal} {Science}\ }\textbf {\bibinfo {volume} {276}},\ \bibinfo
  {pages} {122} (\bibinfo {year} {1997})}\BibitemShut {NoStop}%
\bibitem [{\citenamefont {Pimm}(1982)}]{Pimm1982}%
  \BibitemOpen
  \bibfield  {author} {\bibinfo {author} {\bibfnamefont {S.~L.}\ \bibnamefont
  {Pimm}},\ }\href@noop {} {\emph {\bibinfo {title} {Food Webs}}}\ (\bibinfo
  {publisher} {Chapman and Hall},\ \bibinfo {address} {London},\ \bibinfo
  {year} {1982})\BibitemShut {NoStop}%
\bibitem [{\citenamefont {Chandler}(1962)}]{Chandler1962}%
  \BibitemOpen
  \bibfield  {author} {\bibinfo {author} {\bibfnamefont {A.~D.}\ \bibnamefont
  {Chandler}, \bibfnamefont {Jr.}},\ }\href@noop {} {\emph {\bibinfo {title}
  {Strategy and Structure: Chapters in the History of the {American} Industrial
  Enterprise}}}\ (\bibinfo  {publisher} {MIT Press},\ \bibinfo {address}
  {Cambridge, MA},\ \bibinfo {year} {1962})\BibitemShut {NoStop}%
\bibitem [{\citenamefont {Weber}(1978)}]{Weber1922}%
  \BibitemOpen
  \bibfield  {author} {\bibinfo {author} {\bibfnamefont {M.}~\bibnamefont
  {Weber}},\ }\href@noop {} {\emph {\bibinfo {title} {Economy and Society: An
  Outline of Interpretive Sociology}}},\ edited by\ \bibinfo {editor}
  {\bibfnamefont {G.}~\bibnamefont {Roth}}\ and\ \bibinfo {editor}
  {\bibfnamefont {C.}~\bibnamefont {Wittich}}\ (\bibinfo  {publisher}
  {University of California Press},\ \bibinfo {address} {Berkeley, CA},\
  \bibinfo {year} {1978})\ \bibinfo {note} {english edition}\BibitemShut
  {NoStop}%
\bibitem [{\citenamefont {He}\ \emph {et~al.}(2016)\citenamefont {He},
  \citenamefont {Zhang}, \citenamefont {Ren},\ and\ \citenamefont
  {Sun}}]{He2016}%
  \BibitemOpen
  \bibfield  {author} {\bibinfo {author} {\bibfnamefont {K.}~\bibnamefont
  {He}}, \bibinfo {author} {\bibfnamefont {X.}~\bibnamefont {Zhang}}, \bibinfo
  {author} {\bibfnamefont {S.}~\bibnamefont {Ren}},\ and\ \bibinfo {author}
  {\bibfnamefont {J.}~\bibnamefont {Sun}},\ }\bibfield  {title} {\bibinfo
  {title} {Deep residual learning for image recognition},\ }in\ \href
  {https://doi.org/10.1109/CVPR.2016.90} {\emph {\bibinfo {booktitle} {2016
  {IEEE} Conference on Computer Vision and Pattern Recognition ({CVPR})}}}\
  (\bibinfo {year} {2016})\ pp.\ \bibinfo {pages} {770--778}\BibitemShut
  {NoStop}%
\bibitem [{\citenamefont {Liu}\ \emph {et~al.}(2020)\citenamefont {Liu},
  \citenamefont {Liu}, \citenamefont {Gao}, \citenamefont {Chen},\ and\
  \citenamefont {Han}}]{Liu2020}%
  \BibitemOpen
  \bibfield  {author} {\bibinfo {author} {\bibfnamefont {L.}~\bibnamefont
  {Liu}}, \bibinfo {author} {\bibfnamefont {X.}~\bibnamefont {Liu}}, \bibinfo
  {author} {\bibfnamefont {J.}~\bibnamefont {Gao}}, \bibinfo {author}
  {\bibfnamefont {W.}~\bibnamefont {Chen}},\ and\ \bibinfo {author}
  {\bibfnamefont {J.}~\bibnamefont {Han}},\ }\bibfield  {title} {\bibinfo
  {title} {Understanding the difficulty of training {T}ransformers},\ }in\
  \href {https://doi.org/10.18653/v1/2020.emnlp-main.463} {\emph {\bibinfo
  {booktitle} {Proceedings of the 2020 Conference on Empirical Methods in
  Natural Language Processing (EMNLP)}}}\ (\bibinfo {year} {2020})\ pp.\
  \bibinfo {pages} {5747--5763}\BibitemShut {NoStop}%
\bibitem [{\citenamefont {Doimo}\ \emph {et~al.}(2020)\citenamefont {Doimo},
  \citenamefont {Glielmo}, \citenamefont {Ansuini},\ and\ \citenamefont
  {Laio}}]{Doimo2020Nucleation}%
  \BibitemOpen
  \bibfield  {author} {\bibinfo {author} {\bibfnamefont {D.}~\bibnamefont
  {Doimo}}, \bibinfo {author} {\bibfnamefont {A.}~\bibnamefont {Glielmo}},
  \bibinfo {author} {\bibfnamefont {A.}~\bibnamefont {Ansuini}},\ and\ \bibinfo
  {author} {\bibfnamefont {A.}~\bibnamefont {Laio}},\ }\bibfield  {title}
  {\bibinfo {title} {Hierarchical nucleation in deep neural networks},\ }in\
  \href
  {https://proceedings.neurips.cc/paper_files/paper/2020/hash/54f3bc04830d762a3b56a789b6ff62df-Abstract.html}
  {\emph {\bibinfo {booktitle} {Advances in Neural Information Processing
  Systems}}},\ Vol.~\bibinfo {volume} {33}\ (\bibinfo  {publisher} {Curran
  Associates, Inc.},\ \bibinfo {year} {2020})\ pp.\ \bibinfo {pages}
  {7526--7536},\ \Eprint {https://arxiv.org/abs/2007.03506} {arXiv:2007.03506}
  \BibitemShut {NoStop}%
\bibitem [{\citenamefont {Simon}(1962)}]{Simon1962}%
  \BibitemOpen
  \bibfield  {author} {\bibinfo {author} {\bibfnamefont {H.~A.}\ \bibnamefont
  {Simon}},\ }\bibfield  {title} {\bibinfo {title} {The architecture of
  complexity},\ }\href {https://www.jstor.org/stable/985254} {\bibfield
  {journal} {\bibinfo  {journal} {Proceedings of the American Philosophical
  Society}\ }\textbf {\bibinfo {volume} {106}},\ \bibinfo {pages} {467}
  (\bibinfo {year} {1962})}\BibitemShut {NoStop}%
\bibitem [{\citenamefont {Tsebelis}(2002)}]{Tsebelis2002}%
  \BibitemOpen
  \bibfield  {author} {\bibinfo {author} {\bibfnamefont {G.}~\bibnamefont
  {Tsebelis}},\ }\href@noop {} {\emph {\bibinfo {title} {Veto Players: How
  Political Institutions Work}}}\ (\bibinfo  {publisher} {Princeton University
  Press},\ \bibinfo {address} {Princeton},\ \bibinfo {year} {2002})\BibitemShut
  {NoStop}%
\bibitem [{\citenamefont {North}(1990)}]{North1990}%
  \BibitemOpen
  \bibfield  {author} {\bibinfo {author} {\bibfnamefont {D.~C.}\ \bibnamefont
  {North}},\ }\href {https://doi.org/10.1017/CBO9780511808678} {\emph {\bibinfo
  {title} {Institutions, Institutional Change and Economic Performance}}}\
  (\bibinfo  {publisher} {Cambridge University Press},\ \bibinfo {address}
  {Cambridge},\ \bibinfo {year} {1990})\BibitemShut {NoStop}%
\bibitem [{\citenamefont {Radner}(1993)}]{Radner1993}%
  \BibitemOpen
  \bibfield  {author} {\bibinfo {author} {\bibfnamefont {R.}~\bibnamefont
  {Radner}},\ }\bibfield  {title} {\bibinfo {title} {The organization of
  decentralized information processing},\ }\href
  {https://doi.org/10.2307/2951495} {\bibfield  {journal} {\bibinfo  {journal}
  {Econometrica}\ }\textbf {\bibinfo {volume} {61}},\ \bibinfo {pages} {1109}
  (\bibinfo {year} {1993})}\BibitemShut {NoStop}%
\bibitem [{\citenamefont {Bolton}\ and\ \citenamefont
  {Dewatripont}(1994)}]{Bolton1994}%
  \BibitemOpen
  \bibfield  {author} {\bibinfo {author} {\bibfnamefont {P.}~\bibnamefont
  {Bolton}}\ and\ \bibinfo {author} {\bibfnamefont {M.}~\bibnamefont
  {Dewatripont}},\ }\bibfield  {title} {\bibinfo {title} {The firm as a
  communication network},\ }\href {https://doi.org/10.2307/2118349} {\bibfield
  {journal} {\bibinfo  {journal} {The Quarterly Journal of Economics}\ }\textbf
  {\bibinfo {volume} {109}},\ \bibinfo {pages} {809} (\bibinfo {year}
  {1994})}\BibitemShut {NoStop}%
\bibitem [{\citenamefont {Garicano}(2000)}]{Garicano2000}%
  \BibitemOpen
  \bibfield  {author} {\bibinfo {author} {\bibfnamefont {L.}~\bibnamefont
  {Garicano}},\ }\bibfield  {title} {\bibinfo {title} {Hierarchies and the
  organization of knowledge in production},\ }\href
  {https://doi.org/10.1086/317671} {\bibfield  {journal} {\bibinfo  {journal}
  {Journal of Political Economy}\ }\textbf {\bibinfo {volume} {108}},\ \bibinfo
  {pages} {874} (\bibinfo {year} {2000})}\BibitemShut {NoStop}%
\bibitem [{\citenamefont {Galam}(1986)}]{Galam1986}%
  \BibitemOpen
  \bibfield  {author} {\bibinfo {author} {\bibfnamefont {S.}~\bibnamefont
  {Galam}},\ }\bibfield  {title} {\bibinfo {title} {Majority rule, hierarchical
  structures, and democratic totalitarianism: A statistical approach},\ }\href
  {https://doi.org/10.1016/0022-2496(86)90019-2} {\bibfield  {journal}
  {\bibinfo  {journal} {Journal of Mathematical Psychology}\ }\textbf {\bibinfo
  {volume} {30}},\ \bibinfo {pages} {426} (\bibinfo {year} {1986})}\BibitemShut
  {NoStop}%
\bibitem [{\citenamefont {Galam}(1990)}]{Galam1990}%
  \BibitemOpen
  \bibfield  {author} {\bibinfo {author} {\bibfnamefont {S.}~\bibnamefont
  {Galam}},\ }\bibfield  {title} {\bibinfo {title} {Social paradoxes of
  majority rule voting and renormalization group},\ }\href
  {https://doi.org/10.1007/bf01027314} {\bibfield  {journal} {\bibinfo
  {journal} {Journal of Statistical Physics}\ }\textbf {\bibinfo {volume}
  {61}},\ \bibinfo {pages} {943} (\bibinfo {year} {1990})}\BibitemShut
  {NoStop}%
\bibitem [{\citenamefont {Holling}(1973)}]{Holling1973}%
  \BibitemOpen
  \bibfield  {author} {\bibinfo {author} {\bibfnamefont {C.~S.}\ \bibnamefont
  {Holling}},\ }\bibfield  {title} {\bibinfo {title} {Resilience and stability
  of ecological systems},\ }\href
  {https://doi.org/10.1146/annurev.es.04.110173.000245} {\bibfield  {journal}
  {\bibinfo  {journal} {Annual Review of Ecology and Systematics}\ }\textbf
  {\bibinfo {volume} {4}},\ \bibinfo {pages} {1} (\bibinfo {year}
  {1973})}\BibitemShut {NoStop}%
\bibitem [{\citenamefont {Scheffer}\ \emph {et~al.}(2009)\citenamefont
  {Scheffer}, \citenamefont {Bascompte}, \citenamefont {Brock}, \citenamefont
  {Brovkin}, \citenamefont {Carpenter}, \citenamefont {Dakos}, \citenamefont
  {Held}, \citenamefont {van Nes}, \citenamefont {Rietkerk},\ and\
  \citenamefont {Sugihara}}]{Scheffer2009}%
  \BibitemOpen
  \bibfield  {author} {\bibinfo {author} {\bibfnamefont {M.}~\bibnamefont
  {Scheffer}}, \bibinfo {author} {\bibfnamefont {J.}~\bibnamefont {Bascompte}},
  \bibinfo {author} {\bibfnamefont {W.~A.}\ \bibnamefont {Brock}}, \bibinfo
  {author} {\bibfnamefont {V.}~\bibnamefont {Brovkin}}, \bibinfo {author}
  {\bibfnamefont {S.~R.}\ \bibnamefont {Carpenter}}, \bibinfo {author}
  {\bibfnamefont {V.}~\bibnamefont {Dakos}}, \bibinfo {author} {\bibfnamefont
  {H.}~\bibnamefont {Held}}, \bibinfo {author} {\bibfnamefont {E.~H.}\
  \bibnamefont {van Nes}}, \bibinfo {author} {\bibfnamefont {M.}~\bibnamefont
  {Rietkerk}},\ and\ \bibinfo {author} {\bibfnamefont {G.}~\bibnamefont
  {Sugihara}},\ }\bibfield  {title} {\bibinfo {title} {Early-warning signals
  for critical transitions},\ }\href {https://doi.org/10.1038/nature08227}
  {\bibfield  {journal} {\bibinfo  {journal} {Nature}\ }\textbf {\bibinfo
  {volume} {461}},\ \bibinfo {pages} {53} (\bibinfo {year} {2009})}\BibitemShut
  {NoStop}%
\bibitem [{\citenamefont {Goodwin}(1967)}]{Goodwin1967}%
  \BibitemOpen
  \bibfield  {author} {\bibinfo {author} {\bibfnamefont {R.~M.}\ \bibnamefont
  {Goodwin}},\ }\bibfield  {title} {\bibinfo {title} {A growth cycle},\ }in\
  \href@noop {} {\emph {\bibinfo {booktitle} {Socialism, Capitalism and
  Economic Growth: Essays Presented to {Maurice Dobb}}}},\ \bibinfo {editor}
  {edited by\ \bibinfo {editor} {\bibfnamefont {C.~H.}\ \bibnamefont
  {Feinstein}}}\ (\bibinfo  {publisher} {Cambridge University Press},\ \bibinfo
  {year} {1967})\ pp.\ \bibinfo {pages} {54--58}\BibitemShut {NoStop}%
\bibitem [{\citenamefont {Ezekiel}(1938)}]{Ezekiel1938}%
  \BibitemOpen
  \bibfield  {author} {\bibinfo {author} {\bibfnamefont {M.}~\bibnamefont
  {Ezekiel}},\ }\bibfield  {title} {\bibinfo {title} {The cobweb theorem},\
  }\href {https://doi.org/10.2307/1881734} {\bibfield  {journal} {\bibinfo
  {journal} {The Quarterly Journal of Economics}\ }\textbf {\bibinfo {volume}
  {52}},\ \bibinfo {pages} {255} (\bibinfo {year} {1938})}\BibitemShut
  {NoStop}%
\bibitem [{\citenamefont {Muth}(1961)}]{Muth1961}%
  \BibitemOpen
  \bibfield  {author} {\bibinfo {author} {\bibfnamefont {J.~F.}\ \bibnamefont
  {Muth}},\ }\bibfield  {title} {\bibinfo {title} {Rational expectations and
  the theory of price movements},\ }\href {https://doi.org/10.2307/1909635}
  {\bibfield  {journal} {\bibinfo  {journal} {Econometrica}\ }\textbf {\bibinfo
  {volume} {29}},\ \bibinfo {pages} {315} (\bibinfo {year} {1961})}\BibitemShut
  {NoStop}%
\bibitem [{\citenamefont {Evans}\ and\ \citenamefont
  {Honkapohja}(2001)}]{EvansHonkapohja2001}%
  \BibitemOpen
  \bibfield  {author} {\bibinfo {author} {\bibfnamefont {G.~W.}\ \bibnamefont
  {Evans}}\ and\ \bibinfo {author} {\bibfnamefont {S.}~\bibnamefont
  {Honkapohja}},\ }\href@noop {} {\emph {\bibinfo {title} {Learning and
  Expectations in Macroeconomics}}}\ (\bibinfo  {publisher} {Princeton
  University Press},\ \bibinfo {address} {Princeton},\ \bibinfo {year}
  {2001})\BibitemShut {NoStop}%
\bibitem [{\citenamefont {Mankiw}\ and\ \citenamefont
  {Reis}(2002)}]{Mankiw2002}%
  \BibitemOpen
  \bibfield  {author} {\bibinfo {author} {\bibfnamefont {N.~G.}\ \bibnamefont
  {Mankiw}}\ and\ \bibinfo {author} {\bibfnamefont {R.}~\bibnamefont {Reis}},\
  }\bibfield  {title} {\bibinfo {title} {Sticky information versus sticky
  prices: a proposal to replace the {N}ew {K}eynesian {P}hillips curve},\
  }\href {https://doi.org/10.1162/003355302320935034} {\bibfield  {journal}
  {\bibinfo  {journal} {The Quarterly Journal of Economics}\ }\textbf {\bibinfo
  {volume} {117}},\ \bibinfo {pages} {1295} (\bibinfo {year}
  {2002})}\BibitemShut {NoStop}%
\bibitem [{\citenamefont {Carroll}(2003)}]{Carroll2003}%
  \BibitemOpen
  \bibfield  {author} {\bibinfo {author} {\bibfnamefont {C.~D.}\ \bibnamefont
  {Carroll}},\ }\bibfield  {title} {\bibinfo {title} {Macroeconomic
  expectations of households and professional forecasters},\ }\href
  {https://doi.org/10.1162/00335530360535207} {\bibfield  {journal} {\bibinfo
  {journal} {The Quarterly Journal of Economics}\ }\textbf {\bibinfo {volume}
  {118}},\ \bibinfo {pages} {269} (\bibinfo {year} {2003})}\BibitemShut
  {NoStop}%
\bibitem [{\citenamefont {Coibion}\ \emph {et~al.}(2018)\citenamefont
  {Coibion}, \citenamefont {Gorodnichenko},\ and\ \citenamefont
  {Kamdar}}]{Coibion2018}%
  \BibitemOpen
  \bibfield  {author} {\bibinfo {author} {\bibfnamefont {O.}~\bibnamefont
  {Coibion}}, \bibinfo {author} {\bibfnamefont {Y.}~\bibnamefont
  {Gorodnichenko}},\ and\ \bibinfo {author} {\bibfnamefont {R.}~\bibnamefont
  {Kamdar}},\ }\bibfield  {title} {\bibinfo {title} {The formation of
  expectations, inflation, and the {P}hillips curve},\ }\href
  {https://doi.org/10.1257/jel.20171300} {\bibfield  {journal} {\bibinfo
  {journal} {Journal of Economic Literature}\ }\textbf {\bibinfo {volume}
  {56}},\ \bibinfo {pages} {1447} (\bibinfo {year} {2018})}\BibitemShut
  {NoStop}%
\bibitem [{\citenamefont {Sims}(2003)}]{Sims2003}%
  \BibitemOpen
  \bibfield  {author} {\bibinfo {author} {\bibfnamefont {C.~A.}\ \bibnamefont
  {Sims}},\ }\bibfield  {title} {\bibinfo {title} {Implications of rational
  inattention},\ }\href {https://doi.org/10.1016/s0304-3932(03)00029-1}
  {\bibfield  {journal} {\bibinfo  {journal} {Journal of Monetary Economics}\
  }\textbf {\bibinfo {volume} {50}},\ \bibinfo {pages} {665} (\bibinfo {year}
  {2003})}\BibitemShut {NoStop}%
\bibitem [{\citenamefont {Woodford}(2003)}]{Woodford2002}%
  \BibitemOpen
  \bibfield  {author} {\bibinfo {author} {\bibfnamefont {M.}~\bibnamefont
  {Woodford}},\ }\bibfield  {title} {\bibinfo {title} {Imperfect common
  knowledge and the effects of monetary policy},\ }in\ \href
  {https://doi.org/10.1515/9780691223933-004} {\emph {\bibinfo {booktitle}
  {Knowledge, Information, and Expectations in Modern Macroeconomics: In Honor
  of {Edmund S. Phelps}}}},\ \bibinfo {editor} {edited by\ \bibinfo {editor}
  {\bibfnamefont {P.}~\bibnamefont {Aghion}}, \bibinfo {editor} {\bibfnamefont
  {R.}~\bibnamefont {Frydman}}, \bibinfo {editor} {\bibfnamefont {J.~E.}\
  \bibnamefont {Stiglitz}},\ and\ \bibinfo {editor} {\bibfnamefont
  {M.}~\bibnamefont {Woodford}}}\ (\bibinfo  {publisher} {Princeton University
  Press},\ \bibinfo {year} {2003})\ pp.\ \bibinfo {pages} {23--58}\BibitemShut
  {NoStop}%
\bibitem [{\citenamefont {Malmendier}\ and\ \citenamefont
  {Nagel}(2016)}]{Malmendier2016}%
  \BibitemOpen
  \bibfield  {author} {\bibinfo {author} {\bibfnamefont {U.}~\bibnamefont
  {Malmendier}}\ and\ \bibinfo {author} {\bibfnamefont {S.}~\bibnamefont
  {Nagel}},\ }\bibfield  {title} {\bibinfo {title} {Learning from inflation
  experiences},\ }\href {https://doi.org/10.1093/qje/qjv037} {\bibfield
  {journal} {\bibinfo  {journal} {The Quarterly Journal of Economics}\ }\textbf
  {\bibinfo {volume} {131}},\ \bibinfo {pages} {53} (\bibinfo {year}
  {2016})}\BibitemShut {NoStop}%
\bibitem [{\citenamefont {Wlezien}(1995)}]{Wlezien1995}%
  \BibitemOpen
  \bibfield  {author} {\bibinfo {author} {\bibfnamefont {C.}~\bibnamefont
  {Wlezien}},\ }\bibfield  {title} {\bibinfo {title} {The public as thermostat:
  Dynamics of preferences for spending},\ }\href
  {https://doi.org/10.2307/2111666} {\bibfield  {journal} {\bibinfo  {journal}
  {American Journal of Political Science}\ }\textbf {\bibinfo {volume} {39}},\
  \bibinfo {pages} {981} (\bibinfo {year} {1995})}\BibitemShut {NoStop}%
\bibitem [{\citenamefont {Stimson}\ \emph {et~al.}(1995)\citenamefont
  {Stimson}, \citenamefont {Mackuen},\ and\ \citenamefont
  {Erikson}}]{Stimson1995}%
  \BibitemOpen
  \bibfield  {author} {\bibinfo {author} {\bibfnamefont {J.~A.}\ \bibnamefont
  {Stimson}}, \bibinfo {author} {\bibfnamefont {M.~B.}\ \bibnamefont
  {Mackuen}},\ and\ \bibinfo {author} {\bibfnamefont {R.~S.}\ \bibnamefont
  {Erikson}},\ }\bibfield  {title} {\bibinfo {title} {Dynamic representation},\
  }\href {https://doi.org/10.2307/2082973} {\bibfield  {journal} {\bibinfo
  {journal} {American Political Science Review}\ }\textbf {\bibinfo {volume}
  {89}},\ \bibinfo {pages} {543} (\bibinfo {year} {1995})}\BibitemShut
  {NoStop}%
\bibitem [{\citenamefont {Ferejohn}(1986)}]{Ferejohn1986}%
  \BibitemOpen
  \bibfield  {author} {\bibinfo {author} {\bibfnamefont {J.}~\bibnamefont
  {Ferejohn}},\ }\bibfield  {title} {\bibinfo {title} {Incumbent performance
  and electoral control},\ }\href {https://doi.org/10.1007/BF00124924}
  {\bibfield  {journal} {\bibinfo  {journal} {Public Choice}\ }\textbf
  {\bibinfo {volume} {50}},\ \bibinfo {pages} {5} (\bibinfo {year}
  {1986})}\BibitemShut {NoStop}%
\bibitem [{\citenamefont {Perdomo}\ \emph {et~al.}(2020)\citenamefont
  {Perdomo}, \citenamefont {Zrnic}, \citenamefont {Mendler-D\"unner},\ and\
  \citenamefont {Hardt}}]{Perdomo2020}%
  \BibitemOpen
  \bibfield  {author} {\bibinfo {author} {\bibfnamefont {J.~C.}\ \bibnamefont
  {Perdomo}}, \bibinfo {author} {\bibfnamefont {T.}~\bibnamefont {Zrnic}},
  \bibinfo {author} {\bibfnamefont {C.}~\bibnamefont {Mendler-D\"unner}},\ and\
  \bibinfo {author} {\bibfnamefont {M.}~\bibnamefont {Hardt}},\ }\bibfield
  {title} {\bibinfo {title} {Performative prediction},\ }in\ \href
  {https://proceedings.mlr.press/v119/perdomo20a.html} {\emph {\bibinfo
  {booktitle} {Proceedings of the 37th International Conference on Machine
  Learning}}},\ \bibinfo {series} {Proceedings of Machine Learning Research},
  Vol.\ \bibinfo {volume} {119}\ (\bibinfo {year} {2020})\ pp.\ \bibinfo
  {pages} {7599--7609}\BibitemShut {NoStop}%
\bibitem [{\citenamefont {Shumailov}\ \emph {et~al.}(2024)\citenamefont
  {Shumailov}, \citenamefont {Shumaylov}, \citenamefont {Zhao}, \citenamefont
  {Papernot}, \citenamefont {Anderson},\ and\ \citenamefont
  {Gal}}]{Shumailov2024}%
  \BibitemOpen
  \bibfield  {author} {\bibinfo {author} {\bibfnamefont {I.}~\bibnamefont
  {Shumailov}}, \bibinfo {author} {\bibfnamefont {Z.}~\bibnamefont
  {Shumaylov}}, \bibinfo {author} {\bibfnamefont {Y.}~\bibnamefont {Zhao}},
  \bibinfo {author} {\bibfnamefont {N.}~\bibnamefont {Papernot}}, \bibinfo
  {author} {\bibfnamefont {R.}~\bibnamefont {Anderson}},\ and\ \bibinfo
  {author} {\bibfnamefont {Y.}~\bibnamefont {Gal}},\ }\bibfield  {title}
  {\bibinfo {title} {{AI} models collapse when trained on recursively generated
  data},\ }\href {https://doi.org/10.1038/s41586-024-07566-y} {\bibfield
  {journal} {\bibinfo  {journal} {Nature}\ }\textbf {\bibinfo {volume} {631}},\
  \bibinfo {pages} {755} (\bibinfo {year} {2024})}\BibitemShut {NoStop}%
\bibitem [{\citenamefont {Beer}(1972)}]{Beer1972}%
  \BibitemOpen
  \bibfield  {author} {\bibinfo {author} {\bibfnamefont {S.}~\bibnamefont
  {Beer}},\ }\href@noop {} {\emph {\bibinfo {title} {Brain of the Firm: The
  Managerial Cybernetics of Organization}}}\ (\bibinfo  {publisher} {Allen
  Lane},\ \bibinfo {address} {London},\ \bibinfo {year} {1972})\BibitemShut
  {NoStop}%
\bibitem [{\citenamefont {Conant}\ and\ \citenamefont
  {Ashby}(1970)}]{ConantAshby1970}%
  \BibitemOpen
  \bibfield  {author} {\bibinfo {author} {\bibfnamefont {R.~C.}\ \bibnamefont
  {Conant}}\ and\ \bibinfo {author} {\bibfnamefont {W.~R.}\ \bibnamefont
  {Ashby}},\ }\bibfield  {title} {\bibinfo {title} {Every good regulator of a
  system must be a model of that system},\ }\href
  {https://doi.org/10.1080/00207727008920220} {\bibfield  {journal} {\bibinfo
  {journal} {International Journal of Systems Science}\ }\textbf {\bibinfo
  {volume} {1}},\ \bibinfo {pages} {89} (\bibinfo {year} {1970})}\BibitemShut
  {NoStop}%
\bibitem [{\citenamefont {Odling-Smee}\ \emph {et~al.}(1996)\citenamefont
  {Odling-Smee}, \citenamefont {Laland},\ and\ \citenamefont
  {Feldman}}]{OdlingSmee1996}%
  \BibitemOpen
  \bibfield  {author} {\bibinfo {author} {\bibfnamefont {F.~J.}\ \bibnamefont
  {Odling-Smee}}, \bibinfo {author} {\bibfnamefont {K.~N.}\ \bibnamefont
  {Laland}},\ and\ \bibinfo {author} {\bibfnamefont {M.~W.}\ \bibnamefont
  {Feldman}},\ }\bibfield  {title} {\bibinfo {title} {Niche construction},\
  }\href {https://doi.org/10.1086/285870} {\bibfield  {journal} {\bibinfo
  {journal} {The American Naturalist}\ }\textbf {\bibinfo {volume} {147}},\
  \bibinfo {pages} {641} (\bibinfo {year} {1996})}\BibitemShut {NoStop}%
\bibitem [{\citenamefont {Lenton}\ \emph {et~al.}(2008)\citenamefont {Lenton},
  \citenamefont {Held}, \citenamefont {Kriegler}, \citenamefont {Hall},
  \citenamefont {Lucht}, \citenamefont {Rahmstorf},\ and\ \citenamefont
  {Schellnhuber}}]{Lenton2008}%
  \BibitemOpen
  \bibfield  {author} {\bibinfo {author} {\bibfnamefont {T.~M.}\ \bibnamefont
  {Lenton}}, \bibinfo {author} {\bibfnamefont {H.}~\bibnamefont {Held}},
  \bibinfo {author} {\bibfnamefont {E.}~\bibnamefont {Kriegler}}, \bibinfo
  {author} {\bibfnamefont {J.~W.}\ \bibnamefont {Hall}}, \bibinfo {author}
  {\bibfnamefont {W.}~\bibnamefont {Lucht}}, \bibinfo {author} {\bibfnamefont
  {S.}~\bibnamefont {Rahmstorf}},\ and\ \bibinfo {author} {\bibfnamefont
  {H.~J.}\ \bibnamefont {Schellnhuber}},\ }\bibfield  {title} {\bibinfo {title}
  {Tipping elements in the {E}arth's climate system},\ }\href
  {https://doi.org/10.1073/pnas.0705414105} {\bibfield  {journal} {\bibinfo
  {journal} {Proceedings of the National Academy of Sciences}\ }\textbf
  {\bibinfo {volume} {105}},\ \bibinfo {pages} {1786} (\bibinfo {year}
  {2008})}\BibitemShut {NoStop}%
\bibitem [{\citenamefont {Boettiger}\ and\ \citenamefont
  {Hastings}(2012)}]{Boettiger2012}%
  \BibitemOpen
  \bibfield  {author} {\bibinfo {author} {\bibfnamefont {C.}~\bibnamefont
  {Boettiger}}\ and\ \bibinfo {author} {\bibfnamefont {A.}~\bibnamefont
  {Hastings}},\ }\bibfield  {title} {\bibinfo {title} {Early warning signals
  and the prosecutor's fallacy},\ }\href
  {https://doi.org/10.1098/rspb.2012.2085} {\bibfield  {journal} {\bibinfo
  {journal} {Proceedings of the Royal Society B: Biological Sciences}\ }\textbf
  {\bibinfo {volume} {279}},\ \bibinfo {pages} {4734} (\bibinfo {year}
  {2012})}\BibitemShut {NoStop}%
\bibitem [{\citenamefont {Clements}\ and\ \citenamefont
  {Ozgul}(2018)}]{Clements2016}%
  \BibitemOpen
  \bibfield  {author} {\bibinfo {author} {\bibfnamefont {C.~F.}\ \bibnamefont
  {Clements}}\ and\ \bibinfo {author} {\bibfnamefont {A.}~\bibnamefont
  {Ozgul}},\ }\bibfield  {title} {\bibinfo {title} {Indicators of transitions
  in biological systems},\ }\href {https://doi.org/10.1111/ele.12948}
  {\bibfield  {journal} {\bibinfo  {journal} {Ecology Letters}\ }\textbf
  {\bibinfo {volume} {21}},\ \bibinfo {pages} {905} (\bibinfo {year}
  {2018})}\BibitemShut {NoStop}%
\bibitem [{\citenamefont {Fussmann}\ \emph {et~al.}(2007)\citenamefont
  {Fussmann}, \citenamefont {Loreau},\ and\ \citenamefont
  {Abrams}}]{Fussmann2007}%
  \BibitemOpen
  \bibfield  {author} {\bibinfo {author} {\bibfnamefont {G.~F.}\ \bibnamefont
  {Fussmann}}, \bibinfo {author} {\bibfnamefont {M.}~\bibnamefont {Loreau}},\
  and\ \bibinfo {author} {\bibfnamefont {P.~A.}\ \bibnamefont {Abrams}},\
  }\bibfield  {title} {\bibinfo {title} {Eco-evolutionary dynamics of
  communities and ecosystems},\ }\href
  {https://doi.org/10.1111/j.1365-2435.2007.01275.x} {\bibfield  {journal}
  {\bibinfo  {journal} {Functional Ecology}\ }\textbf {\bibinfo {volume}
  {21}},\ \bibinfo {pages} {465} (\bibinfo {year} {2007})}\BibitemShut
  {NoStop}%
\bibitem [{\citenamefont {Granovetter}(1978)}]{Granovetter1978}%
  \BibitemOpen
  \bibfield  {author} {\bibinfo {author} {\bibfnamefont {M.}~\bibnamefont
  {Granovetter}},\ }\bibfield  {title} {\bibinfo {title} {Threshold models of
  collective behavior},\ }\href {https://doi.org/10.1086/226707} {\bibfield
  {journal} {\bibinfo  {journal} {American Journal of Sociology}\ }\textbf
  {\bibinfo {volume} {83}},\ \bibinfo {pages} {1420} (\bibinfo {year}
  {1978})}\BibitemShut {NoStop}%
\bibitem [{\citenamefont {Schelling}(1971)}]{Schelling1971}%
  \BibitemOpen
  \bibfield  {author} {\bibinfo {author} {\bibfnamefont {T.~C.}\ \bibnamefont
  {Schelling}},\ }\bibfield  {title} {\bibinfo {title} {Dynamic models of
  segregation},\ }\href {https://doi.org/10.1080/0022250X.1971.9989794}
  {\bibfield  {journal} {\bibinfo  {journal} {The Journal of Mathematical
  Sociology}\ }\textbf {\bibinfo {volume} {1}},\ \bibinfo {pages} {143}
  (\bibinfo {year} {1971})}\BibitemShut {NoStop}%
\bibitem [{\citenamefont {Holme}\ and\ \citenamefont
  {Newman}(2006)}]{HolmeNewman2006}%
  \BibitemOpen
  \bibfield  {author} {\bibinfo {author} {\bibfnamefont {P.}~\bibnamefont
  {Holme}}\ and\ \bibinfo {author} {\bibfnamefont {M.~E.~J.}\ \bibnamefont
  {Newman}},\ }\bibfield  {title} {\bibinfo {title} {Nonequilibrium phase
  transition in the coevolution of networks and opinions},\ }\href
  {https://doi.org/10.1103/PhysRevE.74.056108} {\bibfield  {journal} {\bibinfo
  {journal} {Physical Review E}\ }\textbf {\bibinfo {volume} {74}},\ \bibinfo
  {pages} {056108} (\bibinfo {year} {2006})}\BibitemShut {NoStop}%
\bibitem [{\citenamefont {Arrow}(1951)}]{Arrow1951}%
  \BibitemOpen
  \bibfield  {author} {\bibinfo {author} {\bibfnamefont {K.~J.}\ \bibnamefont
  {Arrow}},\ }\href@noop {} {\emph {\bibinfo {title} {Social Choice and
  Individual Values}}}\ (\bibinfo  {publisher} {Wiley},\ \bibinfo {address}
  {New York},\ \bibinfo {year} {1951})\BibitemShut {NoStop}%
\bibitem [{\citenamefont {Estlund}(2008)}]{Estlund2008}%
  \BibitemOpen
  \bibfield  {author} {\bibinfo {author} {\bibfnamefont {D.~M.}\ \bibnamefont
  {Estlund}},\ }\href@noop {} {\emph {\bibinfo {title} {Democratic Authority: A
  Philosophical Framework}}}\ (\bibinfo  {publisher} {Princeton University
  Press},\ \bibinfo {address} {Princeton},\ \bibinfo {year} {2008})\BibitemShut
  {NoStop}%
\bibitem [{\citenamefont {Landemore}(2012)}]{Landemore2012}%
  \BibitemOpen
  \bibfield  {author} {\bibinfo {author} {\bibfnamefont {H.}~\bibnamefont
  {Landemore}},\ }\href@noop {} {\emph {\bibinfo {title} {Democratic Reason:
  Politics, Collective Intelligence, and the Rule of the Many}}}\ (\bibinfo
  {publisher} {Princeton University Press},\ \bibinfo {address} {Princeton},\
  \bibinfo {year} {2012})\BibitemShut {NoStop}%
\bibitem [{\citenamefont {List}\ and\ \citenamefont
  {Goodin}(2001)}]{ListGoodin2001}%
  \BibitemOpen
  \bibfield  {author} {\bibinfo {author} {\bibfnamefont {C.}~\bibnamefont
  {List}}\ and\ \bibinfo {author} {\bibfnamefont {R.~E.}\ \bibnamefont
  {Goodin}},\ }\bibfield  {title} {\bibinfo {title} {Epistemic democracy:
  Generalizing the {C}ondorcet jury theorem},\ }\href
  {https://doi.org/10.1111/1467-9760.00128} {\bibfield  {journal} {\bibinfo
  {journal} {Journal of Political Philosophy}\ }\textbf {\bibinfo {volume}
  {9}},\ \bibinfo {pages} {277} (\bibinfo {year} {2001})}\BibitemShut {NoStop}%
\bibitem [{\citenamefont {Ladha}(1992)}]{Ladha1992}%
  \BibitemOpen
  \bibfield  {author} {\bibinfo {author} {\bibfnamefont {K.~K.}\ \bibnamefont
  {Ladha}},\ }\bibfield  {title} {\bibinfo {title} {The {C}ondorcet jury
  theorem, free speech, and correlated votes},\ }\href
  {https://doi.org/10.2307/2111584} {\bibfield  {journal} {\bibinfo  {journal}
  {American Journal of Political Science}\ }\textbf {\bibinfo {volume} {36}},\
  \bibinfo {pages} {617} (\bibinfo {year} {1992})}\BibitemShut {NoStop}%
\bibitem [{\citenamefont {Grofman}\ \emph {et~al.}(1983)\citenamefont
  {Grofman}, \citenamefont {Owen},\ and\ \citenamefont {Feld}}]{Grofman1983}%
  \BibitemOpen
  \bibfield  {author} {\bibinfo {author} {\bibfnamefont {B.}~\bibnamefont
  {Grofman}}, \bibinfo {author} {\bibfnamefont {G.}~\bibnamefont {Owen}},\ and\
  \bibinfo {author} {\bibfnamefont {S.~L.}\ \bibnamefont {Feld}},\ }\bibfield
  {title} {\bibinfo {title} {Thirteen theorems in search of the truth},\ }\href
  {https://doi.org/10.1007/bf00125672} {\bibfield  {journal} {\bibinfo
  {journal} {Theory and Decision}\ }\textbf {\bibinfo {volume} {15}},\ \bibinfo
  {pages} {261} (\bibinfo {year} {1983})}\BibitemShut {NoStop}%
\bibitem [{\citenamefont {Austen-Smith}\ and\ \citenamefont
  {Banks}(1996)}]{AustenSmith1996}%
  \BibitemOpen
  \bibfield  {author} {\bibinfo {author} {\bibfnamefont {D.}~\bibnamefont
  {Austen-Smith}}\ and\ \bibinfo {author} {\bibfnamefont {J.~S.}\ \bibnamefont
  {Banks}},\ }\bibfield  {title} {\bibinfo {title} {Information aggregation,
  rationality, and the {C}ondorcet jury theorem},\ }\href
  {https://doi.org/10.2307/2082796} {\bibfield  {journal} {\bibinfo  {journal}
  {American Political Science Review}\ }\textbf {\bibinfo {volume} {90}},\
  \bibinfo {pages} {34} (\bibinfo {year} {1996})}\BibitemShut {NoStop}%
\bibitem [{\citenamefont {Kao}\ and\ \citenamefont {Couzin}(2014)}]{Kao2014}%
  \BibitemOpen
  \bibfield  {author} {\bibinfo {author} {\bibfnamefont {A.~B.}\ \bibnamefont
  {Kao}}\ and\ \bibinfo {author} {\bibfnamefont {I.~D.}\ \bibnamefont
  {Couzin}},\ }\bibfield  {title} {\bibinfo {title} {Decision accuracy in
  complex environments is often maximized by small group sizes},\ }\href
  {https://doi.org/10.1098/rspb.2013.3305} {\bibfield  {journal} {\bibinfo
  {journal} {Proceedings of the Royal Society B: Biological Sciences}\ }\textbf
  {\bibinfo {volume} {281}},\ \bibinfo {pages} {20133305} (\bibinfo {year}
  {2014})}\BibitemShut {NoStop}%
\bibitem [{\citenamefont {Conradt}\ and\ \citenamefont
  {Roper}(2003)}]{Conradt2003}%
  \BibitemOpen
  \bibfield  {author} {\bibinfo {author} {\bibfnamefont {L.}~\bibnamefont
  {Conradt}}\ and\ \bibinfo {author} {\bibfnamefont {T.~J.}\ \bibnamefont
  {Roper}},\ }\bibfield  {title} {\bibinfo {title} {Group decision-making in
  animals},\ }\href {https://doi.org/10.1038/nature01294} {\bibfield  {journal}
  {\bibinfo  {journal} {Nature}\ }\textbf {\bibinfo {volume} {421}},\ \bibinfo
  {pages} {155} (\bibinfo {year} {2003})}\BibitemShut {NoStop}%
\bibitem [{\citenamefont {Sumpter}(2010)}]{Sumpter2010}%
  \BibitemOpen
  \bibfield  {author} {\bibinfo {author} {\bibfnamefont {D.~J.~T.}\
  \bibnamefont {Sumpter}},\ }\href@noop {} {\emph {\bibinfo {title} {Collective
  animal behavior}}}\ (\bibinfo  {publisher} {Princeton University Press},\
  \bibinfo {year} {2010})\BibitemShut {NoStop}%
\bibitem [{\citenamefont {Brock}\ and\ \citenamefont
  {Hommes}(1997)}]{BrockHommes1997}%
  \BibitemOpen
  \bibfield  {author} {\bibinfo {author} {\bibfnamefont {W.~A.}\ \bibnamefont
  {Brock}}\ and\ \bibinfo {author} {\bibfnamefont {C.~H.}\ \bibnamefont
  {Hommes}},\ }\bibfield  {title} {\bibinfo {title} {A rational route to
  randomness},\ }\href {https://doi.org/10.2307/2171879} {\bibfield  {journal}
  {\bibinfo  {journal} {Econometrica}\ }\textbf {\bibinfo {volume} {65}},\
  \bibinfo {pages} {1059} (\bibinfo {year} {1997})}\BibitemShut {NoStop}%
\bibitem [{\citenamefont {Brock}\ and\ \citenamefont
  {Hommes}(1998)}]{BrockHommes1998}%
  \BibitemOpen
  \bibfield  {author} {\bibinfo {author} {\bibfnamefont {W.~A.}\ \bibnamefont
  {Brock}}\ and\ \bibinfo {author} {\bibfnamefont {C.~H.}\ \bibnamefont
  {Hommes}},\ }\bibfield  {title} {\bibinfo {title} {Heterogeneous beliefs and
  routes to chaos in a simple asset pricing model},\ }\href
  {https://doi.org/10.1016/S0165-1889(98)00011-6} {\bibfield  {journal}
  {\bibinfo  {journal} {Journal of Economic Dynamics and Control}\ }\textbf
  {\bibinfo {volume} {22}},\ \bibinfo {pages} {1235} (\bibinfo {year}
  {1998})}\BibitemShut {NoStop}%
\bibitem [{\citenamefont {Nerlove}(1958)}]{Nerlove1958}%
  \BibitemOpen
  \bibfield  {author} {\bibinfo {author} {\bibfnamefont {M.}~\bibnamefont
  {Nerlove}},\ }\bibfield  {title} {\bibinfo {title} {Adaptive expectations and
  cobweb phenomena},\ }\href {https://doi.org/10.2307/1880597} {\bibfield
  {journal} {\bibinfo  {journal} {The Quarterly Journal of Economics}\ }\textbf
  {\bibinfo {volume} {72}},\ \bibinfo {pages} {227} (\bibinfo {year}
  {1958})}\BibitemShut {NoStop}%
\bibitem [{\citenamefont {Ashby}(1956)}]{Ashby1956}%
  \BibitemOpen
  \bibfield  {author} {\bibinfo {author} {\bibfnamefont {W.~R.}\ \bibnamefont
  {Ashby}},\ }\href@noop {} {\emph {\bibinfo {title} {An Introduction to
  Cybernetics}}}\ (\bibinfo  {publisher} {Chapman and Hall},\ \bibinfo
  {address} {London},\ \bibinfo {year} {1956})\BibitemShut {NoStop}%
\bibitem [{\citenamefont {Ostrom}(1990)}]{Ostrom1990}%
  \BibitemOpen
  \bibfield  {author} {\bibinfo {author} {\bibfnamefont {E.}~\bibnamefont
  {Ostrom}},\ }\href {https://doi.org/10.1017/CBO9780511807763} {\emph
  {\bibinfo {title} {Governing the Commons: The Evolution of Institutions for
  Collective Action}}}\ (\bibinfo  {publisher} {Cambridge University Press},\
  \bibinfo {address} {Cambridge},\ \bibinfo {year} {1990})\BibitemShut
  {NoStop}%
\bibitem [{\citenamefont {Ostrom}(2010)}]{Ostrom2010}%
  \BibitemOpen
  \bibfield  {author} {\bibinfo {author} {\bibfnamefont {E.}~\bibnamefont
  {Ostrom}},\ }\bibfield  {title} {\bibinfo {title} {Polycentric systems for
  coping with collective action and global environmental change},\ }\href
  {https://doi.org/10.1016/j.gloenvcha.2010.07.004} {\bibfield  {journal}
  {\bibinfo  {journal} {Global Environmental Change}\ }\textbf {\bibinfo
  {volume} {20}},\ \bibinfo {pages} {550} (\bibinfo {year} {2010})}\BibitemShut
  {NoStop}%
\bibitem [{\citenamefont {Mesarovi{\'c}}\ \emph {et~al.}(1970)\citenamefont
  {Mesarovi{\'c}}, \citenamefont {Macko},\ and\ \citenamefont
  {Takahara}}]{Mesarovic1970}%
  \BibitemOpen
  \bibfield  {author} {\bibinfo {author} {\bibfnamefont {M.~D.}\ \bibnamefont
  {Mesarovi{\'c}}}, \bibinfo {author} {\bibfnamefont {D.}~\bibnamefont
  {Macko}},\ and\ \bibinfo {author} {\bibfnamefont {Y.}~\bibnamefont
  {Takahara}},\ }\href@noop {} {\emph {\bibinfo {title} {Theory of
  hierarchical, multilevel, systems}}}\ (\bibinfo  {publisher} {Academic
  Press},\ \bibinfo {year} {1970})\BibitemShut {NoStop}%
\bibitem [{\citenamefont {Salthe}(1985)}]{Salthe1985}%
  \BibitemOpen
  \bibfield  {author} {\bibinfo {author} {\bibfnamefont {S.~N.}\ \bibnamefont
  {Salthe}},\ }\href@noop {} {\emph {\bibinfo {title} {Evolving hierarchical
  systems: their structure and representation}}}\ (\bibinfo  {publisher}
  {Columbia University Press},\ \bibinfo {year} {1985})\BibitemShut {NoStop}%
\bibitem [{\citenamefont {Migdal}(1975)}]{Migdal1975}%
  \BibitemOpen
  \bibfield  {author} {\bibinfo {author} {\bibfnamefont {A.~A.}\ \bibnamefont
  {Migdal}},\ }\bibfield  {title} {\bibinfo {title} {Phase transitions in gauge
  and spin-lattice systems},\ }\href
  {https://www.jetp.ras.ru/cgi-bin/dn/e_042_04_0743.pdf} {\bibfield  {journal}
  {\bibinfo  {journal} {Sov. Phys. JETP}\ }\textbf {\bibinfo {volume} {42}},\
  \bibinfo {pages} {743} (\bibinfo {year} {1975})}\BibitemShut {NoStop}%
\bibitem [{\citenamefont {Kadanoff}(1976)}]{Kadanoff1976}%
  \BibitemOpen
  \bibfield  {author} {\bibinfo {author} {\bibfnamefont {L.~P.}\ \bibnamefont
  {Kadanoff}},\ }\bibfield  {title} {\bibinfo {title} {Notes on {M}igdal's
  recursion formulas},\ }\href {https://doi.org/10.1016/0003-4916(76)90066-x}
  {\bibfield  {journal} {\bibinfo  {journal} {Annals of Physics}\ }\textbf
  {\bibinfo {volume} {100}},\ \bibinfo {pages} {359} (\bibinfo {year}
  {1976})}\BibitemShut {NoStop}%
\bibitem [{\citenamefont {Villegas}\ \emph {et~al.}(2023)\citenamefont
  {Villegas}, \citenamefont {Gili}, \citenamefont {Caldarelli},\ and\
  \citenamefont {Gabrielli}}]{Villegas2023}%
  \BibitemOpen
  \bibfield  {author} {\bibinfo {author} {\bibfnamefont {P.}~\bibnamefont
  {Villegas}}, \bibinfo {author} {\bibfnamefont {T.}~\bibnamefont {Gili}},
  \bibinfo {author} {\bibfnamefont {G.}~\bibnamefont {Caldarelli}},\ and\
  \bibinfo {author} {\bibfnamefont {A.}~\bibnamefont {Gabrielli}},\ }\bibfield
  {title} {\bibinfo {title} {{L}aplacian renormalization group for
  heterogeneous networks},\ }\href {https://doi.org/10.1038/s41567-022-01866-8}
  {\bibfield  {journal} {\bibinfo  {journal} {Nature Physics}\ }\textbf
  {\bibinfo {volume} {19}},\ \bibinfo {pages} {445} (\bibinfo {year}
  {2023})}\BibitemShut {NoStop}%
\bibitem [{\citenamefont {Nurisso}\ \emph {et~al.}(2025)\citenamefont
  {Nurisso}, \citenamefont {Morandini}, \citenamefont {Lucas}, \citenamefont
  {Vaccarino}, \citenamefont {Gili},\ and\ \citenamefont
  {Petri}}]{Nurisso2025}%
  \BibitemOpen
  \bibfield  {author} {\bibinfo {author} {\bibfnamefont {M.}~\bibnamefont
  {Nurisso}}, \bibinfo {author} {\bibfnamefont {M.}~\bibnamefont {Morandini}},
  \bibinfo {author} {\bibfnamefont {M.}~\bibnamefont {Lucas}}, \bibinfo
  {author} {\bibfnamefont {F.}~\bibnamefont {Vaccarino}}, \bibinfo {author}
  {\bibfnamefont {T.}~\bibnamefont {Gili}},\ and\ \bibinfo {author}
  {\bibfnamefont {G.}~\bibnamefont {Petri}},\ }\bibfield  {title} {\bibinfo
  {title} {Higher-order {L}aplacian renormalization},\ }\href
  {https://doi.org/10.1038/s41567-025-02784-1} {\bibfield  {journal} {\bibinfo
  {journal} {Nature Physics}\ }\textbf {\bibinfo {volume} {21}},\ \bibinfo
  {pages} {661} (\bibinfo {year} {2025})}\BibitemShut {NoStop}%
\bibitem [{\citenamefont {Pierson}(1993)}]{Pierson1993}%
  \BibitemOpen
  \bibfield  {author} {\bibinfo {author} {\bibfnamefont {P.}~\bibnamefont
  {Pierson}},\ }\bibfield  {title} {\bibinfo {title} {When effect becomes
  cause: Policy feedback and political change},\ }\href
  {https://doi.org/10.2307/2950710} {\bibfield  {journal} {\bibinfo  {journal}
  {World Politics}\ }\textbf {\bibinfo {volume} {45}},\ \bibinfo {pages} {595}
  (\bibinfo {year} {1993})}\BibitemShut {NoStop}%
\bibitem [{\citenamefont {Goodhart}(1984)}]{Goodhart1984}%
  \BibitemOpen
  \bibfield  {author} {\bibinfo {author} {\bibfnamefont {C.~A.~E.}\
  \bibnamefont {Goodhart}},\ }\bibfield  {title} {\bibinfo {title} {Problems of
  monetary management: The {U.K.} experience},\ }in\ \href
  {https://doi.org/10.1007/978-1-349-17295-5_4} {\emph {\bibinfo {booktitle}
  {Monetary Theory and Practice: The {U.K.} Experience}}}\ (\bibinfo
  {publisher} {Macmillan},\ \bibinfo {address} {London},\ \bibinfo {year}
  {1984})\ pp.\ \bibinfo {pages} {91--121}\BibitemShut {NoStop}%
\bibitem [{\citenamefont {Campbell}(1979)}]{Campbell1976}%
  \BibitemOpen
  \bibfield  {author} {\bibinfo {author} {\bibfnamefont {D.~T.}\ \bibnamefont
  {Campbell}},\ }\bibfield  {title} {\bibinfo {title} {Assessing the impact of
  planned social change},\ }\href
  {https://doi.org/10.1016/0149-7189(79)90048-X} {\bibfield  {journal}
  {\bibinfo  {journal} {Evaluation and Program Planning}\ }\textbf {\bibinfo
  {volume} {2}},\ \bibinfo {pages} {67} (\bibinfo {year} {1979})}\BibitemShut
  {NoStop}%
\bibitem [{\citenamefont {Smith}(2007)}]{SmithDigitalFilters}%
  \BibitemOpen
  \bibfield  {author} {\bibinfo {author} {\bibfnamefont {J.~O.}\ \bibnamefont
  {Smith}, \bibfnamefont {III}},\ }\href
  {https://www.dsprelated.com/freebooks/filters/} {\emph {\bibinfo {title}
  {Introduction to Digital Filters with Audio Applications}}}\ (\bibinfo
  {publisher} {W3K Publishing},\ \bibinfo {year} {2007})\BibitemShut {NoStop}%
\bibitem [{\citenamefont {Hespanha}(2009)}]{Hespanha2009}%
  \BibitemOpen
  \bibfield  {author} {\bibinfo {author} {\bibfnamefont {J.~P.}\ \bibnamefont
  {Hespanha}},\ }\href@noop {} {\emph {\bibinfo {title} {Linear Systems
  Theory}}}\ (\bibinfo  {publisher} {Princeton University Press},\ \bibinfo
  {address} {Princeton},\ \bibinfo {year} {2009})\BibitemShut {NoStop}%
\bibitem [{\citenamefont {Kalman}(1960)}]{Kalman1960}%
  \BibitemOpen
  \bibfield  {author} {\bibinfo {author} {\bibfnamefont {R.~E.}\ \bibnamefont
  {Kalman}},\ }\bibfield  {title} {\bibinfo {title} {On the general theory of
  control systems},\ }\href {https://doi.org/10.1016/s1474-6670(17)70094-8}
  {\bibfield  {journal} {\bibinfo  {journal} {IFAC Proceedings Volumes}\
  }\textbf {\bibinfo {volume} {1}},\ \bibinfo {pages} {491} (\bibinfo {year}
  {1960})}\BibitemShut {NoStop}%
\bibitem [{\citenamefont {Kalman}\ \emph {et~al.}(1963)\citenamefont {Kalman},
  \citenamefont {Ho},\ and\ \citenamefont {Narendra}}]{Kalman1963}%
  \BibitemOpen
  \bibfield  {author} {\bibinfo {author} {\bibfnamefont {R.~E.}\ \bibnamefont
  {Kalman}}, \bibinfo {author} {\bibfnamefont {Y.-C.}\ \bibnamefont {Ho}},\
  and\ \bibinfo {author} {\bibfnamefont {K.~S.}\ \bibnamefont {Narendra}},\
  }\bibfield  {title} {\bibinfo {title} {Controllability of linear dynamical
  systems},\ }\href@noop {} {\bibfield  {journal} {\bibinfo  {journal}
  {Contributions to Differential Equations}\ }\textbf {\bibinfo {volume} {1}},\
  \bibinfo {pages} {189} (\bibinfo {year} {1963})}\BibitemShut {NoStop}%
\bibitem [{\citenamefont {Lee}\ and\ \citenamefont
  {Markus}(1967)}]{LeeMarkus1967}%
  \BibitemOpen
  \bibfield  {author} {\bibinfo {author} {\bibfnamefont {E.~B.}\ \bibnamefont
  {Lee}}\ and\ \bibinfo {author} {\bibfnamefont {L.}~\bibnamefont {Markus}},\
  }\href@noop {} {\emph {\bibinfo {title} {Foundations of Optimal Control
  Theory}}}\ (\bibinfo  {publisher} {John Wiley \& Sons},\ \bibinfo {year}
  {1967})\BibitemShut {NoStop}%
\bibitem [{\citenamefont {Moore}(1981)}]{Moore1981}%
  \BibitemOpen
  \bibfield  {author} {\bibinfo {author} {\bibfnamefont {B.}~\bibnamefont
  {Moore}},\ }\bibfield  {title} {\bibinfo {title} {Principal component
  analysis in linear systems: Controllability, observability, and model
  reduction},\ }\href {https://doi.org/10.1109/tac.1981.1102568} {\bibfield
  {journal} {\bibinfo  {journal} {IEEE Transactions on Automatic Control}\
  }\textbf {\bibinfo {volume} {26}},\ \bibinfo {pages} {17} (\bibinfo {year}
  {1981})}\BibitemShut {NoStop}%
\bibitem [{\citenamefont {Gunderson}\ and\ \citenamefont
  {Holling}(2002)}]{GundersonHolling2002}%
  \BibitemOpen
  \bibinfo {editor} {\bibfnamefont {L.~H.}\ \bibnamefont {Gunderson}}\ and\
  \bibinfo {editor} {\bibfnamefont {C.~S.}\ \bibnamefont {Holling}},\ eds.,\
  \href@noop {} {\emph {\bibinfo {title} {Panarchy: Understanding
  Transformations in Human and Natural Systems}}}\ (\bibinfo  {publisher}
  {Island Press},\ \bibinfo {address} {Washington, DC},\ \bibinfo {year}
  {2002})\BibitemShut {NoStop}%
\bibitem [{\citenamefont {Garnier-Brun}\ \emph {et~al.}(2026)\citenamefont
  {Garnier-Brun}, \citenamefont {Biroli}, \citenamefont {Mézard},\ and\
  \citenamefont {Bouchaud}}]{GarnierBrun2026Speciation}%
  \BibitemOpen
  \bibfield  {author} {\bibinfo {author} {\bibfnamefont {J.}~\bibnamefont
  {Garnier-Brun}}, \bibinfo {author} {\bibfnamefont {G.}~\bibnamefont
  {Biroli}}, \bibinfo {author} {\bibfnamefont {M.}~\bibnamefont {Mézard}},\
  and\ \bibinfo {author} {\bibfnamefont {J.-P.}\ \bibnamefont {Bouchaud}},\
  }\href {https://doi.org/10.48550/arxiv.2611.02955} {\bibinfo {title}
  {Generative diffusion in long-range correlated landscapes: Speciation cascade
  and the foam structure of the score}} (\bibinfo {year} {2026}),\ \Eprint
  {https://arxiv.org/abs/2611.02955} {arXiv:2611.02955} \BibitemShut {NoStop}%
\bibitem [{\citenamefont {Mézard}\ \emph {et~al.}(1984)\citenamefont
  {Mézard}, \citenamefont {Parisi}, \citenamefont {Sourlas}, \citenamefont
  {Toulouse},\ and\ \citenamefont {Virasoro}}]{Mezard1984Nature}%
  \BibitemOpen
  \bibfield  {author} {\bibinfo {author} {\bibfnamefont {M.}~\bibnamefont
  {Mézard}}, \bibinfo {author} {\bibfnamefont {G.}~\bibnamefont {Parisi}},
  \bibinfo {author} {\bibfnamefont {N.}~\bibnamefont {Sourlas}}, \bibinfo
  {author} {\bibfnamefont {G.}~\bibnamefont {Toulouse}},\ and\ \bibinfo
  {author} {\bibfnamefont {M.}~\bibnamefont {Virasoro}},\ }\bibfield  {title}
  {\bibinfo {title} {Nature of the spin-glass phase},\ }\href
  {https://doi.org/10.1103/physrevlett.52.1156} {\bibfield  {journal} {\bibinfo
   {journal} {Physical Review Letters}\ }\textbf {\bibinfo {volume} {52}},\
  \bibinfo {pages} {1156} (\bibinfo {year} {1984})}\BibitemShut {NoStop}%
\bibitem [{\citenamefont {Mézard}\ and\ \citenamefont
  {Parisi}(1991)}]{MezardParisi1991}%
  \BibitemOpen
  \bibfield  {author} {\bibinfo {author} {\bibfnamefont {M.}~\bibnamefont
  {Mézard}}\ and\ \bibinfo {author} {\bibfnamefont {G.}~\bibnamefont
  {Parisi}},\ }\bibfield  {title} {\bibinfo {title} {Replica field theory for
  random manifolds},\ }\href {https://doi.org/10.1051/jp1:1991171} {\bibfield
  {journal} {\bibinfo  {journal} {Journal de Physique I}\ }\textbf {\bibinfo
  {volume} {1}},\ \bibinfo {pages} {809} (\bibinfo {year} {1991})}\BibitemShut
  {NoStop}%
\bibitem [{\citenamefont {Panchenko}(2013)}]{Panchenko2013Ultrametricity}%
  \BibitemOpen
  \bibfield  {author} {\bibinfo {author} {\bibfnamefont {D.}~\bibnamefont
  {Panchenko}},\ }\bibfield  {title} {\bibinfo {title} {The {P}arisi
  ultrametricity conjecture},\ }\href
  {https://doi.org/10.4007/annals.2013.177.1.8} {\bibfield  {journal} {\bibinfo
   {journal} {Annals of Mathematics}\ }\textbf {\bibinfo {volume} {177}},\
  \bibinfo {pages} {383} (\bibinfo {year} {2013})}\BibitemShut {NoStop}%
\bibitem [{\citenamefont {Geshkovski}\ \emph {et~al.}(2025)\citenamefont
  {Geshkovski}, \citenamefont {Letrouit}, \citenamefont {Polyanskiy},\ and\
  \citenamefont {Rigollet}}]{Geshkovski2025}%
  \BibitemOpen
  \bibfield  {author} {\bibinfo {author} {\bibfnamefont {B.}~\bibnamefont
  {Geshkovski}}, \bibinfo {author} {\bibfnamefont {C.}~\bibnamefont
  {Letrouit}}, \bibinfo {author} {\bibfnamefont {Y.}~\bibnamefont
  {Polyanskiy}},\ and\ \bibinfo {author} {\bibfnamefont {P.}~\bibnamefont
  {Rigollet}},\ }\bibfield  {title} {\bibinfo {title} {A mathematical
  perspective on {T}ransformers},\ }\href {https://doi.org/10.1090/bull/1863}
  {\bibfield  {journal} {\bibinfo  {journal} {Bulletin of the American
  Mathematical Society}\ }\textbf {\bibinfo {volume} {62}},\ \bibinfo {pages}
  {427} (\bibinfo {year} {2025})}\BibitemShut {NoStop}%
\bibitem [{\citenamefont {Castin}\ \emph {et~al.}(2026)\citenamefont {Castin},
  \citenamefont {Ablin}, \citenamefont {Carrillo},\ and\ \citenamefont
  {Peyr{\'e}}}]{Castin2025}%
  \BibitemOpen
  \bibfield  {author} {\bibinfo {author} {\bibfnamefont {V.}~\bibnamefont
  {Castin}}, \bibinfo {author} {\bibfnamefont {P.}~\bibnamefont {Ablin}},
  \bibinfo {author} {\bibfnamefont {J.~A.}\ \bibnamefont {Carrillo}},\ and\
  \bibinfo {author} {\bibfnamefont {G.}~\bibnamefont {Peyr{\'e}}},\ }\bibfield
  {title} {\bibinfo {title} {A unified perspective on the dynamics of deep
  {T}ransformers},\ }\href@noop {} {\bibfield  {journal} {\bibinfo  {journal}
  {Foundations of Computational Mathematics}\ } (\bibinfo {year} {2026})},\
  \bibinfo {note} {to appear},\ \Eprint {https://arxiv.org/abs/2501.18322}
  {arXiv:2501.18322} \BibitemShut {NoStop}%
\bibitem [{\citenamefont {Bruno}\ \emph {et~al.}(2025)\citenamefont {Bruno},
  \citenamefont {Pasqualotto},\ and\ \citenamefont {Agazzi}}]{Bruno2025}%
  \BibitemOpen
  \bibfield  {author} {\bibinfo {author} {\bibfnamefont {G.}~\bibnamefont
  {Bruno}}, \bibinfo {author} {\bibfnamefont {F.}~\bibnamefont {Pasqualotto}},\
  and\ \bibinfo {author} {\bibfnamefont {A.}~\bibnamefont {Agazzi}},\
  }\bibfield  {title} {\bibinfo {title} {Emergence of meta-stable clustering in
  mean-field transformer models},\ }in\ \href
  {https://proceedings.iclr.cc/paper_files/paper/2025/hash/164687cb815daae754d33364716e65e6-Abstract-Conference.html}
  {\emph {\bibinfo {booktitle} {Thirteenth International Conference on Learning
  Representations}}}\ (\bibinfo {year} {2025})\ \Eprint
  {https://arxiv.org/abs/2410.23228} {arXiv:2410.23228} \BibitemShut {NoStop}%
\bibitem [{\citenamefont {Karagodin}\ \emph {et~al.}(2025)\citenamefont
  {Karagodin}, \citenamefont {Ge}, \citenamefont {Polyanskiy},\ and\
  \citenamefont {Rigollet}}]{Karagodin2025}%
  \BibitemOpen
  \bibfield  {author} {\bibinfo {author} {\bibfnamefont {N.}~\bibnamefont
  {Karagodin}}, \bibinfo {author} {\bibfnamefont {S.}~\bibnamefont {Ge}},
  \bibinfo {author} {\bibfnamefont {Y.}~\bibnamefont {Polyanskiy}},\ and\
  \bibinfo {author} {\bibfnamefont {P.}~\bibnamefont {Rigollet}},\ }\bibfield
  {title} {\bibinfo {title} {Normalization in attention dynamics},\ }in\ \href
  {https://doi.org/10.52202/085713-4572} {\emph {\bibinfo {booktitle} {Advances
  in Neural Information Processing Systems}}},\ Vol.~\bibinfo {volume} {38}\
  (\bibinfo {year} {2025})\ pp.\ \bibinfo {pages} {151875--151904},\ \Eprint
  {https://arxiv.org/abs/2510.22026} {arXiv:2510.22026} \BibitemShut {NoStop}%
\bibitem [{\citenamefont {Dakos}\ \emph {et~al.}(2008)\citenamefont {Dakos},
  \citenamefont {Scheffer}, \citenamefont {van Nes}, \citenamefont {Brovkin},
  \citenamefont {Petoukhov},\ and\ \citenamefont {Held}}]{Dakos2008}%
  \BibitemOpen
  \bibfield  {author} {\bibinfo {author} {\bibfnamefont {V.}~\bibnamefont
  {Dakos}}, \bibinfo {author} {\bibfnamefont {M.}~\bibnamefont {Scheffer}},
  \bibinfo {author} {\bibfnamefont {E.~H.}\ \bibnamefont {van Nes}}, \bibinfo
  {author} {\bibfnamefont {V.}~\bibnamefont {Brovkin}}, \bibinfo {author}
  {\bibfnamefont {V.}~\bibnamefont {Petoukhov}},\ and\ \bibinfo {author}
  {\bibfnamefont {H.}~\bibnamefont {Held}},\ }\bibfield  {title} {\bibinfo
  {title} {Slowing down as an early warning signal for abrupt climate change},\
  }\href {https://doi.org/10.1073/pnas.0802430105} {\bibfield  {journal}
  {\bibinfo  {journal} {Proceedings of the National Academy of Sciences}\
  }\textbf {\bibinfo {volume} {105}},\ \bibinfo {pages} {14308} (\bibinfo
  {year} {2008})}\BibitemShut {NoStop}%
\bibitem [{\citenamefont {Carpenter}\ and\ \citenamefont
  {Brock}(2006)}]{Carpenter2006}%
  \BibitemOpen
  \bibfield  {author} {\bibinfo {author} {\bibfnamefont {S.~R.}\ \bibnamefont
  {Carpenter}}\ and\ \bibinfo {author} {\bibfnamefont {W.~A.}\ \bibnamefont
  {Brock}},\ }\bibfield  {title} {\bibinfo {title} {Rising variance: a leading
  indicator of ecological transition},\ }\href
  {https://doi.org/10.1111/j.1461-0248.2005.00877.x} {\bibfield  {journal}
  {\bibinfo  {journal} {Ecology Letters}\ }\textbf {\bibinfo {volume} {9}},\
  \bibinfo {pages} {311} (\bibinfo {year} {2006})}\BibitemShut {NoStop}%
\bibitem [{\citenamefont {K{\'e}fi}\ \emph {et~al.}(2014)\citenamefont
  {K{\'e}fi}, \citenamefont {Guttal}, \citenamefont {Brock}, \citenamefont
  {Carpenter}, \citenamefont {Ellison}, \citenamefont {Livina}, \citenamefont
  {Seekell}, \citenamefont {Scheffer}, \citenamefont {van Nes},\ and\
  \citenamefont {Dakos}}]{Kefi2014}%
  \BibitemOpen
  \bibfield  {author} {\bibinfo {author} {\bibfnamefont {S.}~\bibnamefont
  {K{\'e}fi}}, \bibinfo {author} {\bibfnamefont {V.}~\bibnamefont {Guttal}},
  \bibinfo {author} {\bibfnamefont {W.~A.}\ \bibnamefont {Brock}}, \bibinfo
  {author} {\bibfnamefont {S.~R.}\ \bibnamefont {Carpenter}}, \bibinfo {author}
  {\bibfnamefont {A.~M.}\ \bibnamefont {Ellison}}, \bibinfo {author}
  {\bibfnamefont {V.~N.}\ \bibnamefont {Livina}}, \bibinfo {author}
  {\bibfnamefont {D.~A.}\ \bibnamefont {Seekell}}, \bibinfo {author}
  {\bibfnamefont {M.}~\bibnamefont {Scheffer}}, \bibinfo {author}
  {\bibfnamefont {E.~H.}\ \bibnamefont {van Nes}},\ and\ \bibinfo {author}
  {\bibfnamefont {V.}~\bibnamefont {Dakos}},\ }\bibfield  {title} {\bibinfo
  {title} {Early warning signals of ecological transitions: methods for spatial
  patterns},\ }\href {https://doi.org/10.1371/journal.pone.0092097} {\bibfield
  {journal} {\bibinfo  {journal} {PLOS ONE}\ }\textbf {\bibinfo {volume} {9}},\
  \bibinfo {pages} {e92097} (\bibinfo {year} {2014})}\BibitemShut {NoStop}%
\bibitem [{\citenamefont {Ives}(1995)}]{Ives1995}%
  \BibitemOpen
  \bibfield  {author} {\bibinfo {author} {\bibfnamefont {A.~R.}\ \bibnamefont
  {Ives}},\ }\bibfield  {title} {\bibinfo {title} {Measuring resilience in
  stochastic systems},\ }\href {https://doi.org/10.2307/2937138} {\bibfield
  {journal} {\bibinfo  {journal} {Ecological Monographs}\ }\textbf {\bibinfo
  {volume} {65}},\ \bibinfo {pages} {217} (\bibinfo {year} {1995})}\BibitemShut
  {NoStop}%
\bibitem [{\citenamefont {Carpenter}\ \emph {et~al.}(2011)\citenamefont
  {Carpenter}, \citenamefont {Cole}, \citenamefont {Pace}, \citenamefont
  {Batt}, \citenamefont {Brock}, \citenamefont {Cline}, \citenamefont {Coloso},
  \citenamefont {Hodgson}, \citenamefont {Kitchell}, \citenamefont {Seekell},
  \citenamefont {Smith},\ and\ \citenamefont {Weidel}}]{Carpenter2011}%
  \BibitemOpen
  \bibfield  {author} {\bibinfo {author} {\bibfnamefont {S.~R.}\ \bibnamefont
  {Carpenter}}, \bibinfo {author} {\bibfnamefont {J.~J.}\ \bibnamefont {Cole}},
  \bibinfo {author} {\bibfnamefont {M.~L.}\ \bibnamefont {Pace}}, \bibinfo
  {author} {\bibfnamefont {R.}~\bibnamefont {Batt}}, \bibinfo {author}
  {\bibfnamefont {W.~A.}\ \bibnamefont {Brock}}, \bibinfo {author}
  {\bibfnamefont {T.}~\bibnamefont {Cline}}, \bibinfo {author} {\bibfnamefont
  {J.}~\bibnamefont {Coloso}}, \bibinfo {author} {\bibfnamefont {J.~R.}\
  \bibnamefont {Hodgson}}, \bibinfo {author} {\bibfnamefont {J.~F.}\
  \bibnamefont {Kitchell}}, \bibinfo {author} {\bibfnamefont {D.~A.}\
  \bibnamefont {Seekell}}, \bibinfo {author} {\bibfnamefont {L.}~\bibnamefont
  {Smith}},\ and\ \bibinfo {author} {\bibfnamefont {B.}~\bibnamefont
  {Weidel}},\ }\bibfield  {title} {\bibinfo {title} {Early warnings of regime
  shifts: A whole-ecosystem experiment},\ }\href
  {https://doi.org/10.1126/science.1203672} {\bibfield  {journal} {\bibinfo
  {journal} {Science}\ }\textbf {\bibinfo {volume} {332}},\ \bibinfo {pages}
  {1079} (\bibinfo {year} {2011})}\BibitemShut {NoStop}%
\bibitem [{\citenamefont {Coibion}\ and\ \citenamefont
  {Gorodnichenko}(2015)}]{CoibionGorodnichenko2015}%
  \BibitemOpen
  \bibfield  {author} {\bibinfo {author} {\bibfnamefont {O.}~\bibnamefont
  {Coibion}}\ and\ \bibinfo {author} {\bibfnamefont {Y.}~\bibnamefont
  {Gorodnichenko}},\ }\bibfield  {title} {\bibinfo {title} {Information
  rigidity and the expectations formation process: A simple framework and new
  facts},\ }\href {https://doi.org/10.1257/aer.20110306} {\bibfield  {journal}
  {\bibinfo  {journal} {American Economic Review}\ }\textbf {\bibinfo {volume}
  {105}},\ \bibinfo {pages} {2644} (\bibinfo {year} {2015})}\BibitemShut
  {NoStop}%
\bibitem [{\citenamefont {Cohen}\ \emph {et~al.}(2021)\citenamefont {Cohen},
  \citenamefont {Kaur}, \citenamefont {Li}, \citenamefont {Kolter},\ and\
  \citenamefont {Talwalkar}}]{Cohen2021EdgeOfStability}%
  \BibitemOpen
  \bibfield  {author} {\bibinfo {author} {\bibfnamefont {J.~M.}\ \bibnamefont
  {Cohen}}, \bibinfo {author} {\bibfnamefont {S.}~\bibnamefont {Kaur}},
  \bibinfo {author} {\bibfnamefont {Y.}~\bibnamefont {Li}}, \bibinfo {author}
  {\bibfnamefont {J.~Z.}\ \bibnamefont {Kolter}},\ and\ \bibinfo {author}
  {\bibfnamefont {A.}~\bibnamefont {Talwalkar}},\ }\bibfield  {title} {\bibinfo
  {title} {Gradient descent on neural networks typically occurs at the edge of
  stability},\ }in\ \href {https://iclr.cc/virtual/2021/poster/2577} {\emph
  {\bibinfo {booktitle} {Ninth International Conference on Learning
  Representations}}}\ (\bibinfo {year} {2021})\ \Eprint
  {https://arxiv.org/abs/2103.00065} {arXiv:2103.00065} \BibitemShut {NoStop}%
\bibitem [{\citenamefont {Litman}(2026)}]{Litman2026EdgeOfStability}%
  \BibitemOpen
  \bibfield  {author} {\bibinfo {author} {\bibfnamefont {E.}~\bibnamefont
  {Litman}},\ }\href {https://doi.org/10.48550/arxiv.2604.20446} {\bibinfo
  {title} {The origin of edge of stability}} (\bibinfo {year} {2026}),\ \Eprint
  {https://arxiv.org/abs/2604.20446} {arXiv:2604.20446} \BibitemShut {NoStop}%
\bibitem [{\citenamefont {Schmidhuber}(2006)}]{schmidhuber2006godel}%
  \BibitemOpen
  \bibfield  {author} {\bibinfo {author} {\bibfnamefont {J.}~\bibnamefont
  {Schmidhuber}},\ }\href {https://doi.org/10.48550/arXiv.cs/0309048} {\emph
  {\bibinfo {title} {{G}{\"o}del {M}achines: Self-Referential Universal Problem
  Solvers Making Provably Optimal Self-Improvements}}},\ \bibinfo {type} {Tech.
  Rep.}\ \bibinfo {number} {IDSIA-19-03}\ (\bibinfo  {institution} {IDSIA},\
  \bibinfo {year} {2006})\ \bibinfo {note} {version 5},\ \Eprint
  {https://arxiv.org/abs/cs/0309048v5} {arXiv:cs/0309048v5} \BibitemShut
  {NoStop}%
\bibitem [{\citenamefont {Zhang}\ \emph
  {et~al.}(2026{\natexlab{a}})\citenamefont {Zhang}, \citenamefont {Hu},
  \citenamefont {Lu}, \citenamefont {Lange},\ and\ \citenamefont
  {Clune}}]{zhang2025darwin}%
  \BibitemOpen
  \bibfield  {author} {\bibinfo {author} {\bibfnamefont {J.}~\bibnamefont
  {Zhang}}, \bibinfo {author} {\bibfnamefont {S.}~\bibnamefont {Hu}}, \bibinfo
  {author} {\bibfnamefont {C.}~\bibnamefont {Lu}}, \bibinfo {author}
  {\bibfnamefont {R.}~\bibnamefont {Lange}},\ and\ \bibinfo {author}
  {\bibfnamefont {J.}~\bibnamefont {Clune}},\ }\bibfield  {title} {\bibinfo
  {title} {{D}arwin {G}{\"o}del {M}achine: Open-ended evolution of
  self-improving agents},\ }in\ \href
  {https://openreview.net/forum?id=pUpzQZTvGY} {\emph {\bibinfo {booktitle}
  {Fourteenth International Conference on Learning Representations}}}\
  (\bibinfo {year} {2026})\ \Eprint {https://arxiv.org/abs/2505.22954}
  {arXiv:2505.22954} \BibitemShut {NoStop}%
\bibitem [{\citenamefont {Zhang}\ \emph
  {et~al.}(2026{\natexlab{b}})\citenamefont {Zhang}, \citenamefont {Zhao},
  \citenamefont {Yang}, \citenamefont {Foerster}, \citenamefont {Clune},
  \citenamefont {Jiang}, \citenamefont {Devlin},\ and\ \citenamefont
  {Shavrina}}]{zhang2026hyperagents}%
  \BibitemOpen
  \bibfield  {author} {\bibinfo {author} {\bibfnamefont {J.}~\bibnamefont
  {Zhang}}, \bibinfo {author} {\bibfnamefont {B.}~\bibnamefont {Zhao}},
  \bibinfo {author} {\bibfnamefont {W.}~\bibnamefont {Yang}}, \bibinfo {author}
  {\bibfnamefont {J.}~\bibnamefont {Foerster}}, \bibinfo {author}
  {\bibfnamefont {J.}~\bibnamefont {Clune}}, \bibinfo {author} {\bibfnamefont
  {M.}~\bibnamefont {Jiang}}, \bibinfo {author} {\bibfnamefont
  {S.}~\bibnamefont {Devlin}},\ and\ \bibinfo {author} {\bibfnamefont
  {T.}~\bibnamefont {Shavrina}},\ }\bibfield  {title} {\bibinfo {title}
  {{HyperAgents}},\ }in\ \href {https://neurips.cc/virtual/2026/poster/153670}
  {\emph {\bibinfo {booktitle} {Advances in Neural Information Processing
  Systems}}},\ Vol.~\bibinfo {volume} {39}\ (\bibinfo {year} {2026})\ \bibinfo
  {note} {accepted for publication},\ \Eprint
  {https://arxiv.org/abs/2603.19461} {arXiv:2603.19461} \BibitemShut {NoStop}%
\bibitem [{\citenamefont {Paris}(2026)}]{NISTDLMFBeta}%
  \BibitemOpen
  \bibfield  {author} {\bibinfo {author} {\bibfnamefont {R.~B.}\ \bibnamefont
  {Paris}},\ }\href {https://dlmf.nist.gov/8.17.E5} {\bibinfo {title} {{NIST}
  digital library of mathematical functions: Incomplete beta functions}},\
  \bibinfo {howpublished} {National Institute of Standards and Technology}
  (\bibinfo {year} {2026}),\ \bibinfo {note} {sec.~8.17, Eq.~(8.17.5), release
  1.2.8}\BibitemShut {NoStop}%
\bibitem [{\citenamefont {O'Donnell}(2021)}]{ODonnell2021BooleanFunctions}%
  \BibitemOpen
  \bibfield  {author} {\bibinfo {author} {\bibfnamefont {R.}~\bibnamefont
  {O'Donnell}},\ }\href {https://doi.org/10.48550/arxiv.2105.10386} {\bibinfo
  {title} {Analysis of {B}oolean functions}} (\bibinfo {year} {2021}),\
  \bibinfo {note} {revised edition of the 2014 Cambridge University Press
  book},\ \Eprint {https://arxiv.org/abs/2105.10386} {arXiv:2105.10386}
  \BibitemShut {NoStop}%
\bibitem [{\citenamefont {Kaniovski}(2010)}]{Kaniovski2010CorrelatedVotes}%
  \BibitemOpen
  \bibfield  {author} {\bibinfo {author} {\bibfnamefont {S.}~\bibnamefont
  {Kaniovski}},\ }\bibfield  {title} {\bibinfo {title} {Aggregation of
  correlated votes and {C}ondorcet's jury theorem},\ }\href
  {https://doi.org/10.1007/s11238-008-9120-4} {\bibfield  {journal} {\bibinfo
  {journal} {Theory and Decision}\ }\textbf {\bibinfo {volume} {69}},\ \bibinfo
  {pages} {453} (\bibinfo {year} {2010})}\BibitemShut {NoStop}%
\bibitem [{\citenamefont {Rutterford}\ \emph {et~al.}(2015)\citenamefont
  {Rutterford}, \citenamefont {Copas},\ and\ \citenamefont
  {Eldridge}}]{Rutterford2015ClusterTrials}%
  \BibitemOpen
  \bibfield  {author} {\bibinfo {author} {\bibfnamefont {C.}~\bibnamefont
  {Rutterford}}, \bibinfo {author} {\bibfnamefont {A.}~\bibnamefont {Copas}},\
  and\ \bibinfo {author} {\bibfnamefont {S.}~\bibnamefont {Eldridge}},\
  }\bibfield  {title} {\bibinfo {title} {Methods for sample size determination
  in cluster randomized trials},\ }\href {https://doi.org/10.1093/ije/dyv113}
  {\bibfield  {journal} {\bibinfo  {journal} {International Journal of
  Epidemiology}\ }\textbf {\bibinfo {volume} {44}},\ \bibinfo {pages} {1051}
  (\bibinfo {year} {2015})}\BibitemShut {NoStop}%
\bibitem [{\citenamefont {Dahleh}\ \emph {et~al.}(2011)\citenamefont {Dahleh},
  \citenamefont {Dahleh},\ and\ \citenamefont
  {Verghese}}]{Dahleh2011InternalStability}%
  \BibitemOpen
  \bibfield  {author} {\bibinfo {author} {\bibfnamefont {M.}~\bibnamefont
  {Dahleh}}, \bibinfo {author} {\bibfnamefont {M.~A.}\ \bibnamefont {Dahleh}},\
  and\ \bibinfo {author} {\bibfnamefont {G.}~\bibnamefont {Verghese}},\
  }\bibfield  {title} {\bibinfo {title} {Dynamic systems and control: Internal
  stability for {LTI} systems}} (\bibinfo {year} {2011}),\ \bibinfo {note} {mIT
  OpenCourseWare, course 6.241J, Chap.~14}\BibitemShut {NoStop}%
\bibitem [{\citenamefont {Carroll}(2020)}]{Carroll2020EdgeOfChaos}%
  \BibitemOpen
  \bibfield  {author} {\bibinfo {author} {\bibfnamefont {T.~L.}\ \bibnamefont
  {Carroll}},\ }\bibfield  {title} {\bibinfo {title} {Do reservoir computers
  work best at the edge of chaos?},\ }\href {https://doi.org/10.1063/5.0038163}
  {\bibfield  {journal} {\bibinfo  {journal} {Chaos: An Interdisciplinary
  Journal of Nonlinear Science}\ }\textbf {\bibinfo {volume} {30}},\ \bibinfo
  {pages} {121109} (\bibinfo {year} {2020})}\BibitemShut {NoStop}%
\bibitem [{\citenamefont {Boedecker}\ \emph {et~al.}(2012)\citenamefont
  {Boedecker}, \citenamefont {Obst}, \citenamefont {Lizier}, \citenamefont
  {Mayer},\ and\ \citenamefont {Asada}}]{Boedecker2012EdgeOfChaos}%
  \BibitemOpen
  \bibfield  {author} {\bibinfo {author} {\bibfnamefont {J.}~\bibnamefont
  {Boedecker}}, \bibinfo {author} {\bibfnamefont {O.}~\bibnamefont {Obst}},
  \bibinfo {author} {\bibfnamefont {J.~T.}\ \bibnamefont {Lizier}}, \bibinfo
  {author} {\bibfnamefont {N.~M.}\ \bibnamefont {Mayer}},\ and\ \bibinfo
  {author} {\bibfnamefont {M.}~\bibnamefont {Asada}},\ }\bibfield  {title}
  {\bibinfo {title} {Information processing in echo state networks at the edge
  of chaos},\ }\href {https://doi.org/10.1007/s12064-011-0146-8} {\bibfield
  {journal} {\bibinfo  {journal} {Theory in Biosciences}\ }\textbf {\bibinfo
  {volume} {131}},\ \bibinfo {pages} {205} (\bibinfo {year}
  {2012})}\BibitemShut {NoStop}%
\bibitem [{\citenamefont {McCaul}\ \emph
  {et~al.}(2025{\natexlab{a}})\citenamefont {McCaul}, \citenamefont {Tripathy},
  \citenamefont {Marcucci},\ and\ \citenamefont
  {Totero~Gongora}}]{McCaul2025PhotonicReservoirs}%
  \BibitemOpen
  \bibfield  {author} {\bibinfo {author} {\bibfnamefont {G.}~\bibnamefont
  {McCaul}}, \bibinfo {author} {\bibfnamefont {G.}~\bibnamefont {Tripathy}},
  \bibinfo {author} {\bibfnamefont {G.}~\bibnamefont {Marcucci}},\ and\
  \bibinfo {author} {\bibfnamefont {J.~S.}\ \bibnamefont {Totero~Gongora}},\
  }\bibfield  {title} {\bibinfo {title} {Unwrapping photonic reservoirs:
  Enhanced expressivity via random {F}ourier encoding over stretched domains},\
  }\href {https://doi.org/10.1063/5.0283442} {\bibfield  {journal} {\bibinfo
  {journal} {Chaos: An Interdisciplinary Journal of Nonlinear Science}\
  }\textbf {\bibinfo {volume} {35}},\ \bibinfo {pages} {093129} (\bibinfo
  {year} {2025}{\natexlab{a}})}\BibitemShut {NoStop}%
\bibitem [{\citenamefont {McCaul}\ \emph
  {et~al.}(2025{\natexlab{b}})\citenamefont {McCaul}, \citenamefont
  {Totero~Gongora}, \citenamefont {Otieno}, \citenamefont {Savel’ev},
  \citenamefont {Zagoskin},\ and\ \citenamefont
  {Balanov}}]{Weiszmuller2025MinimalReservoirs}%
  \BibitemOpen
  \bibfield  {author} {\bibinfo {author} {\bibfnamefont {G.}~\bibnamefont
  {McCaul}}, \bibinfo {author} {\bibfnamefont {J.~S.}\ \bibnamefont
  {Totero~Gongora}}, \bibinfo {author} {\bibfnamefont {W.}~\bibnamefont
  {Otieno}}, \bibinfo {author} {\bibfnamefont {S.}~\bibnamefont {Savel’ev}},
  \bibinfo {author} {\bibfnamefont {A.}~\bibnamefont {Zagoskin}},\ and\
  \bibinfo {author} {\bibfnamefont {A.~G.}\ \bibnamefont {Balanov}},\
  }\bibfield  {title} {\bibinfo {title} {Minimal quantum reservoirs with
  {H}amiltonian encoding},\ }\href {https://doi.org/10.1063/5.0282921}
  {\bibfield  {journal} {\bibinfo  {journal} {Chaos: An Interdisciplinary
  Journal of Nonlinear Science}\ }\textbf {\bibinfo {volume} {35}},\ \bibinfo
  {pages} {093135} (\bibinfo {year} {2025}{\natexlab{b}})}\BibitemShut
  {NoStop}%
\bibitem [{\citenamefont {Jury}(1962)}]{Jury1962}%
  \BibitemOpen
  \bibfield  {author} {\bibinfo {author} {\bibfnamefont {E.}~\bibnamefont
  {Jury}},\ }\bibfield  {title} {\bibinfo {title} {A simplified stability
  criterion for linear discrete systems},\ }\href
  {https://doi.org/10.1109/jrproc.1962.288193} {\bibfield  {journal} {\bibinfo
  {journal} {Proceedings of the IRE}\ }\textbf {\bibinfo {volume} {50}},\
  \bibinfo {pages} {1493} (\bibinfo {year} {1962})}\BibitemShut {NoStop}%
\bibitem [{\citenamefont {Tresser}\ \emph {et~al.}(2014)\citenamefont
  {Tresser}, \citenamefont {Coullet},\ and\ \citenamefont
  {de~Faria}}]{Tresser2014PeriodDoubling}%
  \BibitemOpen
  \bibfield  {author} {\bibinfo {author} {\bibfnamefont {C.}~\bibnamefont
  {Tresser}}, \bibinfo {author} {\bibfnamefont {P.}~\bibnamefont {Coullet}},\
  and\ \bibinfo {author} {\bibfnamefont {E.}~\bibnamefont {de~Faria}},\
  }\bibfield  {title} {\bibinfo {title} {Period doubling},\ }\href
  {https://doi.org/10.4249/scholarpedia.3958} {\bibfield  {journal} {\bibinfo
  {journal} {Scholarpedia}\ }\textbf {\bibinfo {volume} {9}},\ \bibinfo {pages}
  {3958} (\bibinfo {year} {2014})}\BibitemShut {NoStop}%
\bibitem [{\citenamefont {Galam}(2008)}]{Galam2008Sociophysics}%
  \BibitemOpen
  \bibfield  {author} {\bibinfo {author} {\bibfnamefont {S.}~\bibnamefont
  {Galam}},\ }\bibfield  {title} {\bibinfo {title} {Sociophysics: A review of
  {G}alam models},\ }\href {https://doi.org/10.1142/s0129183108012297}
  {\bibfield  {journal} {\bibinfo  {journal} {International Journal of Modern
  Physics C}\ }\textbf {\bibinfo {volume} {19}},\ \bibinfo {pages} {409}
  (\bibinfo {year} {2008})}\BibitemShut {NoStop}%
\bibitem [{\citenamefont {Boyd}\ and\ \citenamefont
  {Lall}(nd)}]{BoydLallEE263StateTransfer}%
  \BibitemOpen
  \bibfield  {author} {\bibinfo {author} {\bibfnamefont {S.}~\bibnamefont
  {Boyd}}\ and\ \bibinfo {author} {\bibfnamefont {S.}~\bibnamefont {Lall}},\
  }\bibfield  {title} {\bibinfo {title} {Controllability and state transfer}}
  (\bibinfo {year} {n.d.}),\ \bibinfo {note} {stanford University, EE263
  lecture slides}\BibitemShut {NoStop}%
\end{thebibliography}%

\appendix
\section{Majority Compression and the Central Limit Theorem}
\label{app:majority}

Between census points the mandate $\theta_k$ is held fixed while the physical state evolves. We call this the \emph{macroscopic dynamics}: in the scalar model it is the relaxation of the scalar $x$ towards $s\theta_k$, and in the general model its endpoint is $\Phi_T(\mathbf x_k,\theta_k)$ [Eq.~\eqref{eq:normal_x}]. The population observes that endpoint and returns the correction to the mandate. It is this second part of the loop that we calculate here: how does the mean majority signal $R_N(e)$ depend on mismatch? To calculate it, hold the mismatch $e$ fixed for one census. An individual stance has values $+1$ and $-1$, with mean $m(e)$. We assume for now that $m(0)=0$, that the slope at zero is positive, and that the stances are conditionally independent. There are then two collective quantities with quite different sensitivities. First, the fraction of positive responses has expectation $p=p_+(e)=(1+m)/2$ [Eqs.~\eqref{eq:scalar-sign-probability} and~\eqref{eq:agent-mean}], independent of population size. Then the probability that this fraction exceeds a half changes increasingly sharply with $e$. It is this second quantity that enters a majority decision.

For odd $N=2h+1$, let $J$ count the positive inputs. Writing out Eq.~\eqref{eq:mean-majority} gives
\begin{equation}
R_N(e)=2\Pr(J>h\mid e)-1
       =2\sum_{j=h+1}^N {N\choose j}p^j(1-p)^{N-j}-1 .\label{eq:majority-tail}
\end{equation}
The binomial tail is also $I_p(h+1,h+1)$, where $I_p(a,b)$ is the regularised incomplete beta function~\cite[Eq.~(8.17.5)]{NISTDLMFBeta}. This gives a direct way to evaluate the exact finite-population mean in Appendix~\ref{app:simulations}. To see how the mean changes with bias, differentiate the sum. Neighbouring terms cancel, leaving the contribution at the majority threshold:
\begin{equation}
{dR_N\over dm}=a_N(1-m^2)^h,\qquad
 a_N={N{2h\choose h}\over4^h}\sim\sqrt{2N/\pi}.\label{eq:majority-influence}
\end{equation}

The factor $\binom{2h}{h}/4^h$ is the probability that the other $2h$ agents split evenly at zero bias, in which case the remaining agent decides the outcome. There are $N$ such opportunities for a vote to be decisive, which yields the factor $a_N$~\cite[Sec.~5.2]{ODonnell2021BooleanFunctions}. Thus a small change in mismatch gives the exact finite-$N$ gain in Eq.~\eqref{eq:loop-gain}, $K=\alpha a_Nm'(0)$. To find what limits the correction once the mandate starts to alternate, we also need the first nonlinear term. For $m(e)=\tanh(\beta e)$, differentiation and expansion give

\begin{align}
 R_N'(e)&=\beta a_N\operatorname{sech}^{N+1}(\beta e),\\
 R_N(e)&=G_N e-c_Ne^3+O(e^5),\\
 G_N&=\beta a_N,\qquad c_N=\frac{\beta^3 a_N(N+1)}{6}.\label{eq:majority-cubic}
\end{align}

The shape of the microscopic response away from zero can be discarded because the majority changes sign appreciably only while the mean vote count is within a few standard deviations of the threshold. The mean positive count lies $Nm(e)/2$ above $N/2$, while its standard deviation near threshold is $\sqrt N/2$. Thus the transition samples biases of order $N^{-1/2}$. With $e=u/\sqrt N$, the signed gap in standard deviations tends to $m'(0)u$. The central limit theorem puts the standardised count on a unit normal scale, giving $R_N(u/\sqrt N)\to 2\Phi(m'(0)u)-1$, where $\Phi$ is the standard normal cumulative distribution function. The Berry--Esseen theorem bounds the error in replacing the binomial tail by the normal tail~\cite[Sec.~11.5]{ODonnell2021BooleanFunctions}. On bounded $u$-intervals the Bernoulli variance stays away from zero and the third absolute moment stays bounded, so this error is uniformly $O(N^{-1/2})$.

\subsection*{Correlated Responses and $N_{\rm eff}$}

When agents respond to one another or share information, the accuracy of a majority can depend on how their votes are coupled~\cite{Kaniovski2010CorrelatedVotes}. At fixed mismatch let $v$ be the common variance of the individual signs and $\chi>0$ the mean correlation between distinct signs. The variance of their average is $v[1+(N-1)\chi]/N$. The factor $1+(N-1)\chi$ is the familiar design effect for equal-size clusters~\cite[Eq.~(1)]{Rutterford2015ClusterTrials}. Equating this variance to $v/N_{\rm eff}$ defines $N_{\rm eff}=N/[1+(N-1)\chi]$. At fixed positive $\chi$, this effective population cannot grow beyond $1/\chi$.

Does that cap the gain? Suppose all agents copy one common sign with probability $\chi$ and draw independently otherwise, using the same positive-sign probability $p(e)$ in either case. Every pair has correlation $\chi$. The mean majority verdict is then $\chi m(e)+(1-\chi)R_N(e)$, with $R_N$ the independent majority response. The independent part retains a fixed weight, so its slope still grows as $\sqrt N$ for any fixed $\chi<1$. The gain is set by how the dependent count crosses the majority threshold, not by $N_{\rm eff}$, and other correlation structures enter through the same calculation with a different mixture in place of $\chi m(e)+(1-\chi)R_N(e)$.

\section{Scalar Phase Boundary and Loss Onset}
\label{app:scalar_boundary}

The same correction can settle a mismatch or keep it alive by overshooting. To find where one behaviour gives way to the other, we need to follow both the mandate and the state it acts on: when the population next responds, the state may still be relaxing under the previous correction. Appendix~\ref{app:majority} supplies the two properties of the majority response that enter this calculation. Its slope $G_N=R_N'(0)$ determines how strongly a small mismatch is corrected, while its cubic coefficient $c_N$ will determine the amplitude once the correction begins to alternate [Eq.~\eqref{eq:majority-cubic}].

\subsection*{Spectral Radius Determines Stability}

Use primes for quantities at the next census and put $r=\exp(-T/\tau)$, the fraction of the initial displacement retained over the census period. In Eqs.~\eqref{eq:held-scalar-state}--\eqref{eq:scalar-revision}, a repeating mandate requires $x=s\theta$, whereas a vanishing mean revision requires $x=b\theta$. Since $b>s>0$, the only state that meets both conditions is the origin. Displace the system slightly from it. The state first moves to $x'=rx+(1-r)s\theta$, after which the population compares that new reading with $b\theta$. To first order its correction is $K(x'-b\theta)$, where we write $K=\alpha G_N$ for the loop gain at the chosen population [Eq.~\eqref{eq:loop-gain}]. Carrying out these operations in that order gives the one-census return matrix
\begin{align}
 J(K)&=\begin{pmatrix}r&(1-r)s\\Kr&1+K((1-r)s-b)\end{pmatrix},\label{eq:scalar-census-jacobian}\\
 \det J&=r(1-Kb).
\end{align}

Now consider a displacement $\mathbf v=(\delta x,\delta\theta)^\top$ for which one census simply multiplies both components by the same factor. Such a displacement is an eigenmode of Eq.~\eqref{eq:scalar-census-jacobian}: $J\mathbf v=\lambda\mathbf v$, so after $k$ censuses it has become $\lambda^k\mathbf v$. A negative $\lambda$ reverses the disturbance at each step, but it is the magnitude that decides whether the next reversal is larger or smaller. We must test both modes. Writing $\sigma(J)$ for the set of eigenvalues, the largest magnitude is the spectral radius,
\begin{equation}
 \rho(J)=\max_{\lambda\in\sigma(J)}|\lambda|.
 \label{eq:census-spectral-radius}
\end{equation}
All linearised disturbances decay if and only if $\rho(J)<1$~\cite[Ch.~14]{Dahleh2011InternalStability}. This spectral radius condition has parallels in many other fields, including reservoir computing. In that context the spectral radius of the connection weights can be used to tune the response to a drive~\cite{Carroll2020EdgeOfChaos}, and information storage and transfer in recurrent reservoirs peak near the transition from ordered to chaotic dynamics~\cite{Boedecker2012EdgeOfChaos}. Physical reservoirs can also obtain distinguishable outputs without recurrence, from phase wrapping in a feedforward scattering medium~\cite{McCaul2025PhotonicReservoirs} or from delayed samples~\cite{Weiszmuller2025MinimalReservoirs}. In Eq.~\eqref{eq:scalar-census-jacobian} the feedback is already closed, such that changing the state changes the next mandate as well. It is the spectrum of this complete return that sets the boundary in Fig.~\ref{fig:scalar_phase}(c). What it locates is an edge of \emph{stability} rather than of chaos: as the leading multiplier approaches $-1$ the alternating trace of a disturbance lasts longer, and beyond the crossing the scalar model settles into a two-cycle. Feeding the readout of a physical reservoir back into a parameter held between censuses would let one vary the revision gain and measure both the decay of a small pulse and the error on a delayed-input task. Equation~\eqref{eq:census-spectral-radius} predicts where decay ceases. Whether the readout performs best just before that point is a separate, and measurable, question.

The two eigenvalues are the roots of the characteristic polynomial
\begin{align}
 P(\lambda)&=\det(\lambda I-J)
 =\lambda^2-[1+r-KD_1]\lambda+r(1-Kb),\\
 D_1&=b-(1-r)s .
\end{align}
Here $D_1$ is the gap between the reference slope and the response reached in one period starting from zero. For $0<r<1$ and $K>0$ the two roots are real and the discriminant is the squared difference of the diagonal entries of $J$ plus $4Kr(1-r)s$. To decide which root reaches the boundary first, it is easier to evaluate $P$ at the ends of the stable interval than to track the two roots separately. Jury's criterion for a real monic quadratic requires $P(1)>0$, $P(-1)>0$ and $1-\det J>0$~\cite{Jury1962}. In our case these tests reduce to
\begin{align}
 P(1)&=K(1-r)(b-s),\\
 P(-1)&=(1+r)(2-KD),\\
 1-\det J&=1-r+rKb .
\end{align}
Only the middle expression can change sign as the gain rises. The stable interval is therefore
\begin{equation}
 0<K<K_c(T),\qquad K_c(T)=\frac{2}{D},\qquad
 D=b-d,\quad d=s\frac{1-r}{1+r}.
 \label{eq:flip-shortfall}
\end{equation}

The shortfall $D$ has a direct physical meaning. If the mandate alternates between $\pm\Theta$, the state begins the positive hold at $-d\Theta$ and ends it at $+d\Theta$: substituting the endpoints into the relaxation law gives exactly the $d$ in Eq.~\eqref{eq:flip-shortfall}. The reading thus falls short of its reference by $D\Theta$, while the change needed to reverse the mandate is $2\Theta$. Their ratio gives the critical gain in Eq.~\eqref{eq:critical-gain}. It also explains why a longer census interval helps. Since $d=s\tanh(T/2\tau)$ increases with $T$, the state follows more of the swing, leaving a smaller mismatch to drive the reversal.

\subsection*{Flip Bifurcation}

At the boundary the critical eigenvalue is $\lambda_1=-1$. Its companion $\lambda_2=r(b+d)/(b-d)$ is still strictly between zero and one because $s<b$. Differentiating $P(\lambda_1)=0$ gives
\begin{equation}
 \left.\frac{d\lambda_1}{dK}\right|_{K_c}
 =-\frac{(1+r)D}{1+\lambda_2}<0 .
 \label{eq:flip-crossing-slope}
\end{equation}
The matched state therefore loses stability through a reversal, rather than through a disturbance that grows with the same sign. That identifies the linear instability but leaves a physical question open: does a small, stable oscillation take over, or does the system leave the neighbourhood altogether? Both behaviours occur in period doubling~\cite{Tresser2014PeriodDoubling}. For the odd response of Eq.~\eqref{eq:mean-majority}, the same two endpoints let us construct the orbit without linearising. At the positive hold the mismatch is $-D\Theta$, so a correction of $-2\Theta$ requires
\begin{equation}
 2\Theta=\alpha R_N(D\Theta).
 \label{eq:cycle-balance}
\end{equation}
To see whether this can happen, divide by $D\Theta$ and follow the secant slope $R_N(e)/e$ for $e>0$. The finite-majority response is bounded and strictly concave there, so this slope decreases from $G_N$ to zero. It meets $K_c/\alpha$ exactly once for $K>K_c$, and never for $K\le K_c$. At the intersection the tangent is less steep than the secant. Consequently the local gain at either endpoint satisfies
\begin{equation}
 0<\alpha R_N'(D\Theta)
 <\alpha\,\frac{R_N(D\Theta)}{D\Theta}=K_c .
\end{equation}
The two endpoint Jacobians coincide because $R_N'$ is even. The return over a full cycle is then the square of a stable one-step matrix, and hence the cycle attracts nearby disturbances. What grows from zero at onset is the magnitude of the alternating mismatch. Substituting $R_N(e)=G_Ne-c_Ne^3+O(e^5)$ into Eq.~\eqref{eq:cycle-balance} and cancelling the amplitude leaves $\alpha c_N(D\Theta)^2$ to balance the excess gain $K-K_c$. Since both endpoints have the same squared mismatch, at fixed $N$ and $\alpha$ approaching its critical value from above this gives
\begin{equation}
 \overline{\mathcal J}=D^2\Theta^2\sim\frac{K-K_c}{\alpha c_N}.
 \label{eq:scalar-local-loss}
\end{equation}
Thus the amplitude grows as the square root of the excess gain, whereas the loss measured at the census grows linearly, as used in Sec.~\ref{sec:scalar}.

\subsection*{Continuous observation between discrete updates}
\label{subsection:continuous-observation-between-discrete-updates}

So far the population has acted on a single reading at the end of each census period. If agents observe the state throughout the interval, the mandate can still remain fixed between censuses, but the correction need no longer depend on the endpoint alone. Let $m^{\rm mem}$ be a running estimate of mismatch that relaxes towards the instantaneous reading error on time scale $\tau_m$. Integrating $\tau_m\dot m^{\rm mem}=O(x(t))-y_{\rm ref}(\theta_k)-m^{\rm mem}$ through the census period gives
\begin{align}
 m^{\rm mem}_{k+1}&=r_m m^{\rm mem}_k+
 \int_0^T\frac{e^{-(T-u)/\tau_m}}{\tau_m}\,d_k(u)\,du,\\
 d_k(u)&=O(x(kT+u))-y_{\rm ref}(\theta_k),\qquad r_m=e^{-T/\tau_m}.
\end{align}
The integral weights recent observations more heavily, while $r_m$ tells us how much of the preceding census survives. Give this memory a weight $0\le w_{\rm mem}\le1$ by using $(1-w_{\rm mem})e_k+w_{\rm mem}m^{\rm mem}_{k+1}$ as the argument of $R_N$ in Eq.~\eqref{eq:normal_theta}. At zero weight the memory records the mismatch but cannot feed back on it, so the original pair of multipliers is joined by one decaying multiplier $r_m$. For fixed finite $\tau_m$, sufficiently weak coupling leaves the noncritical modes inside the unit circle. The simple crossing in Eq.~\eqref{eq:flip-crossing-slope} can then shift in gain without disappearing. A slower memory or a strong delay is treated by the same eigenvalue calculation on the enlarged state $(x,\theta,m^{\rm mem})$, as for the delayed hierarchy of Eq.~\eqref{eq:hier-delay-polynomial}.

\section{Hierarchical Redescription and Attenuation}
\label{app:hierarchy}

In Sec.~\ref{sec:hierarchy}, the flat majority in Fig.~\ref{fig:hierarchy_postpones} begins to cycle at a population size for which the same inputs, grouped into a tree, still give a decaying disturbance. To explain the shift in onset, we need the gain $K_N^{(L)}$, i.e. how much does the mandate change in response to a small mismatch after the signal has passed through $L$ layers? We calculate it first for the majority tree, where each block replaces its inputs by one sign [Eq.~\eqref{eq:ternary-block}], and then for a flat majority whose verdict passes through stages of fixed transmission. Comparing each gain with $K_c(T)$ in Eq.~\eqref{eq:critical-gain} tells us why regrouping the inputs postpones the crossing, whereas adding transmission stages at fixed $N$ gives the logarithmic depth law of Eq.~\eqref{eq:minimum-depth}.

\begin{figure*}[t]
    \centering
    \includegraphics[width=\linewidth]{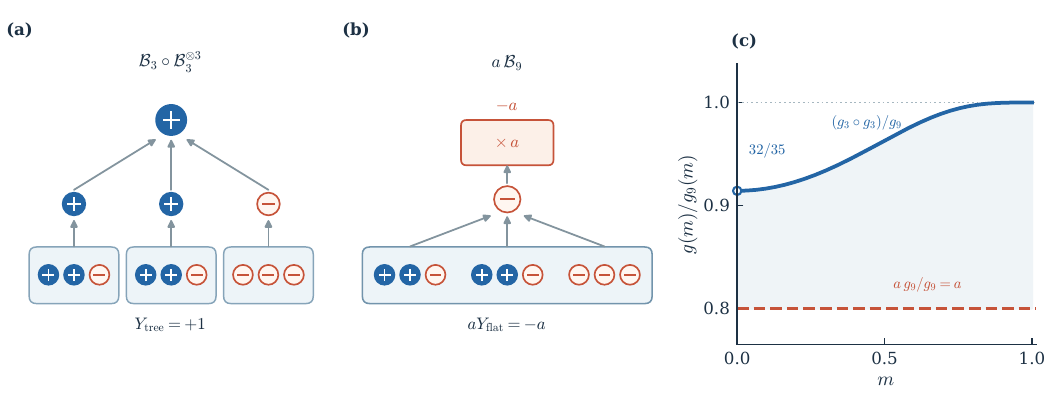}
    \caption{\textbf{Where the majority is taken changes the correction.} Read (a) and (b) from bottom to top. (a) Each boxed triple returns the sign of its sum. The three outputs are $+,+,-$, so a second majority gives $Y_{\rm tree}=+1$. (b) Counting the same four positive and five negative inputs in one block gives $Y_{\rm flat}=-1$. A stage with transmission $a>0$ passes on $-a$. (c) Now vary the mean input sign $m$. Write $g_n(m)$ for the mean verdict of a block of $n$ independent signs [Eq.~\eqref{eq:mean-block-map}]. The solid curve divides the tree response $g_3(g_3(m))$ by the flat response $g_9(m)$. Both vanish at $m=0$, so the open circle marks the ratio of their slopes, $32/35$. At $m=1$ all signs agree and the ratio is one. For the attenuated flat majority (dashed), the corresponding ratio is $a=0.8$ at every nonzero bias.}
    \label{fig:architecture}
\end{figure*}

Figure~\ref{fig:architecture} isolates the difference using nine of the inputs in the hierarchy of Fig.~\ref{fig:recursive-ballot}. A triple passes on its verdict but discards the margin by which it won, so two narrow positive majorities can outweigh a unanimous negative triple. If all nine signs are counted together, that margin still counts, and attenuating the resulting negative verdict cannot make it positive. Panels (a) and (b) show this for one census. To obtain the gain we must repeat the comparison over independent inputs of mean $m$ and ask how the mean output changes as $m$ moves away from zero. Panel (c) shows the two responses relative to the flat majority, and it is their behaviour near zero bias that enters the stability calculation.

For an odd block span $n$, the block operator acts on the actual signs $\sigma_j\in\{-1,+1\}$ and returns one sign:
\begin{equation}
 \mathcal B_n(\sigma_1,\ldots,\sigma_n)
 =\operatorname{sgn}\!\left(\sum_{j=1}^{n}\sigma_j\right).
 \label{eq:odd-majority-block}
\end{equation}
Equation~\eqref{eq:ternary-block} is its three-input instance. Suppose now that the inputs are independent at fixed mismatch and each has mean $m$, so the probability of a positive sign is $p=(1+m)/2$. Although $\mathcal B_n$ still returns only a sign, its mean varies continuously with $m$. Denote that mean by $g_n(m)$. Counting the outcomes with a positive majority as in Eq.~\eqref{eq:majority-tail} gives
\begin{equation}
 g_n(m)=\mathbb E[\mathcal B_n]
 =2\sum_{j=(n+1)/2}^{n}\binom{n}{j}p^j(1-p)^{n-j}-1 .
 \label{eq:mean-block-map}
\end{equation}
The choice of coordinate matters: $g_n$ takes a \emph{bias} as its argument, whereas the population response $R_N(e)$ in Eq.~\eqref{eq:mean-majority} takes a \emph{mismatch}. For a flat majority they are related by $R_N(e)=g_N(m(e))$.

Now follow the signs from the leaves to the root of the hierarchy. Each parent receives one sign from each child, and because the children drew on disjoint independent inputs, their outputs remain independent at fixed $e$. A binary sign is specified by its mean, so the input to the next mean block map is precisely the output bias of the previous one. If every block at level $\ell$ has odd span $n_\ell$, Galam's bottom-up recursion~\cite[Sec.~II]{Galam2008Sociophysics} becomes
\begin{align}
 m_0(e)&=m(e),\qquad m_\ell(e)=g_{n_\ell}\!\left(m_{\ell-1}(e)\right),\\
 K_N^{(L)}&=\alpha m_L'(0)=\alpha m'(0)\prod_{\ell=1}^{L}a_{n_\ell},
 \qquad N=\prod_{\ell=1}^{L}n_\ell .
 \label{eq:disjoint-block-gain}
\end{align}
The product in Eq.~\eqref{eq:disjoint-block-gain} is exact: each block receives unbiased signs at $e=0$, so its contribution to the derivative is $a_{n_\ell}=g_{n_\ell}'(0)$ from Eq.~\eqref{eq:majority-influence}. To understand the origin of the attenuation factor in Sec. \ref{sec:hierarchy}, consider a block with many inputs of common mean $m$. Their sum has mean $n m$ and standard deviation $\sqrt{n(1-m^2)}$. On the majority transition window $m=O(n^{-1/2})$,  and the mean is therefore about $\sqrt n\,m$ standard deviations above the decision threshold at zero. As in Appendix~\ref{app:majority}, the normal approximation gives the probability of a positive verdict. Converting that probability back to a mean sign gives the layer map
\begin{equation}
 m_\ell(e)\simeq 2\Phi\!\left(\sqrt{n_\ell}\,m_{\ell-1}(e)\right)-1
 =\operatorname{erf}\!\left(\sqrt{\frac{n_\ell}{2}}\,m_{\ell-1}(e)\right),
 \label{eq:gaussian-block-recursion}
\end{equation}
where $\Phi$ is the standard normal cumulative distribution function and $m_0(e)=\tanh(\beta e)$. The input to the first layer thus has slope $m_0'(0)=\beta$. Each subsequent layer acts on the bias passed up by the preceding one. Since $\operatorname{erf}'(0)=2/\sqrt{\pi}$, the small-mismatch slope of this Gaussian map propagates as
\begin{equation}
 m_\ell'(0)\simeq
 \frac{2}{\sqrt{\pi}}\sqrt{\frac{n_\ell}{2}}\,m_{\ell-1}'(0)
 =\sqrt{\frac{2n_\ell}{\pi}}\,m_{\ell-1}'(0).
 \label{eq:gaussian-block-slope}
\end{equation}
This is also the large-$n_\ell$ limit of the exact binomial slope in Eq.~\eqref{eq:majority-influence}. Following a perturbation all the way to the root collects one factor from each layer. Because the blocks partition the leaves, $\prod_\ell n_\ell=N$, so the top-level slope and the resulting loop gain are
\begin{align}
 m_L'(0)&\simeq\beta\prod_{\ell=1}^{L}\sqrt{\frac{2n_\ell}{\pi}}
 =\beta\left(\frac{2}{\pi}\right)^{L/2}\sqrt N ,
 \label{eq:large-block-top-slope}\\
 K_N^{(L)}&=\alpha m_L'(0)
 \simeq\alpha\beta\sqrt{\frac{2N}{\pi}}\,c^{L-1}
 \simeq K_N^{(1)}c^{L-1},\qquad c=\sqrt{\frac{2}{\pi}} .
 \label{eq:large-block-relative-gain}
\end{align}
The square roots of the block sizes reproduce the same $\sqrt N$ as a flat count. The difference is how often a sum is replaced by its sign: the tree does this $L$ times, while the flat majority does it once. This is why the gain \emph{relative to the flat majority} carries $L-1$ factors of $c$, even though the slope of every large block exceeds one.

At fixed depth, set Eq.~\eqref{eq:large-block-relative-gain} equal to the critical gain $K_c(T)$. Since the gain grows as the square root of population, compensating for $c^{L-1}$ takes a population larger by its inverse square. In terms of the flat critical size $N_c^{(1)}(T)$ in Eq.~\eqref{eq:critical-size}, the smooth onset estimate is
\begin{equation}
 N_c^{(L)}(T)\simeq N_c^{(1)}(T)\left(\frac{\pi}{2}\right)^{L-1}.
 \label{eq:large-block-critical-size}
\end{equation}
Read the same boundary in the opposite direction, at fixed $N$, and the depth at equality is
\begin{equation}
 L_* = 1+\frac{\log[N/N_c^{(1)}(T)]}{\log(\pi/2)} .
 \label{eq:large-block-depth-boundary}
\end{equation}
The Gaussian estimate places the stable side at $L>L_*$. If depth could be chosen freely, its minimum would be the first admissible integer strictly above this logarithmic boundary. For a majority tree, however, choosing $L$ at fixed $N$ also chooses how to factor $N$ into odd block sizes. The approximation holds at fixed $L$ with every block large, and for small blocks the exact product of Eq.~\eqref{eq:disjoint-block-gain} replaces it. The nine-input example makes the difference measurable. A triple returns a positive sign if two or three of its inputs are positive, giving
\begin{equation}
 g_3(m)=2[3p^2(1-p)+p^3]-1=\frac{3m-m^3}{2}.
 \label{eq:ternary-mean}
\end{equation}
Two rounds therefore have slope $(3/2)^2$, compared with $a_9=315/128$ for the flat vote. Their ratio is $32/35$ at small bias, but at unanimity both verdicts are identical [Fig.~\ref{fig:architecture}(c)]. Repeating triples to depth $L$ gives an absolute slope $(3/2)^L$ for $N=3^L$ inputs. This grows more slowly than the flat slope but still diverges, which is why regrouping only postpones the instability in Fig.~\ref{fig:hierarchy_postpones}. The same holds for any choice of odd spans. Each layer consumes a factor of at least three in population, so a majority tree has depth $L\le\log_3N$, whereas Eq.~\eqref{eq:large-block-depth-boundary} asks for a depth of order $\log N/\log(\pi/2)$, which is larger. Writing $a_n=n^{\nu_n}$, the exponent $\nu_n=\ln a_n/\ln n$ is smallest at $n=3$, so every tree with odd spans $n_\ell\ge3$ obeys $\prod_\ell a_{n_\ell}\ge N^{\ln(3/2)/\ln3}\approx N^{0.37}$, and its gain still grows with $N$.

There is another operation available if we can keep the population fixed and change how its verdict is transmitted. Let a stage pass on a fraction $0<a<1$ of the incoming signal, without holding a new vote or delaying its arrival. It cannot reverse the sign in the nine-input example, but every additional stage now reduces the magnitude of the correction by the same factor. Starting with the flat gain $K_{\rm flat}=K_N$ from Eq.~\eqref{eq:loop-gain}, the gain after transmission is
\begin{equation}
 K_N^{(L)}=K_{\rm flat}a^{L-1}.\label{eq:layer_attenuation}
\end{equation}
To find the least depth that restores stability, compare this product with $K_c(T)$ in Eq.~\eqref{eq:critical-gain}. On taking logarithms the strict inequality becomes
\begin{equation}
 (L-1)|\log a|>\log(K_{\rm flat}/K_c).
\end{equation}
Each stage thus removes a fixed amount of log gain. If the flat gain grows in proportion to $N^\nu$ up to bounded positive factors, and the critical gain stays finite and bounded away from zero, the smallest admissible integer depth satisfies
\begin{equation}
 L_{\min}=\frac{\nu}{|\log a|}\log N+O(1).
 \label{eq:minimum-depth}
\end{equation}
The independent majority has $\nu=1/2$ by Eq.~\eqref{eq:gain-population}. For unequal stage transmissions the log reductions add in the same way: if their cumulative value is $\kappa(L-1)+o(L)$ with $\kappa>0$, the leading coefficient becomes $\nu/\kappa$. For independent random stages $\kappa=-\mathbb E\log a_j$ by the law of large numbers, and correlated stages enter through the growth rate of the product of their transmissions. And if passing the verdict on takes a census interval, multiplying static slopes no longer suffices. The next appendix follows the disturbance through the intervening macroscopic dynamics to find what reaches the next census.

\section{Stability Under Finite Macroscopic Response Time}
\label{app:robustness}

Section~\ref{sec:hierarchy} explains the displacement of the cycling threshold in Fig.~\ref{fig:hierarchy_postpones} by comparing the gain transmitted through a hierarchy, $K_N^{(L)}$, with the scalar critical gain $K_c(T)$. That comparison rests on two timing assumptions - first that the state relaxes on a single time scale during the hold, and second that all the majority rounds finish before the next mandate takes effect. What if the population sees several relaxing components, or receives a correction based on an earlier mandate?

\begin{figure*}[t]
    \centering
    \includegraphics[width=\textwidth]{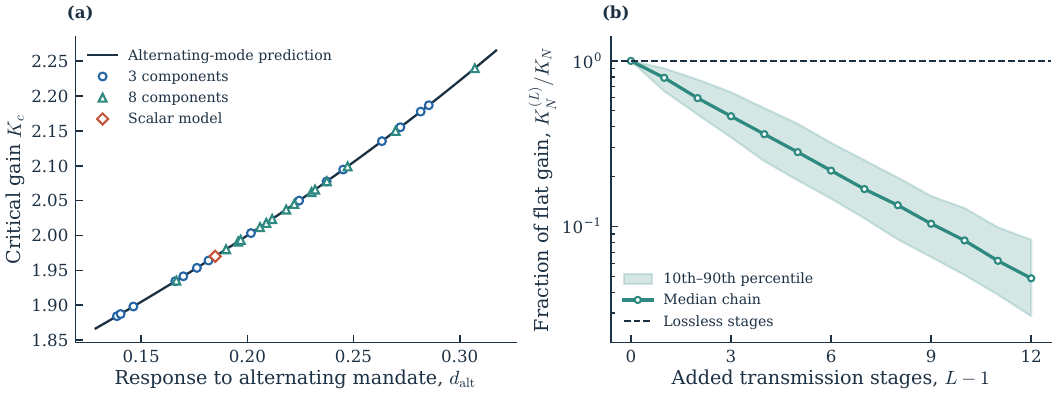}
    \caption{\textbf{Changing the relaxation times shifts the threshold. Attenuating the correction changes the gain.} (a) The line gives $K_c=2/(b-d_{\rm alt})$. Markers locate unit spectral radius independently for the scalar reference ($u=1$) and for 15 draws each of three or eight equally weighted components, with $u_i$ uniform on $[0.25,2.5]$, $b=1.2$ and $s=0.4$. (b) Fraction of the flat-majority gain transmitted by $L-1$ static stages at fixed $N$. The solid curve is the median over 160 chains with independent stage transmissions uniform on $[0.62,0.95]$. Shading spans the tenth to ninetieth percentiles. Lossless stages transmit the whole correction (dashed line).}
    \label{fig:robustness}
\end{figure*}

We return to the census map of Eqs.~\eqref{eq:normal_x}--\eqref{eq:normal_theta} to separate these effects. Appendix~\ref{app:hierarchy} calculates how a small mismatch changes the correction. The other half of the calculation is how much of that correction the population can observe one census later. Figure~\ref{fig:robustness} shows the distinction: changing the relaxation times moves the critical gain, whereas adding instantaneous transmission stages changes the gain compared with that threshold. We then let each layer take one census interval and follow the older signal explicitly in Fig.~\ref{fig:hier-delayed-spectrum}.

Begin with the population response. Its transition width $w_N$ is the mismatch scale over which the mean verdict changes appreciably. Rescale mismatch as $e=w_N u$ and examine the change in response $R_N(w_N u)-R_N(0)$. Suppose this profile and its first derivative in $u$ converge near zero to a profile $q(u)$ with $0<q'(0)<\infty$. Returning from the scaled coordinate to mismatch units gives $R'_N(0)\sim q'(0)/w_N$. The independent majority in Appendix~\ref{app:majority} gives $w_N=N^{-1/2}$. Changing the shape of a smooth individual response away from the threshold leaves this exponent alone when its derivative there remains finite and nonzero.

To see what a more general dynamics might retain, perturb the stable steady state at a census point. Linearise Eqs.~\eqref{eq:normal_x}--\eqref{eq:normal_theta}. The derivatives $A=D_{\mathbf x}\Phi_T$ and $B=D_\theta\Phi_T$ say how the initial state and the held mandate affect the next state. The derivatives $P=D_{\mathbf x}O$ and $Q=D_\theta y_{\rm ref}$ convert those changes into a reading and a reference. The slope $G=D R_N(0)$ then turns their difference into an aggregate response. All derivatives are then evaluated at the matched operating point. A perturbation away from this point will have the form  $A\delta\mathbf x+B\delta\theta$. The complete linear return is therefore
\begin{align}
 \binom{\delta\mathbf x'}{\delta\theta'}&=J_N(T)\binom{\delta\mathbf x}{\delta\theta},\\
 J_N(T)&=\begin{pmatrix}A&B\\\alpha GPA&I+\alpha G(PB-Q)\end{pmatrix}.
 \label{eq:general-spectral-return}
\end{align}
Holding the mandate fixed tells us whether the macroscopic dynamics relax, and a small displacement then evolves under $A$ alone. At a census, however, that displacement changes the reading and hence the next mandate. It is the spectral radius of the complete return $J_N(T)$ in Eq.~\eqref{eq:general-spectral-return} that decides whether all linear disturbances decay, by the criterion in Eq.~\eqref{eq:census-spectral-radius}.

Consider one mandate and one reading, but allow the state $\mathbf x$ to have several components. In this case $B=\mathbf u$ is an input direction and $P=\mathbf p^\top$ a row selecting the reading. For a mode with eigenvalue $\lambda$, the first row of Eq.~\eqref{eq:general-spectral-return} gives $(\lambda I-A)\delta\mathbf x=\mathbf u\delta\theta$. Provided $\lambda$ is not an eigenvalue of $A$, we can solve for $\delta\mathbf x$. The next census reads the advanced state, so the change in its reading per unit change in mandate is $\lambda\mathbf p^\top(\lambda I-A)^{-1}\mathbf u$. Subtracting the reference slope $b$ gives the mismatch that drives the revision. With $K=\alpha G$ (the gain defined in Sec.~\ref{sec:normal_form}), the second row becomes
\begin{equation}
 \lambda-1=K\big[\lambda\mathbf p^\top(\lambda I-A)^{-1}\mathbf u-b\big].
\end{equation}
To find a disturbance that reverses on every census, set $\lambda=-1$. A unit alternating mandate then produces a reading of amplitude $d_{\rm alt}=P(I+A)^{-1}B$ at the end of the period. The shortfall from the reference is $b-d_{\rm alt}$. Reversing the mandate requires a correction of magnitude two, so the gain for this mode is $2/(b-d_{\rm alt})$. This is the calculation of Eq.~\eqref{eq:flip-shortfall} where now the principal concern is how far the measured state can follow an alternating mandate, even when several relaxation times contribute to the reading.

Figure~\ref{fig:robustness}(a) tests this substitution. Each component relaxes towards the same value $s\theta_k$, but retains a fraction $\exp(-u_i)$ of its initial displacement over a census interval, so $u_i$ is that component's relaxation rate multiplied by $T$. The reading is the arithmetic mean of the components. Its alternating response $d_{\rm alt}$ is therefore the mean of the scalar responses $s\tanh(u_i/2)$. In these examples, the gain at which an eigenvalue first reaches unit modulus agrees with the prediction from that mean. A larger alternating response brings the reading closer to its reference, so a larger gain is needed to reverse the mandate.

Panel (b) changes the other half of the loop. Keep the population and macroscopic dynamics fixed, form a flat majority, and pass its signal through additional stages before revising the mandate. Section~\ref{sec:hierarchy} denotes the resulting gain by $K_N^{(L)}$. If stage $j$ transmits a fraction $a_j$ without delay, $K_N^{(L)}/K_N$ is simply the product of those fractions [Eq.~\eqref{eq:layer_attenuation} for equal fractions]. After twelve stages the median chain transmits about five per cent of the original correction. Here the threshold has not changed, but the gain delivered to the loop has fallen.

\subsection*{The spectrum of a delayed hierarchy}
\label{subsection:the-spectrum-of-a-delayed-hierarchy}

What if transmitting the signal takes time? Keep the population fixed, but let each additional layer take one census interval. At depth $L$ the total delay is $h=L-1$ intervals. Write $s_L$ for the change in the eventual reading per unit sustained mandate. This is a sensitivity of the macroscopic reading, whereas $K_N^{(L)}$ in Fig.~\ref{fig:robustness}(b) measures the size of the mandate correction per unit mismatch. In the delayed example the current mandate sets the reference $s_L\theta_k$, while the available reading $y_k$ still responds to earlier mandates. With $\eta$ setting the update strength, the delayed model reads
\begin{align}
 y_{k+1}&=r y_k+(1-r)s_L\theta_{k-h},\label{eq:hier-delayed-readout}\\
 \theta_{k+1}&=\theta_k+\eta s_L(y_k-s_L\theta_k).
 \label{eq:hier-delayed-review}
\end{align}
The correction now uses the reading already available, $y_k$, while $y_{k+1}$ relaxes towards the signal from $h$ intervals ago. If the mandate is held at any constant $\bar\theta$, the reading can settle at $s_L\bar\theta$ with zero mismatch. There is consequently a whole line of matched states, and the question of stability asks whether a mismatch dies away, rather than which particular stable state is reached. To advance one census we must retain the earlier mandates $(\theta_k,\ldots,\theta_{k-h})$ along with $y_k$. For a disturbance proportional to $\lambda^k$, a delay of $h$ intervals multiplies the mandate amplitude by $\lambda^{-h}$. Substitution in Eqs.~\eqref{eq:hier-delayed-readout} and~\eqref{eq:hier-delayed-review} gives the polynomial for the eigenvalues:
\begin{equation}
 \lambda^h(\lambda-r)(\lambda-1+\eta s_L^2)-\eta s_L^2(1-r)=0.
 \label{eq:hier-delay-polynomial}
\end{equation}
One root is always $\lambda=1$. Its disturbance shifts every stored mandate by the same amount and shifts the reading by $s_L$ times that amount. The reference shifts with it, leaving no mismatch. Excluding this neutral direction, let $\rho_\perp$ be the largest modulus of the remaining roots. If $\rho_\perp<1$, every perturbation transverse to the line decays, even though the full spectral radius stays at one. Notably, each layer raises the power of the polynomial equation, meaning that solutions are no longer guaranteed to be strictly real. 

\begin{figure*}[t]
\centering
\includegraphics[width=\textwidth]{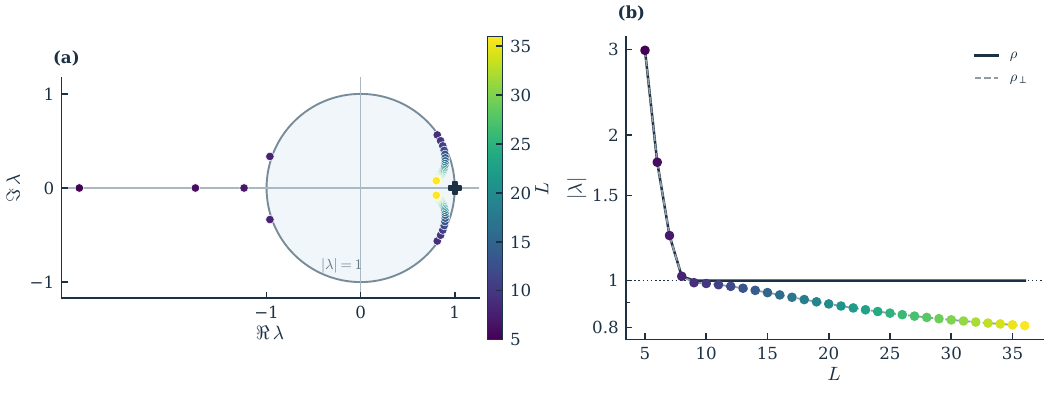}
\caption{\textbf{Stability transverse to a fixed line.} (a) Leading transverse multipliers in the complex $\lambda$ plane, coloured by depth. Both members of a conjugate pair are shown. The cross marks the distinct, exactly neutral mode at $+1$. (b) The full radius $\rho$ and transverse radius $\rho_\perp$ separate once the transverse modes lie inside $|\lambda|=1$. Here $N=10^4$, $r=0.30$, $\eta=0.01$ and $s_L=\beta\sqrt{2N/\pi}\,2^{-(L-1)/4}$ with $\beta=0.5$. Each additional layer weakens the signal and delays it by one census. At depth nine $\rho_\perp=0.9904$, and it remains below one through depth 36.}
\label{fig:hier-delayed-spectrum}
\end{figure*}

Figure~\ref{fig:hier-delayed-spectrum} shows how the mismatch disappears as layers are added. At seven layers the growing disturbance changes sign at every census, while at \emph{eight} layers the leading multipliers form a complex pair, adding a rotational phase to the disturbance. At nine the pair lies inside the unit circle and the mismatch decays. In this example the upper limit on useful depth is a delivery problem, rather than a second source for loss of stability - a correction can decay as it should and still arrive too late (Appendix~\ref{app:cost_functions}).

\section{Cost Functions and Hierarchical Inertia}
\label{app:cost_functions}

Section~\ref{sec:inertia} asks how much hierarchy is worth maintaining once we account for what it costs to send a correction through the layers. Appendix~\ref{app:hierarchy} gives the reduction in feedback gain with depth. We first use that result to calculate the saving in cycling loss from one more layer, then follow a signal from the bottom to the top to find the price of that saving: how much of the signal arrives, and whether it arrives by the deadline at all.

\subsection*{The Cost Minimum}

Consider a population of fixed size whose flat gain $K_N^{(1)}$ lies just above the critical gain $K_c(T)$. Inserting a static stage that transmits a fraction $0<a<1$ of its input reduces the gain to $K_N^{(L)}=K_N^{(1)}a^{L-1}$ at depth $L$ [Eq.~\eqref{eq:layer_attenuation}]. Appendix~\ref{app:hierarchy} obtained the same multiplicative form with $c=\sqrt{2/\pi}$ for the gain of large majority blocks relative to a flat vote [Eq.~\eqref{eq:large-block-relative-gain}]. Here we use $a$ to stand in as a general attenuation factor. Let $A_0>0$ be the proportionality coefficient evaluated at the onset of cycling. The leading estimate for the loss at depth $L$ is then
\begin{equation}
 \overline{\mathcal J}_{\rm dyn}(L)\simeq
 A_0\big[K_N^{(1)}a^{L-1}-K_c(T)\big]_+ ,
 \label{eq:depth-dynamic-loss}
\end{equation}
where the positive part sets the cycling loss to zero once the matched state is stable. The first layer added to a cycling system saves more than a later one: both reduce the gain by the same fraction, but the later layer acts on an already attenuated signal. To locate the cost minimum, take the first two terms of Eq.~\eqref{eq:total-layer-loss} and price each layer at $\mu>0$, so $C(L)=\mu L$. Treating depth temporarily as continuous, we may take derivatives on the cycling branch of Eq. \eqref{eq:depth-dynamic-loss}:
\begin{align}
 \frac{d}{dL}\big(\overline{\mathcal J}_{\rm dyn}+\mu L\big)
 &=\mu-A_0K_N^{(1)}|\ln a|a^{L-1},\label{eq:depth-marginal-cost}\\
 \frac{d^2}{dL^2}\big(\overline{\mathcal J}_{\rm dyn}+\mu L\big)
 &=A_0K_N^{(1)}|\ln a|^2a^{L-1}>0 .
\end{align}
The first derivative compares the maintenance cost with the marginal saving. Where the two are equal, solving Eq.~\eqref{eq:depth-marginal-cost} gives
\begin{equation}
 L_* = 1+\frac{1}{|\ln a|}
 \ln\!\left(\frac{A_0K_N^{(1)}|\ln a|}{\mu}\right).
 \label{eq:continuous-cost-depth}
\end{equation}
If $L_*$ lies between depth one and the stability boundary, it is the unique interior minimum of the two-term approximation. Beyond the boundary the cycling loss is zero and the cost rises with  $\mu$. Thus, if a layer is too expensive even at depth one, the minimum is at one. If adding layers still pays up to the loss of cycling, the minimum is at the boundary. For integer depths the comparison has a particularly simple form: one more layer is cheaper overall precisely when
\begin{equation}
 \overline{\mathcal J}_{\rm dyn}(L)-\overline{\mathcal J}_{\rm dyn}(L+1)>\mu .
 \label{eq:cost-of-one-more-layer}
\end{equation}
Pricing an instability and requiring stability are two different design choices. If the mandate must settle, only depths strictly on the stable side are admissible. In Fig.~\ref{fig:phases}, depths three and four meet that condition and the delivery deadline, so three costs less to maintain. The fifth layer fails even if maintenance is free. To see why, we now have to follow the signal itself.

\subsection*{Transmission Inertia}

In the flat census map, the mismatch $e_k$ produces a mean verdict $R_N(e_k)$, which changes the mandate by $\alpha R_N(e_k)$ [Eqs.~\eqref{eq:mean-majority} and~\eqref{eq:normal_theta}]. It is this verdict that the transmission stages carry. Our concern here is with the signal carrying the verdict, whatever that might be. For the sake of simplicity, we equate the signal to the response $z_k^{(0)}=R_N(e_k)$ for the input and $z_k^{(\ell)}$ for the value held by stage $\ell$ at census $k$, in the same units as the mean verdict. If the mandate is updated from the top stage, its increment is $\alpha z_k^{(L)}$. Whereas the majority blocks in Appendix~\ref{app:hierarchy} form a new verdict from several incoming signs, a transmission stage receives one value from below and retains part of its own previous value. Note that these formulations are not incompatible! The verdict passed up may reduce to a sign in aggregate, but the amplitude of the signal carrying that sign can vary continuously, as can the cost of forming the verdict and transmitting it to the next layer.

Here $L$ counts transmission stages, identified with layers of the hierarchy. In Eq.~\eqref{eq:layer_attenuation} the initial majority is counted as the first layer, leaving $L-1$ transmission stages at total depth $L$. To measure the response of the stages themselves we can prescribe the sequence $z_k^{(0)}$ at the input. A stage with transmission $a$ moves a fraction $\gamma$ of the way towards the attenuated value from its neighbour in one census interval:
\begin{equation}
 z^{(\ell)}_{k+1}=(1-\gamma)z^{(\ell)}_k+\gamma a z^{(\ell-1)}_k,
 \qquad 1\le\ell\le L,\quad 0<\gamma\le1 .
 \label{eq:inertia-layer-update}
\end{equation}
To better measure transmission, let us drive the bottom layer sinusoidally at angular frequency $\omega$ in radians per census. Write the steady response as $z_k^{(\ell)}=\Re[Z_\ell e^{i\omega k}]$. Substitution into Eq.~\eqref{eq:inertia-layer-update} gives $(e^{i\omega}-(1-\gamma))Z_\ell=\gamma a Z_{\ell-1}$. Since each layer passes its output to the next, their complex amplitude ratios multiply~\cite{SmithDigitalFilters}:
\begin{align}
 H_L(\omega)=\frac{Z_L}{Z_0}
 &=\left[\frac{\gamma a e^{-i\omega}}{1-(1-\gamma)e^{-i\omega}}\right]^L ,
 \label{eq:cascade-transfer}\\
 |H_L(\omega)|
 &=\left[\frac{\gamma a}
 {\sqrt{1+(1-\gamma)^2-2(1-\gamma)\cos\omega}}\right]^L .
 \label{eq:cascade-magnitude}
\end{align}
A signal held constant gives every stage time to settle, so the transmitted fraction at $\omega=0$ is $a^L$. For $0<\gamma<1$, the denominator increases from $\omega=0$ to $\pi$. That is, a correction that reverses at every census is attenuated more than a sustained one, as in Fig.~\ref{fig:transmission_inertia}. Taking the negative logarithm in Eq.~\eqref{eq:transmission-inertia} turns this multiplicative attenuation into a cost additive in depth, $\mathcal I_{\rm tr}=-L\log|H_1(\omega)|$. At $\gamma=1$ the stages still attenuate by $a$ but introduce only a shift in time, with no additional frequency dependence.

\subsection*{Transmission time}

Even a verdict transmitted without loss takes time to reach the top. For simplicity we use the same timing assumption as the delayed hierarchy discussed in Sec.~\ref{sec:hierarchy} and modelled in Appendix~\ref{app:robustness} [Eq.~\eqref{eq:hier-delayed-readout}]: transmission from one layer to the next takes one census period $T$. In Eq.~\eqref{eq:inertia-layer-update}, stage $\ell$ reads its neighbour at census $k$ and first passes on that input at $k+1$. A unit pulse applied at the bottom at census zero can therefore first affect stage $\ell$ at census $\ell$. For $\gamma=1$ the whole transmitted pulse travels at this rate. For $\gamma<1$, the first stage returns $\gamma a$ at census one and retains a fraction $1-\gamma$ of that response at each subsequent census, spreading the pulse over later arrival times. Dividing by the total transmitted weight $a$ gives the normalised arrival weights at delays $j=1,2,\ldots$,
\begin{equation}
 h_j=\gamma(1-\gamma)^{j-1},\qquad j\ge1 .
 \label{eq:arrival-weights}
\end{equation}
Passing the pulse through the next layer convolves these weights with those of the layer before. A deadline asks how much of the resulting pulse has reached the top in time, which we calculate in Eq.~\eqref{eq:transport-arrival-fraction}.

\subsection*{Reachability Before a Deadline}

The previous calculations describe how a signal is attenuated and spread out in time. A hard deadline asks a different question: can an intervention have any effect at the top by census $H$? Take the shift chain of Eq.~\eqref{eq:shift-chain} from rest, first with one transmission step per census ($\Delta=T$), so that the deadline allows $H$ steps. Write $\mathbf b_{\rm in}=\mathbf q_1$ for the vector that injects the verdict $z^{(0)}_k$ into the bottom layer, so that $\mathbf z_{k+1}=A\mathbf z_k+\mathbf b_{\rm in}z^{(0)}_k$. The top-layer signal at the deadline is the scalar $\mathbf q_L^\top\mathbf z_H$ in Eq.~\eqref{eq:top-displacement}. Iterating the chain makes $\mathbf q_L^\top\mathbf z_H$ a weighted sum of the inputs $z^{(0)}_0,\ldots,z^{(0)}_{H-1}$. The squared norm of its coefficient vector is $\mathbf q_L^\top W_H \mathbf q_L$, where the finite-horizon reachability Gramian~\cite[slides~8--11]{BoydLallEE263StateTransfer} is
\begin{equation}
 W_H=\sum_{j=0}^{H-1}A^j\mathbf b_{\rm in}\mathbf b_{\rm in}^\top(A^\top)^j .\label{eq:reachability-gramian}
\end{equation}
The sum in Eq.~\eqref{eq:reachability-gramian} has a simple interpretation here. After $j$ shifts the input has reached layer $j+1$ and has been attenuated $j$ times. After $L$ shifts it has left the chain. Thus
\begin{align}
 A^j\mathbf b_{\rm in}
 &=\begin{cases}
 a^j\mathbf q_{j+1}, & 0\le j<L,\\
 \mathbf 0, & j\ge L ,
 \end{cases}\\
 W_H&=\sum_{j=0}^{\min(H-1,L-1)}
 a^{2j}\mathbf q_{j+1}\mathbf q_{j+1}^\top .
 \label{eq:shift-gramian}
\end{align}
The Gramian is diagonal because an input of a given age occupies only one layer, with  its top entry the squared norm of the input coefficients that contribute to the top-layer signal at the deadline. Cauchy--Schwarz bounds that signal by the coefficient norm times the input norm, with equality when the input history is proportional to the coefficient vector~\cite[slides~8--11]{BoydLallEE263StateTransfer}. Reading the top entry of Eq.~\eqref{eq:shift-gramian} therefore gives both Eq.~\eqref{eq:top-displacement} and the cost of a unit top-layer signal in Eq.~\eqref{eq:control-inertia}:
\begin{align}
 z_{\max}(L,H)&=
 \begin{cases}0,& H<L,\\ a^{L-1},&H\ge L,\end{cases}\\
 \mathcal I_{\rm ctrl}(L,H)&=
 \begin{cases}+\infty,&H<L,\\ a^{-2(L-1)},&H\ge L.\end{cases}
 \label{eq:shift-control-wall}
\end{align}
These cases separate a weak signal from an absent one. Here, as in Eq.~\eqref{eq:layer_attenuation}, $L$ counts layers: the verdict enters the bottom layer directly and crosses $L-1$ links, giving $a^{L-1}$. In the cascade of Eq.~\eqref{eq:inertia-layer-update} each of the $L$ stages attenuates its input, which is why Eq.~\eqref{eq:cascade-magnitude} gives $a^L$ at $\omega=0$. Once $H\ge L$, a stronger bottom input can compensate for attenuation, though its required energy grows exponentially with depth for $a<1$. When $H<L$ there is no top coefficient to amplify. This gives the sharp upper depth in Fig.~\ref{fig:phases}, to be compared with the lower depth needed for feedback stability in Appendix~\ref{app:hierarchy}.

The two tests in Fig.~\ref{fig:phases} can now be compared on the same footing. The spectral radius in Eq.~\eqref{eq:census-spectral-radius} tests whether an unforced disturbance of the coupled census map decays. In the static transmission model, insert $K_N^{(L)}$ from Eq.~\eqref{eq:layer_attenuation} into Eq.~\eqref{eq:scalar-census-jacobian} to obtain that return matrix. Equation~\eqref{eq:top-displacement}, by contrast, asks how much of a deliberately applied input reaches a specified coordinate by a specified time~\cite[slides~8--11]{BoydLallEE263StateTransfer}. Indeed the shift matrix $A$ of the delivery experiment satisfies $A^L=0$, so its spectral radius is zero for every depth, including depths at which the top is unreachable by $H$. A modular depth must therefore meet both requirements: decay in the closed loop and delivery in time.

\paragraph{Census-period independent transmission time.}
The one-period convention in Eq.~\eqref{eq:hier-delayed-readout} ties the time available for a correction to the time each layer takes to pass it on. We can release that assumption without changing the census interval $T$. Suppose the layers update simultaneously every $\Delta>0$ units of physical time, reading the values their neighbours held before the update. We launch the signal at time zero from a chain at rest, take the first update at $t=\Delta$, and count an arrival exactly at the deadline as on time. Thus a deadline of $H$ census intervals allows $m=\lfloor HT/\Delta\rfloor$ completed updates. The first response can reach the top at $L\Delta$. For the shift chain, Eqs.~\eqref{eq:shift-gramian} and~\eqref{eq:shift-control-wall} therefore hold with $H$ replaced by $m$. This is the general form of the control inertia in Eq.~\eqref{eq:control-inertia}: the top is unreachable when $L\Delta>HT$, and a unit top-layer signal costs $a^{-2(L-1)}$ otherwise.

To separate travel time from signal strength, we now bound the verdict sent at each transmission step rather than the total energy of Eq.~\eqref{eq:top-displacement}. A population held at mismatch $e>0$ sends at most its mean verdict, $|z^{(0)}_j|\le R_N(e)\le1$. Keep the bottom injection $\mathbf b_{\rm in}=\mathbf q_1$ of the shift chain and the same link transmission $a$. To allow a stage to accumulate its input, let it take up a fraction $\gamma$ per step and retain $1-\gamma$ of its previous value, as in Eq.~\eqref{eq:inertia-layer-update}. Counting transmission steps $j$, the chain obeys
\begin{equation}
 \mathbf z_{j+1}=\big[(1-\gamma)I+\gamma A\big]\mathbf z_j
     +\gamma\mathbf q_1 z^{(0)}_j,\qquad t_j=j\Delta .
 \label{eq:transport-clock-chain}
\end{equation}
The input is injected directly into the first layer, so it crosses $L-1$ attenuating links during its $L$ updates to the top. Holding $\gamma$ fixed while varying $\Delta$ keeps the fraction taken up per update fixed. For delivery by a deadline we need the part of the pulse that has arrived by then. Each of the $L$ stages has the same arrival weights $h_j$ of Eq.~\eqref{eq:arrival-weights}, now counted in transmission steps. Convolving them gives the normalised top response at tick $j$: the last advance is on tick $j$, while the preceding $L-1$ advances can occupy any $L-1$ of the first $j-1$ ticks. Summing the resulting weights gives the fraction delivered by tick $m$,
\begin{equation}
 F_L(m;\gamma)=
 \sum_{j=L}^{m}\binom{j-1}{L-1}\gamma^L(1-\gamma)^{j-L}
 =\Pr\{\operatorname{Bin}(m,\gamma)\ge L\} .
 \label{eq:transport-arrival-fraction}
\end{equation}
The sum is zero for $m<L$. The binomial tail is a convenient expression for the deterministic impulse weights. Because all of those weights are nonnegative, a sustained verdict $z^{(0)}_j=R_N(e)$ attains the largest top-layer signal allowed by this bound. Its response increases up to the deadline, at which it equals $R_N(e)\,a^{L-1}F_L(m;\gamma)$. To deliver a top-layer signal of at least $z_\star>0$ by time $HT$, we therefore require
\begin{equation}
 L\Delta\le HT,\qquad
 R_N(e)\ \ge\ \frac{z_\star}{a^{L-1} F_L(\lfloor HT/\Delta\rfloor;\gamma)} .
 \label{eq:deadline-amplitude-test}
\end{equation}
The second test applies only if the first is met. With immediate uptake ($\gamma=1$) the fraction is then one: a faster clock can bring the verdict forward but cannot restore the amplitude lost at the links. With partial uptake, an earlier first arrival also leaves time for more of the response to accumulate. This is why the amplitude boundary in Fig.~\ref{fig:control_inertia}(c) falls as the clock is sped up. At fixed transmission $a$ the threshold changes only when the deadline permits another whole update, giving the steps in panel (d).

\section{Real-space renormalisation of the block response}
\label{app:rg_universality_proofs}

Section~\ref{sec:rg_universality} uses a block to follow information from one resolution to the next. In the majority tree of Appendix~\ref{app:hierarchy}, the block is a physical operation on a group of signs. In the graph picture of Fig.~\ref{fig:diffusion-resolution}, a block collects sites whose differences have relaxed on the observation scale. We begin with the first construction, where the response to a small input bias can be calculated exactly, and then express the second in terms of the motion and projection of token states.

\subsection*{The General RG Operator}
Suppose a block receives $n$ signs, each of mean $m=m_{\ell-1}(e)$ at mismatch $e$. Let $f_n(x)$ be the conditional mean of the outgoing sign when the arithmetic mean of the inputs is $x$. In a majority block $f_n$ takes the sign of $x$, but a block can also randomise its verdict at a tie or require unanimity. For independent identically distributed binary inputs, the number $j$ of positive signs is binomial with $p=(1+m)/2$, so the exact transformation of the incoming bias profile is
\begin{equation}
 \begin{split}
 m_\ell(e)&=\mathcal R_n[m_{\ell-1}](e)=g_n(m),\\
 g_n(m)&=\sum_{j=0}^n f_n(2j/n-1)\binom{n}{j}p^j(1-p)^{n-j}.
 \end{split}
 \label{eq:rg-functional-exact}
\end{equation}
Here the brackets denote an operator acting on a response \emph{function}, while $g_n$ gives the result at a particular value of its input bias. Equation~\eqref{eq:mean-block-map} is the odd majority case~\cite{Galam2008Sociophysics}. If disjoint blocks again return independent binary signs at fixed $e$, this output supplies the complete one-block input law for the next round.

The sample mean entering $f_n$ has variance $v_n=(1-m^2)/n$. For large spans and a threshold in the central fluctuation window, replacing its distribution by the normal density $\mathcal N(x;m,v_n)$ with that mean and variance gives a useful continuous version of the same operation~\cite[Sec.~11.5]{ODonnell2021BooleanFunctions}:
\begin{equation}
 \mathcal R_n[m_{\ell-1}](e)\simeq
 \int_{-\infty}^{\infty} f_n(x)\,\mathcal N(x;m,v_n)\,dx .
 \label{eq:rg-gaussian-operator}
\end{equation}
For the sign rule this integral gives the error function in Eq.~\eqref{eq:gaussian-block-recursion} near zero bias. With the same rule and span at every layer, a fixed bias profile $m_*$ obeys $\mathcal R_n[m_*]=m_*$. A sign-symmetric rule has the unbiased profile $m_*=0$. If we perturb it by a small profile $\delta m(e)$, the first-order output is $g_n'(0)\delta m(e)$. Writing $g=g_n$ for this repeated block gives
\begin{equation}
 \lambda_0=g'(0),\qquad (g^{\circ L})'(0)=\lambda_0^L .
 \label{eq:block-multiplier}
\end{equation}
For monotone blocks the slope may exceed, equal or fall below one, which classifies the transmission of small biases under repeated blocking. Closed-loop stability then follows by inserting the transmitted gain into the census map, Eq.~\eqref{eq:critical-gain}. At unit slope the cubic term of $g$ sets the nonlinear behaviour, as $c_N$ does in Eq.~\eqref{eq:scalar-local-loss}.

\subsection*{Unanimous Block Map ($n=3$)}
For a block of three inputs, replace the majority rule of Eq.~\eqref{eq:ternary-mean} by unanimity. An all-positive triple returns $+1$ and an all-negative triple returns $-1$. On a mixed triple use an independent fair sign, whose mean is zero. Only the two unanimous events contribute to the expected output, so substituting $p=(1+m)/2$ gives
\begin{align}
 g_{\rm un}(m)&=p^3-(1-p)^3
 =\frac{(1+m)^3-(1-m)^3}{8},\\
 &=\frac34m+\frac14m^3,\qquad
 \lambda_0=g_{\rm un}'(0)=\frac34 .
 \label{eq:unanimous-bias}
\end{align}
A mixed triple is more common than a unanimous one near zero bias. The fair fallback thus washes out part of the incoming preference: for $0<|m|<1$ its magnitude is multiplied by $(3+m^2)/4<1$ at each round, and tends to zero under iteration. If mixed triples pass on an abstention instead, the first-round mean is unchanged but the next round also needs the probability of a zero input. For larger blocks with a fixed supermajority threshold above one half, it is the probability in the tail beyond that threshold that controls transmission. A threshold within the shrinking transition window of Appendix~\ref{app:majority} probes a different part of the distribution.

\subsection*{Concentrated aggregation weights}
Unequal weights change which fluctuations a threshold sees. Let the positive weights sum to one, and let $S_N=\sum_i w_i\sigma_i$ be the signal whose sign is reported. Conditional independence gives $\mathbb E[S_N]=m$ and
\begin{equation}
 v_N(m)=\operatorname{Var}(S_N)=(1-m^2)\sum_{i=1}^N w_i^2 .
 \label{eq:weighted-sign-variance}
\end{equation}
For example, take $w_i=i^{-\xi}/\sum_{j=1}^N j^{-\xi}$, with rank exponent $\xi>1$. The normalisation and the sum of squares then both have finite limits, giving
\begin{equation}
 v_N(0)=\frac{\sum_{i=1}^N i^{-2\xi}}{(\sum_{i=1}^N i^{-\xi})^2}
 \longrightarrow\frac{\zeta(2\xi)}{\zeta(\xi)^2}>0 ,
 \label{eq:rank-weight-variance}
\end{equation}
where $\zeta(q)=\sum_{i\ge1}i^{-q}$ for $q>1$. This differs from equal weighting, for which $v_N(0)=1/N$. Where the individual weights are small compared with $\sqrt{v_N(0)}$, a normal approximation to the sign response gives the local estimate
\begin{equation}
 g_{\rm G}(m)=\operatorname{erf}\!\left(\frac{m}{\sqrt{2v_N(m)}}\right),
 \qquad g_{\rm G}'(0)=\sqrt{\frac{2}{\pi v_N(0)}} .
 \label{eq:weighted-gaussian-response}
\end{equation}
The small-weight condition is what permits the Gaussian replacement~\cite[Sec.~5.2]{ODonnell2021BooleanFunctions}. For the rank law above, the largest weight divided by the square root of the sum of squared weights tends to $1/\sqrt{\zeta(2\xi)}$, rather than to zero. In the extreme concentration limit the answer is simpler to get directly. As soon as $w_1>1/2$, even unanimous opposition from every other input cannot overturn the first sign. The outgoing sign is then $\sigma_1$, hence
\begin{equation}
 g(m)=m,\qquad g'(0)=1.\label{eq:lambda_oligarchy}
\end{equation}
This is the marginal case of Eq.~\eqref{eq:block-multiplier}: the incoming bias is transmitted unchanged. Its gain in mismatch units is $\alpha m'(0)$, which can be compared directly with $K_c(T)$ in Eq.~\eqref{eq:critical-gain}.

\paragraph{The corresponding construction for token states.}
For the geometric example in Sec.~\ref{sec:rg_universality}, a useful starting point is one residual attention head in continuous depth. Arrange the $n_{\rm tok}$ token vectors of dimension $d$ as the rows of $X(t)$, let $A(X)$ be the matrix of attention weights with each row summing to one, and let $W_V$ act on the feature coordinates. In a relaxation model where one step moves a token towards the weighted values of its neighbours, the small-step limit is
\begin{equation}
 \frac{dX}{dt}=A(X)XW_V-X .
 \label{eq:token-attention-example}
\end{equation}
Here $t$ measures depth. The equation describes interacting token trajectories, as in continuous-depth models of attention~\cite{Geshkovski2025,Castin2025,Bruno2025}. To track direction separately from length, write the transpose of row $i$ of $X$ in Eq.~\eqref{eq:token-attention-example} as a column $\mathbf x_i=r_i \mathbf u_i$, where $r_i=\|\mathbf x_i\|>0$ and $\mathbf u_i$ is a unit vector. Differentiating gives
\begin{equation}
 \dot{\mathbf u}_i=\frac{1}{r_i}(I-\mathbf u_i \mathbf u_i^\top)\dot{\mathbf x}_i .
 \label{eq:token-angular-speed}
\end{equation}
The projection removes radial motion. For a given tangential velocity, a larger radius means a smaller change in direction per layer, which is the physical content of a speed factor. Its depth dependence is set by the update: for example, the comparison of Karagodin et al.~\cite{Karagodin2025} assigns unit speed to ordinary Post-LN and a decreasing speed to a separately scaled update.

To represent a group of rapidly mixing sites by one variable, consider the symmetric graph diffusion used in Fig.~\ref{fig:diffusion-resolution}~\cite{Villegas2023,Nurisso2025}. For a fixed symmetric stochastic attention matrix $A$, the relaxation in Eq.~\eqref{eq:token-attention-example} with $W_V=I$ is graph diffusion, $\dot X=-\mathcal L X$ with $\mathcal L=I-A$. The fast modes describe differences that disappear first, just as in Fig.~\ref{fig:diffusion-resolution}(a). For this undirected Laplacian let the rows of $P$ be normalised, mutually orthogonal indicators of disjoint blocks. On the nonconstant modes $\mathcal L^\dagger$ weights each eigenmode by its relaxation time, the reciprocal of its positive eigenvalue. Projecting this response onto the block coordinates gives one effective operator,
\begin{equation}
 \mathcal L_{\rm block}=(P\mathcal L^\dagger P^\top)^\dagger .
 \label{eq:block-projected-relaxation}
\end{equation}
For a connected undirected graph, write $\lambda_2(\mathcal L)$ for the slowest nonzero relaxation rate. Since $PP^\top=I$, the largest eigenvalue of $P\mathcal L^\dagger P^\top$ is at most $1/\lambda_2(\mathcal L)$. With the constant mode retained by the block partition, inversion on the nonconstant block modes gives $\lambda_2(\mathcal L_{\rm block})\ge\lambda_2(\mathcal L)$ for two or more blocks. This comparison concerns the time taken for a distinction to relax within the graph. The natural feedback experiment suggested by the main text is to let a token readout revise a parameter held between observations. Then the relevant spectrum is that of the complete return in Eq.~\eqref{eq:general-spectral-return}, just as it is for the population census.

\section{Simulation Details}
\label{app:simulations}

\subsection*{Parameter Choices}
Figure~\ref{fig:scalar_phase} makes two comparisons. The contour predicts where a small perturbation stops decaying, as calculated from the multipliers of Eq.~\eqref{eq:scalar-census-jacobian}. To check it independently, we iterate the nonlinear map and measure the mismatch that persists after the transient. We feed back the expected signed majority from Eq.~\eqref{eq:majority-tail}, and treat sampled verdicts separately (Appendix~\ref{app:stochastic_validation}). Use $\tau$ as the time unit. The reference run holds the mandate for $T=\tau$, takes $b=1.2$, $s=0.4$ and $\beta=0.6$, and starts from $(x_0,\theta_0)=(0.1,0.15)$. At step size $\alpha=0.35$, the continuous large-population estimate in Eq.~\eqref{eq:critical-size} is $N_c\simeq138.25$. The exact majority gain $\alpha\beta a_N$ first exceeds $K_c(T)$ at the odd population size $N=139$.

\subsection*{Simulation Procedure}
That odd-size scan locates the onset but cannot approach it continuously. To measure the local loss law, vary $\alpha$ instead at fixed $N$ and use the final 2,000 of 24,000 iterates to average the loss after the initial transient. At $K/K_c-1=10^{-3}$ the ratios to the local loss in Eq.~\eqref{eq:scalar-local-loss} were $1.00094,\ 1.00091,\ 1.00090$ for $N=15,71,251$. Further from threshold the $O(e^5)$ terms of Eq.~\eqref{eq:majority-cubic} enter and the curves depart from the leading prediction.

Two checks do not use the Gaussian representative. For $N=3,\ \beta=1,\ r=1/2,\ b=3/2,\ s=1/2$, direct iteration of the mean response of the ternary majority in Eq.~\eqref{eq:ternary-block} gives $\overline{\mathcal J}/(\alpha-1)=1.50008$ at $\alpha=1.001$, against the limiting value $3/2$. Next replace $\tanh(\beta e)$ by $(2/\pi)\arctan(\pi\beta e/2)$. These laws have the same linear slope but different cubic terms. Repeating the experiment at $N=15$, with the appropriate cubic in each denominator, gives normalised losses $1.00188$ and $1.00224$ at relative excess gain $0.002$. The exponent survives without requiring the amplitudes to be equal.

\subsection*{Hierarchical Simulations}
For depth $L=1,2,3,4$ we vary the bottom block span $n$ over odd integers $n\ge3$ such that $N=n3^{L-1}\le731$. At each census we apply the binomial mean [Eq.~\eqref{eq:majority-tail} with $N=n$] to $\tanh(\beta e_k)$, then apply the ternary polynomial $(3m-m^3)/2$ another $L-1$ times (Appendix~\ref{app:hierarchy}). For Fig.~\ref{fig:hierarchy_postpones} we use the scalar parameters above at $\alpha=0.35$. The gain predicts first supercritical populations of $139,183,243,351$ in order of depth. Separately we average $e_k^2$ over the last 2,000 of 18,000 updates. At each depth the first mean above $10^{-5}$ occurs at the predicted size. In the separate static design, extra stages with transmission $a=0.8$ [Eq.~\eqref{eq:layer_attenuation}] may be inserted without changing $N$. Its stability condition and the delivery requirement $H=4$ are evaluated separately in Fig.~\ref{fig:phases}. Placing the transport delays inside the loop instead gives the delayed spectrum of Eq.~\eqref{eq:hier-delay-polynomial}.

\subsection*{Checks on the predictions}
To compare the spectral prediction with the measured onset in Fig.~\ref{fig:scalar_phase}(c), we first calculated both multipliers for odd $N$ from 3 to 401 at 100 periods in $0.18\le T/\tau\le3.1$. At every period the first odd population size with $\rho(J)\ge1$ matched the finite-majority gain criterion. We then located onset without using the Jacobian: after 5,000 updates, average $e_k^2$ over the final 500 and select the first odd population size for which it exceeds $10^{-4}$. At $T/\tau=0.5,1,2$ the two procedures gave the same sizes, $117,139,179$, respectively (Fig.~\ref{fig:scalar_phase}c).

For each response law the loss is measured from the endpoint mismatch in Eq.~\eqref{eq:scalar-mismatch}, before updating the parameter. The reference in Fig.~\ref{fig:scalar_phase}(a) is $(K-K_c)/(\alpha c_N)$ at the same $N$ and step size [Eq.~\eqref{eq:scalar-local-loss}]. For the leaky cascade the simulated impulse response and the transfer function in Eq.~\eqref{eq:cascade-transfer} differed by less than $10^{-14}$ on the tested frequency grid.

\section{Stochastic Finite-$N$ Validation}
\label{app:stochastic_validation}

The iterations in Appendix~\ref{app:simulations} use the expected sign from Eq.~\eqref{eq:majority-tail} at each census. A finite population instead supplies a sampled verdict. We distinguish the number $N$ of binary inputs in each majority from the number $M$ of conditionally independent verdicts averaged before the mandate changes. Each verdict uses its own $N$ inputs, drawn independently at the same mismatch, so one census samples $MN$ responses and $M$ is not limited by $N$. Figure~\ref{fig:empirical_signatures} tests how that sampling changes the onset of alternating mismatch.

\subsection*{Microscopic Model}
At zero mismatch the fraction of positive responses fluctuates on the scale $N^{-1/2}$, while the majority of an odd number of inputs returns either sign with equal probability. At mismatch $e$, the positive-majority probability is $q_N(e)=[1+R_N(e)]/2$ [Eq.~\eqref{eq:majority-tail}], distinct from the single-agent probability $p_+(e)$ in Eq.~\eqref{eq:scalar-sign-probability}. The variance of the aggregate sign is $1-R_N^2$. The conditional variance of the parameter increment for an average of $M$ independent aggregate signs is
\begin{equation}
 \operatorname{Var}(\theta_{k+1}-\theta_k\mid e_k)
 =\frac{\alpha^2}{M}\big[1-R_N(e_k)^2\big].
 \label{eq:sampled-increment-variance}
\end{equation}
Averaging more independent aggregates suppresses this source of noise even at zero mismatch, while adding more inputs to one aggregate does not.

\paragraph{The noisy flip.}

Keep $N$ fixed and vary $\alpha$ through $\alpha_c=K_c/R_N'(0)$. Write $\epsilon=K_N/K_c-1$. The critical eigenvalue of the census Jacobian $J$ [Eq.~\eqref{eq:scalar-census-jacobian}] is $-1$ at $\epsilon=0$ and moves as $-1-\psi\epsilon+O(\epsilon^2)$, where $\psi=2(1+r)/(1+\lambda_2)>0$ and $\lambda_2$ is the other multiplier (Appendix~\ref{app:scalar_boundary}). Use the alternating response $d$ and reference shortfall $D=b-d$ of Eq.~\eqref{eq:flip-shortfall}. The critical right eigenvector of Eq.~\eqref{eq:scalar-census-jacobian} is $\mathbf v=(-d,1)^\top$. If $\boldsymbol{\ell}^\top$ is the left eigenvector normalised by $\boldsymbol{\ell}^\top\mathbf v=1$, the projected mismatch $\tilde e_k=-D\boldsymbol{\ell}^\top(x_k,\theta_k)^\top$ equals $e_k$ on that eigendirection. The factor $-D$ follows from sampling the state at the end of the hold. The noise enters only through the mandate component of this projection.

Taking out the sign alternation with $w_k=(-1)^k \tilde e_k$ leaves a slowly varying amplitude. Keeping the leading odd nonlinearity and the leading variance at zero mismatch gives
\begin{equation}
 \begin{gathered}
 w_{k+1}-w_k \simeq \psi\epsilon w_k-\tilde c\,w_k^3+\frac{\sigma}{\sqrt M}\eta_k,\\
 \tilde c=\frac{\psi\alpha_c c_N}{K_c},\qquad
 \sigma=D|\ell_\theta|\alpha_c .
 \end{gathered}
 \label{eq:noisy-flip-step}
\end{equation}
Here $\eta_k$ denotes a centred noise of leading variance one, and $\ell_\theta$ is the scalar mandate component of $\boldsymbol{\ell}$. The cubic comes from projecting $-\alpha_c c_N e_k^3$ in the mandate update [Eqs.~\eqref{eq:scalar-revision} and~\eqref{eq:majority-cubic}] and using $\psi=K_c D\ell_\theta$. Without noise the positive branch satisfies $w^2=\psi\epsilon/\tilde c$, recovering the local squared mismatch in Eq.~\eqref{eq:scalar-local-loss}.

On the slow time scale approximating the increments in Eq.~\eqref{eq:noisy-flip-step} by drift and diffusion with zero stationary probability current gives a density proportional to $\exp[M(\psi\epsilon w^2-\tilde c\,w^4/2)/\sigma^2]$. Put
\begin{equation}
 w_0=\left(\frac{\sigma^2}{\tilde cM}\right)^{1/4},
 \qquad u=\frac{\psi\epsilon\sqrt M}{\sigma\sqrt{\tilde c}}.
 \label{eq:noisy-flip-scales}
\end{equation}
The density in the coordinate $y=w/w_0$ is then proportional to $\exp(uy^2-y^4/2)$. Figure~\ref{fig:empirical_signatures}(b) compares its root mean square with that of the actual mismatch, without fitting either axis. At $u=0$ the predicted root mean square is $w_0[\,\sqrt 2\,\Gamma(3/4)/\Gamma(1/4)\,]^{1/2}$. Here $\Gamma$ denotes the gamma function. The quartic density changes shape through $u=0$, but its second moment remains smooth there. The reduction holds for large $M$ and small $|\epsilon|$. At small $M$ the same calculation can be carried out with the exact binomial transition kernel behind Eq.~\eqref{eq:sampled-increment-variance} in place of its diffusion limit.

\paragraph{Sampling the rounded boundary.}

For Fig.~\ref{fig:empirical_signatures}(a) we used $T/\tau=1$, $\alpha=0.35$, $b=1.2$, $s=0.4$ and $\beta=0.6$. Each pair $(N,M)$ has 80 independent trajectories with 6,144 censuses discarded and 8,192 retained. The grid uses odd $N$ from 55 to 327 with additional sizes around 139. For panel (b) we hold $N=139$, take 22 equally spaced $u$ values from $-3$ to $4$ and vary $\alpha$. We use 96 paths per setting, discarding 8,192 censuses and retaining 16,384. Panel (c) uses a separate ensemble of 112 paths per setting, with 16,384 discarded and 32,768 retained. At each step we draw the binomial count conditional on the current mismatch. All paths enter the averages. Each path starts from $x_0=0.1$ with an initial mandate offset uniform on $[-\alpha/M,\alpha/M]$ about $0.15$, which covers one period of the two-census mandate lattice.

The pointwise intervals use the standard error of independent path means of $e_k^2$ and are then transformed to root mean square. To bracket a crossing in panel (a), we use the sampled sizes on either side at which the difference of mean squared mismatches has opposite signs and exceeds $1.96$ combined standard errors in magnitude. The gain width in the local sweep is the first positive $\epsilon$ at which the measured RMS reaches $1.4$ times its value at $\epsilon=0$, interpolated between gain samples. Fitting its logarithm against $\log M$ at the four largest $M$ gives a slope of $-0.496$. Fitting the longer-run amplitudes at zero detuning over the same $M$ gives $-0.251$ (95\% path-bootstrap interval $[-0.253,-0.249]$). At zero detuning, the RMS estimates from the two halves of the longer runs differ by at most $0.43\%$, and the estimates from the two independent ensembles differ by less than $0.4\%$. The mean square of $(e_{k+1}-e_k)/2$ accounts for more than $96\%$ of the raw second moment at the largest four $M$ over tested points with $|u|\le2$, so the variance sits in the alternating component.

\end{document}